\documentclass[aps,showpacs,nofootinbib]{revtex4-1}
\usepackage{amsmath}
\usepackage{amsfonts}
\usepackage{amssymb}
\usepackage{graphicx}
\usepackage{float}
\usepackage{epsfig}
\usepackage{hyperref}
\usepackage[section]{placeins}
\newcommand{\be}{\begin{eqnarray}}
\newcommand{\ee}{\end{eqnarray}}
\newcommand\del{\partial}
\newcommand{\nn}{\nonumber }

\newcommand{\Str}{{\rm Str}\,}
\newcommand{\Sdet}{{\rm Sdet}\,}
\newcommand{\bmat}{\left ( \begin{array}{cc} 
}
\newcommand{\emat}{\end{array} \right )}

 \newcommand{\Ort}{{\rm O}}
\newcommand{\UOSp}{{\rm UOSp}}

\newcommand{\USp}{{\rm Sp}}
\newcommand{\U}{{\rm U}}
\newcommand{\SU}{{\rm SU}}
\newcommand{\Gl}{{\rm Gl}}
\newcommand{\CSE}{{\rm CSE}}
\newcommand{\GOE}{{\rm GOE}}
\newcommand{\chiGOE}{{\rm chGOE}}
\newcommand{\Self}{{\rm Self}}
\newcommand{\Sym}{{\rm Sym}}
\newcommand{\sign}{{\rm sign}}

\newcommand{\Pf}{\textrm{Pf}}

\newcommand{\tr}{\textrm{tr}\,}

\newcommand{\diag}{\textrm{diag}}

\newcommand{\eins}{\leavevmode\hbox{\small1\kern-3.8pt\normalsize1}}
\newcommand{\vect}{\left ( \begin{array}{c}}
\newcommand{\evect}{ \end{array} \right )}

\begin{document}
 \title{Dirac Spectrum of the Wilson Dirac Operator for QCD with Two Colors}
\author{Mario Kieburg$^{1,2}$}
\email{mkieburg@physik.uni-bielefeld.de}
\author{Jacobus J. M. Verbaarschot$^{1}$}
\email{jacobus.verbaarschot@stonybrook.edu}
\author{Savvas Zafeiropoulos$^{1,3}$}
\email{zafeiro@th.physik.uni-frankfurt.de}
\affiliation{$^{1}$ Department of Physics and Astronomy, Stony Brook University, Stony Brook, NY 11794-3800, USA}
\affiliation{$^{2}$ Fakult\"at f\"ur Physik, Universit\"at Bielefeld, Postfach 100131, 33501 Bielefeld, Germany}
\affiliation{$^{3}$Institut f\"ur Theoretische Physik, Goethe-Universit\"at Frankfurt,
Max-von-Laue-Str.~1, 60438 Frankfurt am Main, Germany}


\begin{abstract}
We study the lattice artefacts of the Wilson Dirac operator for 
QCD with two colors and  fermions in the fundamental representation from the viewpoint of chiral perturbation theory. These effects are studied with the help of the following spectral observables: the level density of the Hermitian Wilson Dirac operator, the distribution of chirality over the real eigenvalues, and the chiral condensate for the quenched as well as for the unquenched theory. We provide analytical expressions for all these quantities.
 Moreover we derive constraints for the level density 
of the real eigenvalues of the non-Hermitian Wilson Dirac operator and the number of
 additional real modes. The latter is a good measure for the strength 
of lattice artefacts. All computations are confirmed by 
Monte Carlo simulations of the corresponding random matrix 
theory which agrees with chiral perturbation theory of 
two color QCD with Wilson fermions.
\end{abstract}

\maketitle

\section{Introduction}\label{sec:intro}

The breaking of chiral symmetry is an  essential feature of the 
vacuum structure of QCD. It is a  non-perturbative phenomenon that can
only be studied from first principles by means of lattice QCD simulations.
Because of the Nielsen-Ninomya theorem, exact continuum type chiral symmetry on the lattice
necessarily  implies the presence of doublers which have to be eliminated. Two
widely used methods to do so are staggered fermions, which break the
flavor symmetry but preserve a ``chiral'' symmetry,  
and Wilson fermions which preserve the flavor symmetry but 
 explicitly break the chiral symmetry.
 
 An important ingredient of 
the vacuum structure is the topology of the gauge field configurations and its
relation to the zero modes of the Dirac operator according to the 
Atiyah-Singer index theorem. Strictly speaking there is no 
topology at  non-zero lattice spacing.
For the Wilson Dirac operator one can obtain the topological charge by the spectral flow in the quark mass \cite{flow}. Other ways to define the topological charge on the lattice is through the index of the overlap operator \cite{Hasenfratz}as well as through spectral projectors \cite{sp1, sp2}. One can equivalently employ purely gluonic methods where one measures a lattice discretized version of the volume integral of the topological charge density. The most common methods involve the gradient flow \cite{wilsonFlow}, cooling \cite{cool1, cool2, cool3, cool4} and APE/HYP smearing \cite{ape1, ape2}. Of course all these methods agree up to cut-off effects but it is only the spectral flow and the index of the overlap operator that give an integer result.
We refer the reader to \cite{KCproc} for a critical comparison of all methods. 

The order parameter for chiral symmetry breaking is the chiral condensate which,
according to the Banks-Casher formula, is directly related to
 the spectral density of the smallest Dirac eigenvalues. It is therefore
important to have a detailed understanding of discretization effects on
the Dirac spectrum and in particular on the topological zero modes. This
problem was studied by means of chiral Lagrangians 
for the Wilson Dirac operator for 
QCD with three or more colors in the fundamental
representation \cite{DSV10,ADSV10}. Using mean field theory it was shown that the gap of 
the Hermitian Wilson Dirac operator closes when entering the Aoki phase.
The microscopic Dirac spectrum was evaluated in great detail by means 
of a supersymmetric extension of the chiral Lagrangian and exploiting its equivalence with
 chiral random matrix theory~\cite{AkeNag,Kie}. One of the main results  is that   the Gaussian broadening
of the topological zero modes scales as $ \sqrt {V}\widetilde{a}=a$ for small values of the lattice spacing $a$ which was first observed
by lattice simulations \cite{Luscher}. In the thermodynamic limit  the real eigenvalues develop a band with a width proportional to $VW_8\widetilde{a}^2=a^2$.
A second important result is that the first order scenario \cite{ShaSin98}
can only occur
in  the presence of dynamical quarks while in the quenched case 
at non-zero lattice spacing we necessarily
have a transition to the Aoki phase when approaching the chiral 
limit \cite{KSV}. 
We have also shown  that the low-energy constants can be determined in a simple way by
the properties of the Dirac spectrum \cite{KVZ-prl}.

In the present article we consider the spectrum of the Wilson Dirac operator for QCD 
with quarks in the fundamental representation of $\SU(2)$. This theory reaches 
the conformal window for a smaller number of flavors as compared to three color QCD \cite{sannino,Ohki:2010sr} and is of
interest for modeling physics beyond the standard model, see for example two interesting reviews by Rummukainen and Kuti \cite{kari,kuti}. 
What distinguishes this theory from QCD with
three or more colors is that $\SU(2)$ is pseudo-real, and so it is possible 
to construct a gauge field independent basis for which the Dirac operator
becomes real. For QCD at non-zero chemical potential this gives rise to 
a fermion determinant that is real so that the theory can be simulated
for an even number of pairwise degenerate quark flavors, and  is 
frequently studied as
a model for QCD at non-zero chemical potential \cite{adam,KSTVZ, Hands, JOA}.  
More physically, this implies that quarks and conjugate quarks
are in the same flavor multiplet resulting in a doubling of the flavor
symmetry group.
We expect that the 
discretization effects for two colors  are qualitatively the
same as for QCD with three or more colors. The main difference is
the  repulsion between the eigenvalues of the Dirac operator which is linear for small spacings.

We compute the microscopic spectral density $\rho_5$ of the Hermitian Wilson
Dirac operator starting
from a supersymmetric 
extension  of the chiral Lagrangian for 
two color Wilson fermions \cite{sannino-su2} in the epsilon regime. 
The calculation
is much more complicated than for QCD with three colors, and it is not clear
whether the density of the complex and real eigenvalues of the Wilson Dirac
operator can be obtained analytically. Moreover we compute  the distribution $\rho_\chi$ of chirality over the real spectrum of the non-Hermitian Wilson-Dirac operator and the mass dependence of the level density $\rho_5$ at the origin. Both quantities provide lower and upper
bounds for the level density $\rho_{\rm real}$ of the real eigenvalues of the non-Hermitian Wilson Dirac operator. We also consider the chiral condensate in the theory with and without dynamical quarks.

Before calculating the observables mentioned above we briefly recall the properties of the Wilson Dirac operator for two colors and the definitions of the spectral observables in Sec.~\ref{sec:object}. In Sec.~\ref{sec:chi-Lag} we discuss the chiral Lagrangian for fermionic and bosonic  quarks.
The supersymmetric partition function for the quenched theory is evaluated and discussed in Sec.~\ref{sec:que-part}. Thereby we consider the continuum limit as well as the limit of a very coarse lattice and the related thermodynamic limit. The exact results and the detailed discussions of the single spectral observables are summarized in Sec.~\ref{sec:spectrum}. Concluding remarks are made in Sec.~\ref{sec:conclusio}. In the appendices we briefly recall some properties of Bessel functions, present the detailed computation of the quenched partition function, and briefly rederive some spectral observables of continuum QCD with two colors and the Gaussian orthogonal ensemble.
 
Some of the results that appear in this article were first presented at the conference LATTICE 2012 \cite{lat12}.

\section{QCD with two Colors}\label{sec:object}

In subsection~\ref{sec:Dirac-op} we briefly recall the properties of the four dimensional QCD Dirac operator for the two color theory and fermions in the fundamental representation. Additionally we explain how chiral perturbation theory and the corresponding symmetry breaking pattern result from these properties. The variables we consider will be introduced in subsection~\ref{sec:observe}.

\subsection{Dirac Operator for QCD with Two Colors}\label{sec:Dirac-op}

For four dimensional QCD with two colors, the massless Dirac operator $D$ 
anti-commutes with $\gamma_5$ implying chiral symmetry. Moreover, $D$ commutes with an anti-unitary operator,
\cite{V}
\be\label{AUS}
[D,K\tau_2 C] = 0,
\ee
where $K$ is the complex conjugation operator, $C\equiv \gamma_2\gamma_4$ 
the charge conjugation matrix, and
$\tau_2$ the second Pauli matrix acting in color space. Because $(K\tau_2 C)^2 =1$ it can be shown
that there exists a gauge independent basis for which the matrix elements of
$D$ are real. In the Dyson classification of random matrix ensembles this symmetry
corresponds to  $\beta_{\rm D} = 1$.

Let us consider $N_{\rm f}$ quarks with a diagonal  mass matrix $\widetilde{m}=\diag(\widetilde{m}_1,\ldots,\widetilde{m}_{N_{\rm f}})$. Then the fermionic part of the Euclidean  Lagrangian is given by
\be\label{Lag}
{\cal L}=\vect \psi_1^* \\ \psi_2^* \evect^T \bmat \eins\otimes\widetilde{m} & i\sigma_\mu d_\mu\otimes\eins_{N_{\rm f}} \\
i(\sigma_\mu d_\mu)^\dagger\otimes\eins_{N_{\rm f}} & \eins\otimes\widetilde{m} \emat \vect \psi_1 \\ \psi_2 \evect,
\ee
where
\be
d_\mu = \del_\mu +i A_\mu
\ee
is the covariant derivative, $\sigma_4=-i\eins_2$ and $\sigma_a$ are the two-dimensional unit matrix and the three Pauli matrices, respectively, acting in Dirac space. The transformation property of the vector field under complex conjugation, $A_\mu^*=-\tau_2A_\mu\tau_2$ is the reason for the anti-unitary symmetry~\eqref{AUS}. It also implies $(\sigma_\mu d_\mu)^*=-\sigma_2\tau_2\sigma_\mu d_\mu\sigma_2\tau_2$. This symmetry can be used to rewrite the two terms of the Lagrangian~\eqref{Lag},
\cite{adam}
\be
{\cal L} = {\cal L}_0 +{\cal L}_m ,
\ee
 as
\be
\label{L0}
{\cal L}_0= \vect \psi_1^* \\ \sigma_2\tau_2\psi_1  \evect^T
\bmat 0 & i\sigma_\mu d_\mu \\ i\sigma_\mu d_\mu & 0 \emat\otimes\eins_{N_{\rm f}}
 \vect \sigma_2\tau_2 \psi_2^* \\ \psi_2 \evect
=\vect \psi_1^* \\ \sigma_2\tau_2\psi_1  \evect ^T
i\sigma_\mu d_\mu\otimes\eins_{2N_{\rm f}}
\vect  \psi_2  \\ \sigma_2\tau_2 \psi_2^* \evect 
\ee
and
\be
{\cal L}_m &=&\frac {1}{2} \vect \psi_1^* \\ \sigma_2\tau_2\psi_1  \evect ^T \sigma_2\tau_2
\otimes\bmat 0 & \widetilde{m} \\ -\widetilde{m} & 0 \emat 
\vect \psi_1^* \\ \sigma_2\tau_2\psi_1  \evect + \frac 12 \vect \psi_2 \\ \sigma_2\tau_2\psi_2^*  \evect^T \sigma_2\tau_2
\otimes\bmat 0 & \widetilde{m} \\ -\widetilde{m} & 0 \emat 
\vect \psi_2 \\  \sigma_2\tau_2 \psi_2^*\evect.
\label{massterm}
\ee
Equation~\eqref{L0} implies
that the flavor symmetry group is enhanced to $\U(2N_{\rm f})$ given by the transformation
\begin{equation}\label{ferm-trafo}
\vect  \psi_1^*  \\ \sigma_2\tau_2 \psi_1 \evect \rightarrow \eins\otimes U^*\vect  \psi_1^*  \\ \sigma_2\tau_2 \psi_1 \evect,\ \vect  \psi_2  \\ \sigma_2\tau_2 \psi_2^* \evect \rightarrow \eins\otimes U\vect  \psi_2  \\ \sigma_2\tau_2 \psi_2^* \evect
\end{equation}
with $U\in\U(2N_{\rm f})$. The flavor
symmetry is spontaneously broken by the formation of a chiral condensate
which is given by the mass derivative of the partition function. Since
the flavor symmetry group does not act on $\sigma_2$ and $\tau_2$ in Eq.~\eqref{massterm}, the symmetry is broken to  $\USp(2N_{\rm f})$~
\footnote{The unitary symplectic group $\USp(2N_{\rm f})$ is the section of the unitary group and the symplectic group which is sometimes also denoted as ${\rm USp}(2N_{\rm f})$. It should not be confused with the symplectic group itself which is non-compact and is sometimes denoted by $\USp(2N_{\rm f})$, too.} resulting in the symmetry breaking
pattern $\U(2N_{\rm f}) \to \USp(2N_{\rm f})$ \cite{Peskin,shifman3}. After taking into account
the axial anomaly from configurations with a non-trivial topological charge $\nu\neq0$, the Goldstone manifold is given by $\SU(2N_{\rm f})/\USp(2N_{\rm f})$ with $2N_{\rm f}^2-N_{\rm f}-1$ Goldstone bosons.

When introducing the lattice spacing $\widetilde{a}$ in the Dirac operator $D_{\rm W}=D-\widetilde{a}\Delta$ one has to choose one of many fermionic discretizations. We are interested in the lattice artefacts of the Wilson Dirac operator,
where the additional term in 
  the Lagrangian~\eqref{Lag} is given by~\cite{Wilson}
\begin{eqnarray}\label{Wilsonterm}
{\cal L}_{\rm W} &=&-\widetilde{a}\vect \psi_1^* \\ \psi_2^* \evect^T \bmat \Delta & 0 \\ 0 & \Delta \emat\otimes\eins_{N_{\rm f}}  \vect \psi_1 \\ \psi_2 \evect\\
&=&-\frac{\widetilde{a}}{2}\left[ \vect \psi_1^* \\ \sigma_2\tau_2\psi_1  \evect ^T \Delta\sigma_2\tau_2\otimes\bmat 0 & \eins_{N_{\rm f}} \\ -\eins_{N_{\rm f}} & 0 \emat 
\vect \psi_1^* \\ \sigma_2\tau_2\psi_1  \evect + \vect \psi_2^* \\ \sigma_2\tau_2\psi_2  \evect^T 
\Delta\sigma_2\tau_2\otimes\bmat 0 & \eins_{N_{\rm f}} \\ -\eins_{N_{\rm f}} & 0 \emat 
\vect \psi_2 \\  \sigma_2\tau_2 \psi_2^*\evect\right].\nonumber
\end{eqnarray}
Note that the four-dimensional Laplacian $\Delta=d_\mu^2$ satisfies the symmetries $\Delta^*=\tau_2\Delta\tau_2$ and $[\Delta,\sigma_2]=0$. Hence the Wilson term~\eqref{Wilsonterm} transforms in the same way as the mass term~\eqref{massterm}.

\subsection{Spectral Observables} \label{sec:observe}

The central object of our studies is the partially quenched partition function
\begin{equation}\label{part-SUSY-start}
Z_\nu(\widetilde{M},\widetilde{m},\widetilde{m}',\widetilde{x}_0,\widetilde{x}_1,\widetilde{a})=\left\langle\frac{\det(-\widetilde{a}\Delta+\gamma_\mu d_\mu+\widetilde{m}\eins+ \widetilde{x}_0\gamma_5)}{\det(-\widetilde{a}\Delta+\gamma_\mu d_\mu+\widetilde{m}'\eins+ \widetilde{x}_1\gamma_5)}\right\rangle_{N_{\rm f}}, 
\end{equation}
where $\left\langle.\right\rangle_{N_{\rm f}}$ is the average of two-color QCD with $N_{\rm f}$ dynamical quarks. The masses of the $N_{\rm f}$ dynamical quarks are encoded in the matrix $\widetilde{M}$ which may have also non-diagonal elements. The axial mass $ \widetilde{x}_1$ has a non-vanishing imaginary part $i \varepsilon$ to guarantee convergence of the integrals.

The generating function~\eqref{part-SUSY-start} allows us to derive two Green functions depending on whether we differentiate with respect to $\widetilde{m}'$ or $\widetilde{x}_1$. The differentiation in $\widetilde{x}_1$ yields the Green function corresponding to the Hermitian Wilson Dirac operator $D_5=\gamma_5 D_{\rm W}$, i.e.
\be\label{G5-def}
G_5(\widetilde{M},\widetilde{m},\widetilde{\lambda},\widetilde{a}) =\left.\partial_{\widetilde{x}_1}Z_\nu(\widetilde{M},\widetilde{m},\widetilde{m}',\widetilde{x}_0,\widetilde{x}_1,\widetilde{a})\right|_{\substack{\widetilde{m}=\widetilde{m}'\\ \widetilde{x}_0=\widetilde{x}_1=\widetilde{\lambda}}}= \frac{1}{V} \left\langle{\tr} \frac{1}{D_5+\widetilde{m}\gamma_5+ \widetilde{\lambda}\eins}\right\rangle_{N_{\rm f}}.
\ee
Its discontinuity yields the level density of $D_5$,
\be\label{rho5-def}
\rho_5(\widetilde{M},\widetilde{m},\widetilde{\lambda},\widetilde{a})= \frac{1}{\pi}\lim_{\varepsilon\to0} {\rm Im}\, G_5(\widetilde{M},\widetilde{m},\widetilde{\lambda}+i\varepsilon,\widetilde{a})= \frac{1}{V} \left\langle\sum_k \delta[\lambda^{5}_k(\widetilde{m})-\widetilde{\lambda}]\right\rangle_{N_{\rm f}},
\ee
where $\lambda^{5}_k(\widetilde{m})$ is an eigenvalue of $D_5+\widetilde{m}\gamma_5$ with respect to the eigenvector $|k\rangle$.

The level density $\rho_5$ satisfies  an inequality with respect to the level density $\rho_{\rm real}$ of the real eigenvalues of the non-Hermitian Wilson Dirac operator $D_{\rm W}$. This can be seen by considering the level density~\eqref{rho5-def} at $\widetilde{\lambda}=0$,
\be\label{rho5-a}
\rho_5(\widetilde{M},\widetilde{m},\widetilde{\lambda}=0,\widetilde{a})= \frac{1}{V} \left\langle\sum_k \delta[\lambda^{5}_k(\widetilde{m})]\right\rangle_{N_{\rm f}}.
\ee
The corresponding eigenvalue equation reads
\begin{equation}\label{eigenval-eq}
(D_5+\widetilde{m}\gamma_5)|k\rangle=\lambda^{5}_k(\widetilde{m})|k\rangle\ \Longleftrightarrow\ (D_{\rm W}-\lambda^{5}_k(\widetilde{m})\gamma_5)|k\rangle=-\widetilde{m}|k\rangle.
\end{equation}
Hence the value $-\widetilde{m}$ is a real eigenvalue of $D_{\rm W}$ when $\lambda^{5}_k(\widetilde{m})=0$ which we denote by $\lambda_k^{\rm W}$. Moreover taking the scalar product with the bra vector $\langle k|$ in the first equality of Eq.~\eqref{eigenval-eq} and differentiating with respect to the mass $\widetilde{m}$ yields
\begin{equation}\label{eigenval-eq-a}
\partial_{\widetilde{m}}\lambda^{5}_k(\widetilde{m})|_{\lambda^{5}_k(\widetilde{m})=0}=\left.\partial_{\widetilde{m}}\langle k|(D_5+\widetilde{m}\gamma_5)|k\rangle\right|_{\lambda^{5}_k(\widetilde{m})=0}=\langle k|\gamma_5|k\rangle.
\end{equation}
The derivative of the vectors $\langle k |$ and $|k\rangle$ drops out because of the eigenvalue equations $\langle k|(D_5+\widetilde{m}\gamma_5)=0$ and $(D_5+\widetilde{m}\gamma_5)|k\rangle=0$. The expectation value $\langle k|\gamma_5|k\rangle$ is the chirality of the vector $|k\rangle$. Thus the slope of the spectral flow with respect to the quark mass at its zeros gives the chiralities of the real modes~\cite{flow}.

Expressing the eigenvalues $\lambda^{5}_k(\widetilde{m})$ in terms of $\lambda_k^{\rm W}$ and plugging Eq.~\eqref{eigenval-eq-a} into Eq.~\eqref{rho5-a} we find the relation
\be\label{rho5-b}
\rho_5(\widetilde{M},\widetilde{m},\widetilde{\lambda}=0,\widetilde{a})= \frac{1}{V} \left\langle\sum_k \delta[\lambda^{5}_k(\widetilde{m})]\right\rangle_{N_{\rm f}}= \frac{1}{V}
\left \langle \sum_{\lambda^{\rm W}_k\in\mathbb{R}}\frac{\delta[\lambda^{\rm W}_k+\widetilde{m}]}{|\partial_{\widetilde{m}}\lambda^{5}_k(\widetilde{m})|_{\lambda^{5}_k(\widetilde{m})=0}| }\right\rangle_{N_{\rm f}}= \frac{1}{V}
\left \langle \sum_{\lambda^{\rm W}_k\in\mathbb{R}}\frac{\delta[\lambda^{\rm W}_k+\widetilde{m}]}{|\langle k|\gamma_5|k\rangle| }\right\rangle_{N_{\rm f}}.
\ee
Since the modulus of the chirality of a normalized vector, $\langle k|k \rangle=1$, is always less or equal to $1$ we obtain the inequality
\be\label{inequ-a}
\rho_5(\widetilde{M},\widetilde{m},\widetilde{\lambda}=0,\widetilde{a})=  \frac{1}{V}
\left \langle \sum_{\lambda^{\rm W}_k\in\mathbb{R}}\frac{\delta[\lambda^{\rm W}_k+\widetilde{m}]}{|\langle k|\gamma_5|k\rangle| }\right\rangle_{N_{\rm f}}\geq\frac{1}{V}
\left \langle \sum_{\lambda^{\rm W}_k\in\mathbb{R}}\delta[\lambda^{\rm W}_k+\widetilde{m}]\right\rangle_{N_{\rm f}}=\rho_{\rm real}(\widetilde{M},\widetilde{m},\widetilde{a}),
\ee
where the right hand side is indeed the level density of the real eigenvalues of $D_{\rm W}$.

The derivative of the partially quenched partition function with respect to the mass $\widetilde{m}'$ yields another Green function
\be\label{G-def}
G'(\widetilde{M},\widetilde{m},\widetilde{\lambda},\widetilde{a}) &=\left.\partial_{\widetilde{m}'}Z_\nu(\widetilde{M},\widetilde{m},\widetilde{m}',\widetilde{x}_0,\widetilde{x}_1,\widetilde{a})\right|_{\substack{\widetilde{m}=\widetilde{m}'\\ \widetilde{x}_0=\widetilde{x}_1=\widetilde{\lambda}}}= \displaystyle\frac{1}{V} \left\langle{\tr} \frac{1}{D_{\rm W}+\widetilde{m}\eins+ \widetilde{\lambda}\gamma_5}\right\rangle_{N_{\rm f}}.
\ee
Setting $\widetilde{\lambda}=i\varepsilon$ in the limit $\varepsilon\to0$ we can take the real part and the imaginary part.
The real part is equal to the mass-dependent chiral condensate
\be\label{sigma-def}
\Sigma(\widetilde{M},\widetilde{m},\widetilde{a}) =-\frac{1}{V}\left \langle\tr \frac {1}{D_{\rm W}+\widetilde{m}\eins}\right \rangle_{N_{\rm f}}.
\ee
In the case that $\widetilde{m}$ equals one of the eigenvalues $\widetilde{m}_1,\ldots,\widetilde{m}_{N_{\rm f}}$ of the mass matrix $\widetilde{M}$, e.g.\ say $\widetilde{m}=\widetilde{m}_1$, one does not need the partially quenched partition function~\eqref{part-SUSY-start}
 but the result can be immediately derived from the partition function
\begin{equation}\label{part-start}
Z_\nu^{N_{\rm f}}(\widetilde{M},\widetilde{a})=\left\langle\prod\limits_{j=1}^{N_{\rm f}}\det(D_{\rm W}+\widetilde{m}_j\eins)\right\rangle.
\end{equation}
Then the chiral condensate is given by
\be\label{sigma-def-unquenched}
\Sigma^{N_{\rm f}}(\widetilde{M},\widetilde{m}_1,\widetilde{a}) =-\frac{1}{V}\partial_{m_1}{\rm ln}\,Z_\nu^{N_{\rm f}}(\widetilde{M},\widetilde{a})=-\frac{1}{V}\left \langle\tr \frac {1}{D_{\rm W}+\widetilde{m}_1\eins}\right \rangle_{N_{\rm f}}.
\ee

The imaginary part of $G'$ yields the distribution of chirality over the real eigenvalues
\be
\rho_\chi(\widetilde{M},\widetilde{m},\widetilde{a})&=& -\frac{1}{\pi}\lim_{\varepsilon\to0} {\rm Im}\, G'(\widetilde{M},\widetilde{m},i\varepsilon,\widetilde{a})\nn\\
&=&\frac{1}{2\pi i V}\lim_{\varepsilon\to0} \left\langle{\tr} \left(\frac{1}{D_{\rm W}+\widetilde{m}\eins- i \varepsilon\gamma_5}-\frac{1}{D_{\rm W}+m\eins+  i \varepsilon\gamma_5}\right)\right\rangle_{N_{\rm f}}\nn\\
&=&\frac{1}{2\pi i V}\lim_{\varepsilon\to0} \left\langle{\tr} \gamma_5\left(\frac{1}{D_5+\widetilde{m}\gamma_5- i \varepsilon\eins}-\frac{1}{D_5+m\gamma_5+  i \varepsilon\eins}\right)\right\rangle_{N_{\rm f}}\nn\\
&=& \frac{1}{V} \left\langle\sum_k \delta[\lambda^{5}_k(\widetilde{m})]|\langle k|\gamma_5|k \rangle|\right\rangle_{N_{\rm f}}.\label{rhochi-def}
\ee
Here we employ the same relation between $\lambda^{5}_k$ and $\lambda^{\rm W}_k$ as for the level density $\rho_5$ such that we have
\be\label{rhochi-a}
\rho_\chi(\widetilde{M},\widetilde{m},\widetilde{a})= \frac{1}{V} \left\langle\sum_k \delta[\lambda^{\rm W}_k+\widetilde{m}]{\rm sign}(\langle k|\gamma_5|k \rangle)\right\rangle_{N_{\rm f}}.
\ee
This expression explains the name of this quantity since it is the average sign of chirality at a fixed real eigenvalue $\lambda_k^{\rm W}=-\widetilde{m}$. Understanding $\rho_\chi$ in this way has two implications.
First the distribution of chirality over the real eigenvalues is normalized with the index $\nu$,
\be\label{index-rhochi}
\int_\mathbb{R} d\widetilde{m} \rho_\chi(\widetilde{m}) = \sum_{\lambda^{\rm W}_k\in {\mathbb R}}  
{\rm sign}[\langle k |\gamma_5|k \rangle]=\nu.
\ee
We recall that the index $\nu$ is a topological invariant such that also this normalization is well-defined for each single configuration.
It counts the total number of spectral flow lines that start at $-\infty$ for $\widetilde{m} \to -\infty$ and end at $+\infty$ for $ \widetilde{m} \to +\infty$. Note that $\rho_\chi(\widetilde{m})$ does not have to be necessarily positive definite.

The second implication of Eq.~\eqref{rhochi-a} is another inequality with the level density of the real eigenvalues of $D_{\rm W}$,
\be\label{inequ-b}
|\rho_\chi(\widetilde{M},\widetilde{m},\widetilde{a})|=  \frac{1}{V} \left|\left\langle\sum_k \delta[\lambda^{\rm W}_k+\widetilde{m}]{\rm sign}(\langle k|\gamma_5|k \rangle)\right\rangle_{N_{\rm f}}\right|\leq\frac{1}{V}
\left \langle \sum_{\lambda^{\rm W}_k\in\mathbb{R}}\delta[\lambda^{\rm W}_k+\widetilde{m}]\right\rangle_{N_{\rm f}}=\rho_{\rm real}(\widetilde{M},\widetilde{m},\widetilde{a}).
\ee
Combining this inequality with Eq.~\eqref{inequ-a} we have an upper and lower bound for the level density $\rho_{\rm real}$,
\be 
\rho_\chi(\widetilde{M},\widetilde{m},\widetilde{a})\leq|\rho_\chi(\widetilde{M},\widetilde{m},\widetilde{a})| \leq \rho_{\rm real}(\widetilde{M},\widetilde{m},\widetilde{a}) \leq \rho_5(\widetilde{M},\widetilde{m},\widetilde{\lambda}=0,\widetilde{a}).
\label{inequ-c}
\ee
This inequality can be integrated over $\widetilde{m}$ such that we find an estimate for the average number of  additional real modes  for the set of configurations with a fixed index $\nu$ which is
\be  \label{inequ-d}
0\le N_{\rm add}=N_{\rm real} -|\nu |\le
\frac{1}{V}
\left \langle\sum_{\lambda^{\rm W}_k\in {\mathbb R}}  
\frac {1- \langle k |\gamma_5|k \rangle }{|\langle k |\gamma_5|k \rangle|}\right\rangle_{N_{\rm f}}.
\ee
These estimates encode the only information available about the level density $\rho_{\rm real}$ as long as there is no closed expression for this quantity. Such a closed expression was indeed derived for QCD with three colors in \cite{Kie,KVZ-prl} but seems to be much harder to obtain for the two color case. 

\section{Chiral Lagrangian and Partition Function}\label{sec:chi-Lag}

We consider the epsilon regime of lattice QCD. In this regime the quark mass $\widetilde{m}$, the level spacing $\widetilde{a}$, the momentum of the Goldstone bosons $p$, and the amplitude of the Goldstone bosons $\Pi$ scale with the volume $V$ as follows~\cite{GL_thermo, ONS1, ONS2}
\begin{equation}\label{scaling}
 \widetilde{m}\sim\widetilde{a}^2\sim p^4\sim\Pi^4\sim\frac{1}{V}.
\end{equation}
In this particular scaling limit the Goldstone bosons with momentum $p\neq0$ decouple from those with zero momentum such that it is legitimate to consider only the effective action of the latter. The corresponding chiral Lagrangian $\mathcal{L}_\chi$ is space-time independent. Hence the integral (sum) over the lattice volume $V$ can be performed yielding a prefactor of the chiral Lagrangian.

It is clear that the epsilon regime is not the most physical regime. 
However, the mass scale of the epsilon domain is the one
of lowest eigenvalues which are most sensitive to lattice artefacts.
Using the leading order epsilon domain chiral Lagrangian one can also 
determine the region   where the artificial phase structures like 
the Aoki-phase~\cite{ShaSin98} manifest themselves, 
see also subsection~\ref{sec:thermo-dyn}. Another important application
is the  measurement of  physical parameters such as the chiral condensate, the pion decay constant as well as the electroweak effective couplings from lattice simulations in the epsilon regime \cite{GiustiEpsilon1, GiustiEpsilon2, AokiEpsilon3}.

\subsection{Chiral Lagrangian for the Fermionic Quarks}\label{sec:ferm}

 The chiral Lagrangian for QCD with two colors follows from the requirements
that the effective partition function should have the same flavor transformation
properties as the QCD partition function in the full theory. Similar to three color QCD
the partition function in the epsilon domain is given by an integral over
the Goldstone manifold $\U(2N_{\rm f})/\USp(2N_{\rm f})$.
 A matrix $U\in \U(2N_{\rm f})/\USp(2N_{\rm f})$ is an antisymmetric unitary 
matrix parametrized by $U = V I V^T$ with $I$ defined by
\be\label{symp-unit}
I = \bmat 0 & \eins_{N_{\rm f}} \\ -\eins_{N_{\rm f}} & 0 \emat
\ee
The chiral partition function for a fixed topological charge $\nu$ is given by \cite{adam,sannino-su2,magnea-o,dalmazi-jv},
\begin{eqnarray}
Z_{\nu}^{N_{\rm f}}(M,a)&=&\int_{\U(2N_{\rm f})/\USp(2N_{\rm f})} d\mu({U}) {\det}^{\nu/2}{U}
 \exp\left[ \frac{ 1}{2} \tr (MI{U} + (MI)^\dagger {U}^{-1}) -a^2 \tr 
\left((I {U})^2 + (I {U})^{-2}\right)\right].
\label{zchi}
\ee
Since the integrand is a class function, the integration can be extended to
$\U(N_{\rm f})$ with the integration measure given by the Haar measure.
Without restriction of generality we can assume that $\nu\geq0$ is non-negative. To keep the notation simple we have introduced the dimensionless quantities,
\begin{equation}\label{dim-less}
 M=V\Sigma\widetilde{M}\ {\rm and}\ a^2=VW_8\widetilde{a}^2.
\end{equation}
Thus the low energy constants $\Sigma$ (chiral condensate) and $W_8$ are
 absorbed in these new quantities. Moreover we wish to recall that 
the low energy constant $W_8$ can be \textit{a priori} positive or 
negative such that the effective lattice spacing $a$ may also appear as 
an imaginary number even though the physical lattice spacing 
$\widetilde{a}$ is always real and positive. However, as is the case for fundamental Wilson quarks for QCD with three or more colors,
we find that  $W_8>0$ in the
case of two colors.  This is discussed  
in subsections~\ref{sec:bos} and~\ref{sec:partition}.
The general mass matrix $MI$ is antisymmetric, but we restrict ourselves
to a mass matrix such that
$M=(m+ix_0)\eins_{2N_{\rm f}}$ proportional to the identity matrix. Here we have to distinguish between the real quark mass $m$ and the real axial mass $x_0$. The latter is introduced as a source for generating particular observables, see subsection~\ref{sec:observe}.

The chiral Lagrangian in Eq.~\eqref{zchi} is not the most general lowest 
order Lagrangian in the epsilon regime. We can add the terms 
$VW_6\widetilde{a}^2\tr^2 I(U - U^{-1})$ and 
$VW_7\widetilde{a}^2\tr^2 I(U + U^{-1})$, which are of the same 
order as the other terms. As in the case of three color QCD these two 
terms have the same symmetry properties as the $W_8$ term, and they 
have to be included \textit{a priori}. However, based on large $N_{\rm c}$
arguments \cite{KL}, we expect that the strongest effect results from the 
$W_8$ term. Moreover the $W_6$ and $W_7$ term can be easily introduced 
by a convolution with a Gaussian in the quark mass $m$ and in 
the axial mass $x_0$, respectively, see \cite{KVZ-prl}. 
Therefore we stick to the partition function $Z_{\nu}^{N_{\rm f}}(M,a)$.

The coset of antisymmetric unitary matrices is equivalent to 
 the circular symplectic ensemble $\CSE(2N_{\rm f})$ which is the ensemble of
 self-dual unitary matrices. Matrices in this ensemble satisfy the
relation $U^T=-IUI$ and are related
to antisymmetric unitary matrices via the scaling $ U \rightarrow IU$. This set of matrices was first studied by Dyson~\cite{Dyson}.
Thus the eigenvalues of $ U$ are pure phases and are Kramers degenerate.
In terms of the  $\CSE(2N_{\rm f})$ the chiral partition function~\eqref{zchi} reads
\begin{eqnarray}
Z_{\nu}^{N_{\rm f}}(M,a)&=&\int_{\CSE(2N_{\rm f})} d\mu(U) {\det}^{\nu/2}U
 \exp\left[ \frac{ 1}{2} \tr (MU +M^\dagger U^{-1}) -a^2 \tr (U^2 + U^{-2})\right],
\label{zchi2}
\end{eqnarray}
 which may be more familiar to some readers. 
The measure $d\mu(U)$ on the Goldstone manifold $\CSE(2N_{\rm f})$ is the normalized Haar measure induced from the Haar measure on $\U(2N_{\rm f})$ and can
be extended to an integration over $\U(2N_f)$ as in case of the representation~(\ref{zchi}). 
The measure can also be expressed in terms of the flat measure $dU$ which is the product of differentials of all independent matrix entries, i.e. the measure is~\cite{hua}
\be
d\mu(U)\propto\frac {dU}{{\det}^{N_{\rm f}-1/2} U}.\label{fermionmeasure}
\ee
It is invariant under the group action $U\to IV^TIUV$ for any $V\in\U(2N_{\rm f})$ because the Jacobian of this transformation is $d(IV^TIUV)={\det}^{2N_{\rm f}-1} V dU$. When diagonalizing $U=S\diag(e^{i\varphi_1} ,\ldots ,e^{i\varphi_{N_{\rm f}}} )\otimes\eins_2 S^\dagger$ with $S\in\USp(2N_{\rm f})$ the flat measure
reads
\be 
dU \propto \prod_{1\leq k<l\leq N_{\rm f}} (e^{i\varphi_k} -e^{i\varphi_l})^4 d\mu(S) \prod_{k=1}^{N_{\rm f}}  de^{i\varphi_k} 
\ee
with $d\mu(S)$ the normalized Haar measure of the coset $\USp(2N_{\rm f})/\USp^{N_{\rm f}}(2)$.

To interpret the $a^2$ term in Eq.~\eqref{zchi2} as a diffusive process, see \cite{split-dyn} for three color QCD, it is quite convenient to  express also this term as a mass term convoluted with a Gaussian,
\begin{eqnarray}\label{Gauss-conv}
\exp\left[-a^2 \tr(U^2 + U^{-2})\right] &=&\exp\left[-a^2\tr(U - U^{-1})^2 -4N_{\rm f} a^2\right]\\
&=&C_{N_{\rm f}}e^{-4N_{\rm f} a^2}\int_{\Self(2N_{\rm f})} d\sigma \exp\left[-\frac{1}{16 a^2} \tr\sigma^2+\frac {i}{2}\tr\sigma(U-U^{-1})\right]\nonumber
\end{eqnarray}
with the constant $C_{N_{\rm f}}=2^{N_{\rm f}/2}(8a^2\pi)^{-N_{\rm f}(2N_{\rm f}-1)/2}$. The matrix $\sigma$ is self-dual and Hermitian $\sigma =\sigma^\dagger= -I \sigma^T I$ and the set of those matrices is denoted by $\Self(2N_{\rm f})$.
Thus the partition function~\eqref{zchi2} simplifies to
\be
Z_\nu^{N_{\rm f}}(M,a)&=& C_{N_{\rm f}}e^{-4N_{\rm f} a^2}\int_{\Self(2N_{\rm f})} d\sigma \int_{\CSE(2N_{\rm f})} d\mu(U) {\det}^{\nu/2}U
 \exp\left[ \frac{1}{2} \tr((M+i\sigma)U + (M+i\sigma)^\dagger U^{-1}) -\frac{1}{16 a^2}\tr\sigma^2 \right]\nonumber\\
 &=& C_{N_{\rm f}}e^{-4N_{\rm f} a^2}\int_{\Self(2N_{\rm f})} d\sigma
 \exp\left[-\frac{1}{16 a^2}\tr\sigma^2\right]Z_{\nu}^{N_{\rm f}}(M+i\sigma, a=0),
\label{z-sig-u}
\ee 
which is expressed in terms of the partition function at $a=0$, see~\cite{split-dyn,dalmazi-jv}. This expression has some interesting relations to the microscopic partition function of the chiral Gaussian orthogonal random matrix ensemble ($\chiGOE$) which is up to a factor the same as the partition function $Z_{\nu}^{N_{\rm f}}(M+i\sigma, a=0)$,
\begin{equation}\label{part-chiGOE}
Z_{\chiGOE}^{(\nu)}([M+i\sigma][M^\dagger-i\sigma])={\det}^{-\nu/2}(M^\dagger-i\sigma) Z_{\nu}(M+i\sigma, a=0),
\end{equation}
and of the Gaussian orthogonal random matrix ensemble ($\GOE$) of finite matrix dimension $\nu\times\nu$,
\begin{equation}\label{part-GOE}
Z_{\GOE}^{(\nu)}\left(\frac{M^\dagger}{4a}\right)=C_{N_{\rm f}}e^{-4N_{\rm f} a^2}\int_{\Self(2N_{\rm f})} d\sigma {\det}^{\nu/2}(M^\dagger-i\sigma)\exp\left[-\frac{1}{16 a^2}\tr\sigma^2\right].
\end{equation}
In subsection~\ref{sec:small-a} we will argue that in the continuum limit the partition function~\eqref{z-sig-u} factorizes into these two partition functions, i.e.
\begin{equation}
Z_\nu^{N_{\rm f}}(M,a)\overset{|a|\ll1}{=}Z_{\GOE}^{(\nu)}\left(\frac{M^\dagger}{4a}\right)Z_{\chiGOE}^{(\nu)}(MM^\dagger).
\end{equation}
Thus the $\nu$ eigenvalues of the finite dimensional $\GOE$ lie on the scale $a$ and shrink to a single Dirac delta function at the origin showing that they are the former $\nu$ zero modes of continuum QCD with two colors.

Since the matrix $\sigma$ is self-dual its eigenvalues are Kramers degenerate.  The flat measure $d\sigma$ can be rephrased in terms of these eigenvalues via a diagonalization $\sigma=SsS^\dagger =S {\rm diag} (s_1,\ldots, s_{N_{\rm f}})\otimes\eins_2S^\dagger$ with $S\in\USp(2N_{\rm f})$ yielding
\be
d\sigma\propto \prod_{1\leq k<l\leq N_{\rm f}} (s_k -s_l)^4 d\mu(S) \prod_{k=1}^{N_{\rm f}} d s_k.
\ee
When additionally assuming that the mass matrix $M=(m+ix_0)\eins_{2N_{\rm f}}$ is proportional to the identity matrix the diagonalizing matrix $S$ drops out in Eq.~\eqref{z-sig-u}. Note that $S$ can be absorbed by $U$ because $U$ is Haar distributed. Then the partition function~\eqref{z-sig-u} reduces to
\be
Z_\nu^{N_{\rm f}}(M,a)\propto\prod_{k=1}^{N_{\rm f}}\left( \int_{-\infty}^{\infty} ds_k (m+ix_0+is_k)^\nu \exp\left[-\frac{s_k^2}{8a^2}\right]\right)\prod_{1\leq k<l\leq N_{\rm f}} (s_k -s_l)^4 Z_{\chiGOE}^{(\nu)}\left([m+ix_0]\eins_{2N_{\rm f}}+is\right).
\label{zfermfinal}
\ee
Therefore the chiral partition function at finite lattice spacing can be understood as a convolution of each single quark mass with a Gaussian. Since these Gaussians are not completely statistical independent we can understand such a convolution as collective fluctuations as we have already seen those for  three color QCD, cf.~\cite{KSV,KVZ-prl}.

\subsection{Chiral Lagrangian for Bosonic Quarks}\label{sec:bos}

To find  the bosonic chiral Lagrangian, we start from the observation that it is possible to find a gauge independent basis in 
which the massive Hermitian Wilson Dirac operator of $N_{\rm b}$ bosonic flavors, $D_5=\gamma_5D_{\rm W}=\gamma_5\gamma_\mu d_\mu\otimes\eins_{N_{\rm b}}+\gamma_5\otimes\widetilde{m}+ \eins\otimes\widetilde{x}-\widetilde{a}\gamma_5\Delta\otimes\eins_{N_{\rm b}}$, is real and symmetric
because of the anti-unitary symmetry. We need to generate its inverse and, hence,  its matrix Green function such that the axial mass $\widetilde{x}$ must have a non-vanishing imaginary part which we assume to have the signature $L=\diag(+\eins_p,-\eins_{N_{\rm b}-p})$ with $0\leq p\leq N_{\rm b}$ positive signs.  Then the inverse square root of  the determinant of $D_5$ can be represented as an integral over real bosonic fields $(\phi_1^T,\phi_2^T)$,
\begin{equation} \label{det-Gaus}
\frac {1}{\sqrt{\det D_5}} = C\int  d\phi_1d\phi_2\exp\left[i\int_V d^4x\vect \phi_1 \\ \phi_2 \evect^T\left(\begin{array}{cc} \eins\otimes L\widetilde{x}+\eins\otimes L\widetilde{m}-\widetilde{a}\Delta\otimes L & i\sigma_\mu d_\mu\otimes L \\ -i(\sigma_\mu d_\mu)^\dagger\otimes L & \eins\otimes L\widetilde{\lambda}-\eins\otimes L\widetilde{m}+\widetilde{a}\Delta\otimes L \end{array}\right) \vect \phi_1 \\ \phi_2 \evect\right]
\end{equation}
with $C$ a normalization constant. The multiplication of the matrix $iL$ guarantees the existence of the integral.

To eliminate the square root we have to double the number of flavors $N_{\rm b}\to2N_{\rm b}$. The Dirac term of the Lagrangian then reads
\begin{equation}
L_0=\phi_1^T \left( i\sigma_\mu d_\mu -i(\sigma_\mu d_\mu)^*\right)\otimes \widetilde{L}\phi_2,
\end{equation}
with $\widetilde{L}=L\otimes\eins_2$ where the two-dimensional unit matrix reflects the doubling of the number of flavors. This term is invariant under
\be 
\phi_1
\to \eins\otimes U \phi_1,\ 
\phi_2
\to \eins\otimes[ \widetilde{L} (U^{-1})^T] \phi_2
\ee
with $U \in \Gl(2N_{\rm b},\mathbb{R})$ because the bosonic fields are real. With 
maximum breaking of chiral symmetry \cite{Peskin,shifman3}, the set of transformations that leaves
the ``bosonic'' chiral condensate,
\begin{equation}
\Sigma\propto\left\langle\int_V d^4x\vect \phi_1 \\ \phi_2 \evect^T\left(\begin{array}{cc} \eins\otimes \widetilde{L} & 0 \\ 0 & -\eins\otimes \widetilde{L} \end{array}\right) \vect \phi_1 \\ \phi_2 \evect\right\rangle\neq0,
\end{equation}
invariant reduces to
the set of matrices satisfying
 $ U^T\widetilde{L}U=\widetilde{L}$ and $U\in \Gl(2N_{\rm b},\mathbb{R})$. This set is known as the non-compact orthogonal group $\Ort(2p,2N_{\rm b}-2p)$. The Lorentz group $\Ort(3,1)$ is a particular example of this kind of non-compact groups.
 
 The pattern of spontaneous symmetry breaking is given by  $\Gl(2N_{\rm b},\mathbb{R}) \to \Ort(2p,2N_{\rm b}-2p)$ and the
resulting Goldstone manifold 
$\Sym_+(2p,2N_{\rm b}-2p)=\Gl(2N_{\rm b},\mathbb{R})/\Ort(2p,2N_{\rm b}-2p)$ is the set of real matrices $U\in\mathbb{R}^{2N_{\rm b}\times2N_{\rm b}}$ with the symmetry $U^T=\widetilde{L} U \widetilde{L}$ and $2p$ positive real eigenvalues and $2N_{\rm b}-2p$ negative real eigenvalues.
Such matrices can be parametrized by $U=LWW^T$ with $W\in\Gl(2N_{\rm b},\mathbb{R})$. Another parametrization is via a diagonalization $U=O\diag(e^{s_1},\ldots,e^{s_{2p}},-e^{s_{2p+1}},-e^{s_{2N_{\rm b}}})O^{-1}$ with $O\in \Ort(2p,2N_{\rm b}-2p)$ and $s_j\in\mathbb{R}$. The corresponding invariant measure $d\mu(U)$ is given by
\begin{equation}\label{inv-non-com-meas}
 d\mu(U)\propto\frac{dU}{{\det}^{N_{\rm b}+1/2}U},\ {\rm with}\ dU\propto \prod_{1\leq k<l\leq 2p}\left|e^{s_k}-e^{s_l}\right|\prod_{\substack{1\leq k\leq 2p \\ 2p+1\leq l\leq2N_{\rm b}}}(e^{s_k}+e^{s_l})\prod_{2p+1\leq k<l\leq 2N_{\rm b}}\left|e^{s_k}-e^{s_l}\right| d\mu(O) \prod_{k=1}^{2N_{\rm b}}de^{s_k}
\end{equation}
and $d\mu(O)$ the invariant measure on $\Ort(2p,2N_{\rm b}-2p)$. Hence $d\mu(U)$ is indeed invariant under the group action $U\to LV^TLUV$ with $V\in\Gl(2N_{\rm b},\mathbb{R})$ because the flat measure (product of the differentials of all independent matrix entries) transforms as $d(LV^TLUV)={\det}^{2N_{\rm b}+1}V dU$.

Let us first discuss the bosonic partition function for $p= 0$ 
or $p=2N_{\rm b}$. In these cases
the imaginary part of $x$ has only positive or negative entries, 
i.e. $L\to L\eins_{N_{\rm b}}$ with $L=\pm1$. 
Then the non-compact orthogonal group reduces to the compact 
$\Ort(2N_{\rm b})$ and the coset is equal to the set of all 
real symmetric positive definite matrices $\Sym_+(2N_{\rm b})$.
There are no integrability issues, even for $a=0$, since the convergence of 
the integrals is assured by the imaginary part of $\widetilde{x}$.  For $a\ne 0$, the convergence
of the integral is already guaranteed by the $a^2$ term if $p=0, 2N_{\rm b}$.

For $p=0, 2N_{\rm b}$ the chiral partition function is given by
\begin{equation}\label{chi-part-bos-b}
Z_\nu(m,x,a)=\int_{\Sym_+(2N_{\rm b})} d\mu(U) {\det}^{-\nu/2}U
 \exp\left[ \frac{iL}{2} \tr m(U - U^{-1}) +\frac{iL}{2} \tr x(U + U^{-1}) -a^2 \tr (U^2 + U^{-2})\right].
\end{equation}
The sign $L=\pm1$ does not drop out although the imaginary part 
of $x$ can be dropped for finite $a$. Because the  eigenvalues 
of $U$ are positive definite, the transformation $U\to-U$ is not allowed. 
However we can absorb the sign in front of the mass by changing $U\to U^{-1}$ which yields a change of the topological charge $\nu\to-\nu$.
There is some freedom in the choice of the signs and phase factors
of the terms in the action, but the sign of
the $a^2$ term is fixed by the physics of the problem. To have convergent
integrals for $a^2 \ne 0$ we necessarily need  that the sign of this
term is negative. We also require that the bosonic and fermionic terms
in the action can be combined into a supertrace. This is achieved by
rotating the fermionic and bosonic variables by a factor $i$ and 
including an overall minus sign in the bosonic action which gives
the representation of Eq.~(\ref{chi-part-bos-b}), cf. the fermionic partition function~\eqref{zchi2}.

For $p \ne 0, 2N_b$ the integral over ${\rm O}(2p,2N_b-2p)$ is non-compact such that the measure~(\ref{inv-non-com-meas})
cannot be normalized even by group invariant potentials. We always have to include an infinitesimal increment in $x$, even for
$a^2\ne 0$, in order to get a convergent integral.
 The bosonic partition function is then given by
\begin{equation}\label{chi-part-bos}
Z_\nu(m,\lambda,a)=\int_{\Sym_+(2p,2N_{\rm b}-2p)} d\mu(U) {\det}^{-\nu/2}U
 \exp\left[ \frac{i}{2} \tr mL(U - U^{-1}) +\frac{i}{2} \tr (xL+i\epsilon)(U + U^{-1}) -a^2 \tr (U^2 + U^{-2})\right].
\end{equation}

\subsection{Supersymmetric Partition Function and Quenched Theory}\label{sec:part-SUSY}

Having discussed the bosonic and fermionic partition function we are now ready to formulate
the supersymmetric partition function that generates the microscopic Dirac spectrum. Especially we focus on the average of the Green function which can be traced back to derivatives of the partially quenched partition function~\eqref{part-SUSY-start}. Hence we put $N_{\rm f}\to N_{\rm f}+1$ in Eq.~\eqref{zchi2} and set $N_{\rm b}=1$ in Eq.~\eqref{chi-part-bos-b}. Then the combined chiral partition function reads
\begin{equation}\label{chi-part-SUSY}
Z_\nu(\widehat{M},\widehat{X},a)=C^{-1}\int\limits_{\Sigma(2N_{\rm f}+2|2)} d\mu(U) {\Sdet}^{\nu/2}U
 \exp\left[ -\frac{iL}{2} \Str (\widehat{M}U - \widehat{M}^\dagger U^{-1}) -\frac{iL}{2} \Str \widehat{X}(U + U^{-1}) +a^2 \Str (U^2 + U^{-2})\right]
\end{equation}
with
\begin{equation}
\widehat{M}=\left(\begin{array}{ccc} M & 0 & 0 \\ 0 & m\eins_2 & 0 \\ 0 & 0 & m'\eins_2 \end{array}\right),\ \widehat{X}=\left(\begin{array}{ccc} 0 & 0 & 0 \\ 0 & x_0\eins_2 & 0 \\ 0 & 0 & x_1\eins_2 \end{array}\right),
\end{equation}
and
\begin{equation}
C=\left.\int_{\Sigma(2N_{\rm f}+2|2)} d\mu(U) {\Sdet}^{\nu/2}U
 \exp\left[ -\frac{iL}{2} \Str (\widehat{M}U - \widehat{M}^\dagger U^{-1}) -\frac{iL}{2} \Str \widehat{X}(U + U^{-1}) +a^2 \Str (U^2 + U^{-2})\right]\right|_{\substack{m=m'\\x_0=x_1}}.
\end{equation}
The mass matrix of the dynamical quarks again satisfies $M=-IM^TI$. The normalization constant guarantees that the partition function is equal to $1$ if $m=m'$ and $x_0=x_1$ because the determinants in Eq.~\eqref{part-SUSY-start} cancel. A supermatrix $U$ in the Goldstone manifold $\Sigma(2N_{\rm f}+2|2)=\U(2N_{\rm f}+2|2)/\UOSp(2N_{\rm f}+2|2)$ has the block structure
\begin{equation}
U= \bmat U_{\rm f}  & -Ig^T \\ g & U_{\rm b} \emat,
\end{equation}
where $ U_{\rm f} \in \CSE(2N_{\rm f}+2)$,  $ U_{\rm b} \in \Sym_+(2)$, and $g$ is a $2\times(2N_{\rm f}+2)$ matrix whose matrix entries are independent Grassmann (anti-commuting) variables. The whole matrix fulfills the symmetry
\begin{equation}\label{real-sym}
U^T=  \bmat U_{\rm f}^T  & -g^T \\ gI & U_{\rm b}^T \emat=\bmat -I\eins_{2N_{\rm f}+2}  & 0 \\ 0 & \eins_2 \emat\bmat U_{\rm f}  & -Ig^T \\ g & U_{\rm b} \emat\bmat  I\eins_{2N_{\rm f}+2}  & 0 \\ 0 & \eins_2 \emat=\bmat -I\eins_{2N_{\rm f}+2}  & 0 \\ 0 & \eins_2 \emat U\bmat I\eins_{2N_{\rm f}+2}  & 0 \\ 0 & \eins_2 \emat.
\end{equation}
We recall that ``$T$" acts  on a supermatrix as the supertransposition which has a slightly different action on the off-diagonal blocks (the sign of the upper right block changes). Moreover, we choose the following notation of the supertrace and the superdeterminant
\begin{equation}
\Str U=\tr U_{\rm f}-\tr U_{\rm b},\ \Sdet U=\frac{\det(U_{\rm f}+iIg^TU_{\rm b}^{-1}g)}{\det U_{\rm b}}.
\end{equation}
This choice differs with respect to some other works~\cite{Berezin-book,Efe-book} where the sign of the supertrace is reversed and the superdeterminant is the inverse of the definition here. The invariant measure $d\mu(U)$ is given by
\begin{equation}
d\mu(U)=\frac{dU_{\rm f}dU_{\rm b}dg}{\Sdet^{N_{\rm f}-1/2} U},
\end{equation}
where $dU_{\rm f}$, $dU_{\rm b}$, and $dg$ are the flat measures meaning the product of the differentials of all independent matrix entries. It naturally appears in superbosonization~\cite{Som,LSZ,KSG} which can be used to directly map random matrix theory to a finite dimensional version of the chiral partition function~\cite{KKG14}.

For the quenched theory where $N_{\rm f}=0$, the partition function drastically simplifies
\begin{equation}\label{zgen}
Z_\nu(\widehat{M},\widehat{X},a)=C^{-1}\int\limits_{\Sigma(2|2)} dU_{\rm f}dU_{\rm b}dg {\Sdet}^{(\nu+1)/2}U
 \exp\left[ -\frac{iL}{2} \Str\widehat{M} (U - U^{-1}) -\frac{iL}{2} \Str \widehat{X}(U + U^{-1}) +a^2 \Str (U^2 + U^{-2})\right],
\end{equation}
where we already wrote the measure in terms of the flat one. The matrices in this expression are
\begin{equation}\label{U-mat}
 \widehat{M}=\left(\begin{array}{cc} m\eins_2 & 0 \\ 0 & m'\eins_2 \end{array}\right),\ \widehat{X}=\left(\begin{array}{cc} x_0\eins_2 & 0 \\ 0 & x_1\eins_2 \end{array}\right),\ U=\diag(\eins_2,O)\left(\begin{array}{cccc} e^{i\varphi} & 0 & \alpha^* & \beta^* \\ 0 & e^{i\varphi} & -\alpha & -\beta \\ \alpha & \alpha^*  & e^{s_1} & 0 \\ \beta & \beta^* & 0 & e^{s_2} \end{array}\right)\diag(\eins_2,O^T)
\end{equation}
with $O\in\Ort(2)$. The orthogonal matrix $O$ drops out and yields a constant. Thus the remaining integration is over a phase $e^{i\varphi}$ with $\varphi\in[-\pi,\pi]$, two positive variables $e^{s_1}$ and $e^{s_2}$ with $s_1,s_2\in\mathbb{R}$, and four Grassmann variables $\alpha,\,\alpha^*,\,\beta,\,\beta^*$. The measure becomes
\begin{equation}
dU_{\rm f}dU_{\rm b}dg\rightarrow de^{i\varphi} \left|e^{s_1}-e^{s_2}\right|de^{s_1} de^{s_2}d\alpha d\alpha^*d\beta d\beta^*=i e^{i\varphi+s_1+s_2}\left|e^{s_1}-e^{s_2}\right| d\varphi ds_1 ds_2d\alpha d\alpha^*d\beta d\beta^*.
\end{equation}
We recall that the integration over Grassmann variables is defined as $\int d\alpha=0$ and $\int \alpha d\alpha=1$ and similar for the other Grassmann variables. Integrations of higher orders are not needed to be defined since Grassmann variables are nilpotent because of their anti-commuting nature.

As for the partition function of fermionic quarks~\eqref{zchi2} we can linearize the term in $a^2$ by introducing an auxiliary supermatrix $\sigma$,
\begin{eqnarray}
Z_\nu(\widehat{M},\widehat{X},a)&=&C^{-1}\int_{\widetilde{\Sigma}(2|2)}d\sigma\int_{\Sigma(2|2)} d\mu(U) {\Sdet}^{\nu/2}U
 \exp\left[\frac{1}{16a^2}\Str\sigma^2 -\frac{iL}{2} \Str(\widehat{M}-\sigma) (U - U^{-1})\right]\nn\\
 &&\left.\times\exp\left[-\frac{iL}{2} \Str \widehat{X}(U + U^{-1})  \right]\right/\int_{\widetilde{\Sigma}(2|2)}d[\sigma]\exp\left[\frac{1}{16a^2}\Str\sigma^2\right] .\label{chi-part-SUSY-b}
\end{eqnarray}
The set $\widetilde{\Sigma}(2|2)$ consists of $(2|2)\times(2|2)$ supermatrices which  can be parametrized as follows
\begin{equation}\label{para-sigma}
\sigma=\diag(\eins_2,\widetilde{O})\left(\begin{array}{cccc} iu & 0 & \eta^* & \chi^* \\ 0 & iu & -\eta & -\chi \\ \eta & \eta^*  & v_1 & 0 \\ \chi & \chi^* & 0 & v_2 \end{array}\right)\diag(\eins_2,\widetilde{O}^T)
\end{equation}
with $\widetilde{O}\in{\rm O}(2)$, $u,v_1,v_2\in\mathbb{R}$ and $\eta,\eta^*,\chi,\chi^*$ four independent Grassmann variables. Additionally, the supermatrix $\sigma$ fulfills the symmetry~\eqref{real-sym}. With the help of the partition function of the chiral Gaussian orthogonal ensemble,
\begin{equation}\label{part-chiGOE-SUSY}
Z_{\chiGOE}^{(\nu)}([\widehat{M}+\widehat{X}-\sigma][\widehat{M}-\widehat{X}-\sigma])={\Sdet}^{-\nu/2}(\widehat{M}-\widehat{X}-\sigma) Z_\nu(\widehat{M}-\sigma,\widehat{X},a=0),
\end{equation}
we can rewrite the quenched partition function as a convolution of the $Z_{\chiGOE}^{(\nu)}$ with a Gaussian
\begin{eqnarray}
Z_\nu(\widehat{M},\widehat{X},a)&=&\frac{C(a=0)}{C(a\neq0)}\int_{\widetilde{\Sigma}(2|2)}d\sigma
 \exp\left[\frac{1}{16a^2}\Str\sigma^2\right]{\Sdet}^{\nu/2}(\widehat{M}-\widehat{X}-\sigma)\nn\\
 &&\left.Z_{\chiGOE}^{(\nu)}([\widehat{M}+\widehat{X}-\sigma][\widehat{M}-\widehat{X}-\sigma])\right/\int_{\widetilde{\Sigma}(2|2)}d[\sigma]\exp\left[\frac{1}{16a^2}\Str\sigma^2\right].\label{chi-part-SUSY-c}
\end{eqnarray}
This representation shows the diffusive character of the effect of a finite lattice spacing $a$ on the spectrum of the Dirac operator, see \cite{Guhr} for the diffusive approach in supersymmetric spaces. Thus it exhibits a similar effect as already found for lattice QCD with three colors of the Wilson Dirac operator \cite{split-dyn,KVZ-prl}.  Interestingly, when omitting the term $Z_{\chiGOE}^{(\nu)}$ in the integral~\eqref{chi-part-SUSY-c} we obtain the finite dimensional quenched partition function of an orthogonal Gaussian random matrix ensemble,
\begin{equation}\label{part-GOE-SUSY}
Z_{\GOE}^{(\nu)}\left(\frac{\widehat{M}-\widehat{X}}{4a}\right)=\left.\int_{\widetilde{\Sigma}(2|2)}d\sigma
 \exp\left[\frac{1}{16a^2}\Str\sigma^2\right]{\Sdet}^{\nu/2}(\widehat{M}-\widehat{X}-\sigma)\right/\int_{\widetilde{\Sigma}(2|2)}d[\sigma]\exp\left[\frac{1}{16a^2}\Str\sigma^2\right].
\end{equation}
We recall that the real eigenvalues of this ensemble scale with the lattice spacing $a$ which can be also seen in this expression. The normalization constant $C$ is independent of $a$ because the partition function should be normalized for any choice of $\widehat{M}=m\eins_4$ and $\widehat{X}=x\eins_4$, cf. Eq.~\eqref{part-SUSY-start} where we started from. Thus the ratio is $C(a=0)/C(a\neq0)=1$.

Now we have all ingredients to calculate and discuss the quenched partition function.

\section{Quenched Partition Function}\label{sec:que-part}

In subsection~\ref{sec:eval} we present the result for the quenched partition function at finite lattice spacing, quark mass and axial mass. We reduce this result to a compact integral which represents the fermionic valence quark and a non-compact two-fold integral which reflects the nature of the bosonic valence quark. The continuum limit $a\to0$ is derived  in subsection~\ref{sec:small-a}. The limit of a very coarse lattice $|a|\gg1$ 
and the related thermodynamic limit are discussed in subsections~\ref{sec:large-a} and \ref{sec:thermo-dyn}, respectively.
 
\subsection{Discussion of the Partition Function at finite $a$}\label{sec:eval}

The evaluation of the partition function~\eqref{zgen} is straightforward but tedious.
Since we have only four different Grassmann variables, the Grassmann integrals
can be evaluated by a brute force expansion. This is worked out in appendix~\ref{app:Grass}.
 The final result for the generating function is given by
\be 
Z_\nu(\widehat{M},\widehat{X},a)&=&\frac{1}{4}\int_{-\pi}^{\pi} \frac{d\varphi}{2\pi} \int_{-\infty}^{\infty}ds_1 \int_{-\infty}^{\infty} ds_2\left|\sinh\frac{s_1-s_2}{2}\right | 
e^{\nu (2i\varphi-s_1-s_2)/2} \exp\left[2Lm\sin\varphi+iLm'(\sinh s_1+\sinh s_2)\right]\nn\\
&&\times\exp\left[-2iLx_0\cos\varphi+iLx_1(\cosh s_1+\cosh s_2)+4a^2\cos2\varphi-2a^2(\cosh2s_1+\cosh2s_2)\right]\nn\\
&&\times \left [(4a^2(e^{-2i\varphi}+e^{-2s_1}+e^{i\varphi+s_1}+e^{-i\varphi-s_1})+iL(m-x_0)e^{-i\varphi} 
+ iL(m'-x_1) e^{-s_1}-\nu-1) \right .\nn \\
&& \hspace*{0.5cm}\times (4a^2(e^{-2i\varphi}+e^{-2s_2}+e^{i\varphi+s_2}+e^{-i\varphi-s_2})+iL(m-x_0)e^{-i\varphi} 
+ iL(m'-x_1) e^{-s_2}-\nu-1)\nn \\
&& \left .+ 4a^2(3e^{-2i\varphi}+e^{-s_1-s_2}+ e^{-2s_1}+e^{-2s_2}+
2e^{-i\varphi-s_1}+2e^{-i\varphi-s_2})
 \right . \nn \\
&& \left .+iL(2(m-x_0)e^{-i\varphi}+ (m'-x_1) (e^{-s_1}+ e^{-s_2}))-\nu -1 \right ]\nn\\
&=&16a^4(\Phi_{\nu-4}S_{\nu,0}+\Phi_{\nu}S_{\nu+4,0}+\Phi_{\nu+2}S_{\nu-2,0}-\Phi_{\nu-2}S_{\nu+2,0} +4\Phi_{\nu-2}S_{\nu+2,2}+2\Phi_{\nu-1}S_{\nu-1,1}\nn\\
&&+2\Phi_{\nu-3}S_{\nu+1,1}+8\Phi_{\nu+1}S_{\nu+1,3}-6 \Phi_{\nu+1}S_{\nu+1,1}+2\Phi_{\nu-1}S_{\nu+3,1}+4\Phi_{\nu}S_{\nu,2}-2\Phi_{\nu}S_{\nu,0})\nn\\
&&+4a^2((2\nu-1)\Phi_{\nu-2}S_{\nu,0}-(2\nu+1)\Phi_{\nu}S_{\nu+2,0}+4\nu\Phi_{\nu}S_{\nu+2,2} +2(\nu+1)\Phi_{\nu+1}S_{\nu-1,1}+2(\nu-1)\Phi_{\nu-1}S_{\nu+1,1})\nn\\
&&-8a^2(m-x_0)(\Phi_{\nu-3}S_{\nu,0}+2\Phi_{\nu-1}S_{\nu+2,2}-\Phi_{\nu-1}S_{\nu+2,0}+\Phi_{\nu}S_{\nu-1,1}+\Phi_{\nu-2}S_{\nu+1,1})\nn\\
&&-8a^2(m'-x_1)(\Phi_{\nu-2}S_{\nu+1,1}+\Phi_{\nu}S_{\nu+3,1}+2\Phi_{\nu+1}S_{\nu,2}-\Phi_{\nu+1}S_{\nu,0}+\Phi_{\nu-1}S_{\nu+2,0})\nn\\
&&+(m-x_0)^2\Phi_{\nu-2}S_{\nu,0}+2(m-x_0)(m'-x_1)\Phi_{\nu-1}S_{\nu+1,1}-2\nu(m-x_0)\Phi_{\nu-1}S_{\nu,0}\nn\\
&&+(m'-x_1)^2\Phi_{\nu}S_{\nu+2,0}-2\nu(m'-x_1)\Phi_{\nu}S_{\nu+1,1}+(\nu+1)\nu\Phi_{\nu}S_{\nu,0}.\label{final}
\ee
Each term factorizes into a non-compact two-dimensional 
integral $S_{\mu,\alpha}$ and a compact one-dimensional integral $\Phi_\mu$ which are defined by
\begin{eqnarray}
\Phi_\mu(m,x_0,a)&=&(-iL)^\mu\int_{-\pi}^\pi\frac{d\varphi}{2\pi}e^{i\mu\varphi}\exp\left[2Lm\sin\varphi-2iLx_0\cos\varphi+4a^2\cos2\varphi\right]\nn\\
&=&(-i)^\mu\int_{-\pi}^\pi\frac{d\varphi}{2\pi}e^{i\mu\varphi}\exp\left[2m\sin\varphi-2ix_0\cos\varphi+4a^2\cos2\varphi\right]
\nn\\
 &=&\frac{(m^2-x_0^2)^{\mu/2}}{(m+x_0)^\mu}\sum_{l=-\infty}^\infty\left(\frac{x_0-m}{x_0+m}\right)^lI_{l}\left(4a^2\right)I_{\mu+2l}\left(2\sqrt{m^2-x_0^2}\right)\label{comp-int}
\end{eqnarray}
and
\begin{eqnarray}
S_{\mu,\alpha}(m',x_1,a)&=&\frac{(iL)^\mu}{4}\int_{-\infty}^\infty ds_1\int_{-\infty}^\infty ds_2 \left|\sinh\frac{s_1-s_2}{2}\right |  e^{-\mu(s_1+s_2)/2} \cosh^\alpha\frac{s_1-s_2}{2}\nn\\
 &&\times\exp\left[iLm'(\sinh s_1+\sinh s_2)+iLx_1(\cosh s_1+\cosh s_2)-2a^2(\cosh2s_1+\cosh2s_2)\right]\nn\\
 &&\times\exp\left[im'(\sinh s_1+\sinh s_2)+ix_1(\cosh s_1+\cosh s_2)-2a^2(\cosh2s_1+\cosh2s_2)\right]\nn\\
 &=&(iL)^\mu\int_{-\infty}^\infty ds\int_{1}^\infty dy  e^{-\mu s} y^\alpha\exp\left[2iLm'y\sinh s+2iLx_1y\cosh s-4a^2(2y^2-1)\cosh  2s\right],\label{non-comp-int}
\end{eqnarray}
where we have performed the substitution $y=\cosh(s_1-s_2)/2$ and $s=(s_1+s_2)/2$ in the second line of Eq.~\eqref{non-comp-int}. The function $I_l$ is the modified Bessel function of the first kind.

The partition function is correctly normalized which can be readily checked either by setting $a^2=m=m'=0$ and $x_0=x_1=iL\varepsilon$ with $\varepsilon\to\infty$ or by setting $m=m'=x_0=x_1=0$ and $a^2\to\infty$. In the latter case where we take $a\to\infty$ we have to distinguish between even and odd $\nu$ because of the following asymptotics,
\begin{eqnarray}
\Phi_\mu(m=0,x_0=i\epsilon\to\infty,a=0)\to\frac{(-iL)^\mu e^{2\varepsilon}}{2\sqrt{\pi\varepsilon}},&\quad&\Phi_\mu(m=0,x_0=0,a\to\infty)\to\frac{(iL)^\mu e^{2a^2}}{\sqrt{8\pi}\,a}\delta_{{\rm mod}_2(\nu),0},\nn\\
S_{\mu,\alpha}(m=0,x_0=i\epsilon\to\infty,a=0)\to\frac{\sqrt{\pi}(iL)^\mu e^{-2\varepsilon}}{2\varepsilon^{3/2}},&\ &S_{\mu,\alpha}(m=0,x_0=0,a\to\infty)\to\frac{\sqrt{\pi}(-iL)^\mu e^{-2a^2}}{2^{11/2}a^3}.\label{asy}
\end{eqnarray}
The partition function should be also unity when $m=m'$, $x_0=x_1$, and $a$ finite. However this general situation is quite non-trivial to verify. It goes back to Cauchy-like integrals of superfunctions invariant under supersymmetric groups~\cite{PS,Efetov,Con,Con-Gro,Efe-book,KKG} which were first derived in a general form by Wegner~\cite{Wegner}. Indeed the integrand in Eq.~\eqref{zgen} is invariant under the supergroup $\UOSp(2|2)$ when $m=m'$ and $x_0=x_1$.

We emphasize that the representation of the compact integral~\eqref{comp-int} as a sum is highly convergent for $m, x_0, a^2$ small enough  since the Bessel function behaves as $I_l\propto1/l!$ for large order. This cannot be said about the non-compact twofold integral~\eqref{non-comp-int} which causes some trouble if the lattice spacing $a$ is too large or too small or when the quark mass $m'$ or the axial mass $x_1$ are too large. Then we have large cancellations which challenge the numerical integration. 
Hence it is advantageous to improve this integral. Luckily we are either interested in its value for $x_1=0$ or in its imaginary part. In appendix~\ref{app:Im} we calculate both expressions and find
\begin{eqnarray}
S_{\mu,\alpha}(m',x_1=0,a)&=&\frac{[\sign(\mu)]^\mu}{\sqrt{2\pi a^2}(|\mu|-1)!}\int_{1}^\infty dy\int_{0}^\infty dt \frac{y^{\alpha+1}}{\sqrt{2y^2-1}}\exp\left[-4a^2(2y^2-1)\right]\label{non-comp-int-x0}\\
&&\times\frac{1}{t}\partial_{t'}^{|\mu|-1}\left[|t'+t|^{|\mu|}\exp\left[-\frac{y^2}{8a^2(2y^2-1)}(t'+t-m')^2\right]K_{\mu}\left(2y|t'+t|\right)-\{t\to-t\}\right]_{t'=0}\nn
\end{eqnarray}
and
\begin{eqnarray}
&&\mathcal{S}_{\mu,\alpha}(m',x_1,a)=\frac{1}{\pi}\lim_{\varepsilon\to0}{\rm Im}\,S_{\mu,\alpha}(m',x_1+i\varepsilon,a) \label{non-comp-int-Im}\\
&=&-(-\sign\,\mu)^{\mu}\sqrt{\frac{\pi}{8 a^2}}\int_{1}^\infty dy\int_{-\infty}^\infty dt \frac{y^{\alpha-|\mu|+1}}{\sqrt{2y^2-1}}\exp\left[-\frac{y^2}{8a^2(2y^2-1)}(t-\sign(\mu)m')^2-4a^2(2y^2-1)\right]\nn\\
&&\times\sum_{k=0}^{|\mu|-1}\frac{(|\mu|-k-1)!}{(|\mu|-1)!k!}[y^2(x_1^2-t^2)]^k\delta^{(|\mu|-1)}(t-x_1)\nn\\
&&+(-1)^{\mu}\sign\, x_1\sqrt{\frac{\pi}{8 a^2}}\int_{1}^\infty dy\int_{-\infty}^\infty dt \frac{y^{\alpha+1}}{\sqrt{2y^2-1}}\exp\left[-\frac{y^2}{8a^2(2y^2-1)}(t-m')^2-4a^2(2y^2-1)\right]\nn\\
&&\times J_{\mu}\left(2y\sqrt{x_1^2-t^2}\right)\left(\frac{x_1+t}{x_1-t}\right)^{\mu/2}\Theta(x_1^2-t^2).\nn
\end{eqnarray}
The Heaviside step function $\Theta(x_1^2-t^2)$ restricts the integral over $t$ to a compact interval in the second term while in the first term the integral can be simply evaluated by a Taylor expansion in $t$ and has to be skipped for $\mu=0$. For $\mu=0$ the term $1/t$ and the derivative have to be omitted  and the minus sign in front of $\{t\to-t\}$ becomes a plus sign in Eq.~\eqref{non-comp-int-x0}.
 The functions $J_\mu$ and $K_\mu$ are the Bessel function of the first kind and the modified Bessel function of the second kind.

\subsection{Continuum Limit ($|a|\ll1$)}\label{sec:small-a}

To understand what happens when taking the continuum limit $|a|\to0$ it is 
useful to consider a random matrix theory which is equivalent to the same chiral partition function as two color QCD with Wilson fermions. This random matrix ensemble consists of random matrices of the form
\begin{equation}\label{RMT}
 D_{\rm W}=\left(\begin{array}{cc} aA & W \\ -W^T & aB\end{array}\right)
\end{equation}
with $A=A^T$ and $B=B^T$ real symmetric matrices of size  $n\times n$ and $(n+\nu)\times(n+\nu)$, respectively, and $W$ an $n\times(n+\nu)$ real rectangular matrix. The random matrix~\eqref{RMT} belongs to the symmetry class ${\rm 19 QC_+^*}$ in the classification scheme of non-Hermitian matrices by Magnea~\cite{Magnea}. When we choose the Gaussian
\begin{equation}\label{RMT-Prob}
 P(D_{\rm W})=\exp\left[-\frac{1}{16}(\tr A^2+\tr B^2)-\frac{1}{2n}\tr WW^T\right]
\end{equation}
as  the distribution of $D_{\rm W}$, we do not have to unfold its spectrum, i.e. in the limit of large matrix dimension $n\to\infty$ the smallest eigenvalues of $D_{\rm W}$ and $D_5=\gamma_5D_{\rm W}$ (with $\gamma_5=\diag(\eins_n,-\eins_{n+\nu})$ in its matrix form) around the origin are 
of order $\mathcal{O}(1)$. We have chosen the Gaussian distribution for simplicity. 
Due to universality other probability distributions  give rise to 
the same chiral Lagrangian.

The random matrix~\eqref{RMT} can be brought into the form 
\be
D'_{\rm W} = \bmat O_1^T &0 \\0 & O_2^T \emat   \left ( \begin{array}{ccc}
 A & C & 0 \\
 C^T & B' & f\\
0 & f^T & b\\ 
\end{array} \right )
\bmat O_1 &0 \\0 & O_2 \emat,
\ee
where $A$, $B'$ , $f$ and $b$ are of $O(a)$, $O_1\in\Ort(n)$, and $O_2\in\Ort(n+\nu)$.
At vanishing lattice spacing $a$ the matrix $D'_{\rm W}$ has $\nu$ zero modes 
and $n$ pairs of imaginary eigenvalues $\pm iy$. This ensemble is known as the chiral Gaussian orthogonal ensemble~\cite{mehta,magnea-o,dalmazi-jv}. For small lattice spacing $|a|\ll1$ we can apply first order perturbation theory.
To lowest order in $a$, the secular equation factorizes as
\be
\det (D_{\rm W}- \lambda\eins_{2n+\nu}) = 
\det \left [ \bmat -\lambda\eins_{n} & C\\ C^T & -\lambda\eins_{n} \emat \right ]
\det (b -\lambda\eins_{\nu}).
\ee
Therefore to leading order the effect of a non-zero lattice spacing is
the broadening of the zero modes to the spectral density of the
ensemble the matrix $b$ is drawn from.
By the central limit theorem, the distributions of the matrix elements
of $b$ are Gaussian even if the matrix elements of $B$ do not have
a Gaussian distribution. Hence the distribution of $b$ is a $\nu$ dimensional $\GOE$. The variance of $b$ is equal to the one
of $B$ so that the distribution of $b$ is given by
\be\label{Gauss}
p(b) \propto \exp\left[-\frac N{4\tilde a^2} {\rm tr} b^2\right]=\exp\left[-\frac V{32 a^2 W_8} {\rm tr} b^2\right],
\ee
where we used the relation between the dimensionless lattice spacing $a$ and the physical one when identifying $N \sim V$, i.e.
\be
a^2 V W_8 = N \tilde a^2/8.
\ee
For $\nu=1$ the distribution is a simple Gaussian
\be
p(b) = \exp\left[-\frac V{32 a^2 W_8}  b^2\right].
\ee
In applying the central limit theorem, we have assumed that the average of the matrix elements of $O_2 B O_2^T$ vanishes. This is not the
case if the average $\langle A_{kl} \rangle,\langle B_{kl} \rangle \sim a\delta_{kl}$ which indeed is the situation in lattice simulations, e.g. see \cite{DHS}. However, 
such term is exactly of the form of a mass term and can be eliminated by a redefinition of the mass justifying that the central limit theorem
 gives a centered Gaussian~\eqref{Gauss}.

Indeed the splitting of the spectrum into to one of   $\chiGOE$ and one of $\GOE$ 
can be also seen at the level of chiral perturbation theory which also applies to the epsilon domain of QCD. 
\FloatBarrier\noindent
Considering the quenched partition function~\eqref{zgen} we have shown that it can 
be rewritten as the convolution of the partition function of continuum QCD without zero modes 
and a  Gaussian function of the  supermatrix $\sigma$, see Eq.~\eqref{chi-part-SUSY-c}. 
Rescaling $\sigma\to a\sigma$ we can drop the $\sigma$ dependence in the partition function $Z_{\chiGOE}^{(\nu)}([\widehat{M}+\widehat{X}-a\sigma][\widehat{M}-\widehat{X}-a\sigma])$ since the function remains finite at $a=0$. Hence the quenched partition function factorizes for $|a|\ll1$,
\begin{eqnarray}
Z_\nu(\widehat{M},\widehat{X},a)&\overset{a\ll1}{\approx}&Z_{\chiGOE}^{(\nu)}\left(\widehat{M}^2-\widehat{X}^2\right)Z_{\GOE}^{(\nu)}\left(\frac{\widehat{M}-\widehat{X}}{4a}\right).\label{chi-part-SUSY-small-a}
\end{eqnarray}
Both partition functions and their resulting spectral observables are summarized in appendix~\ref{app:cont-limit}. This factorization is a also shown in Fig.~\ref{fig:comp}. 
The comparison of the approximation~\eqref{chi-part-SUSY-small-a} 
with Monte Carlo simulations of the random matrix model~\eqref{RMT} and the  full analytical result~\eqref{final} confirms that this factorization applies quite well for $|a|<0.1$. 
Only for small indices, namely $\nu=0,1,2$, the deviations are persistent. 
The reason is the non-analytic nature of the spectrum of the continuum QCD Dirac operator at the spectral edge $\lambda=\pm m\neq0$. The first derivative of the level density of the Hermitian Dirac operator $D_5$ diverges at the edge for 
$\nu=0,1,2$. Thus the limit $a\to0$ cannot be interchanged with the integral over the eigenvalues when averaging over the spectrum implying that the continuum limit is not uniform in these cases. If the quark mass vanishes, only the case $\nu=0$ is non-uniform. We return to this point in subsection~\ref{sec:rho5}.

\begin{figure}[h!]
\centerline{\includegraphics[width=0.47\textwidth]{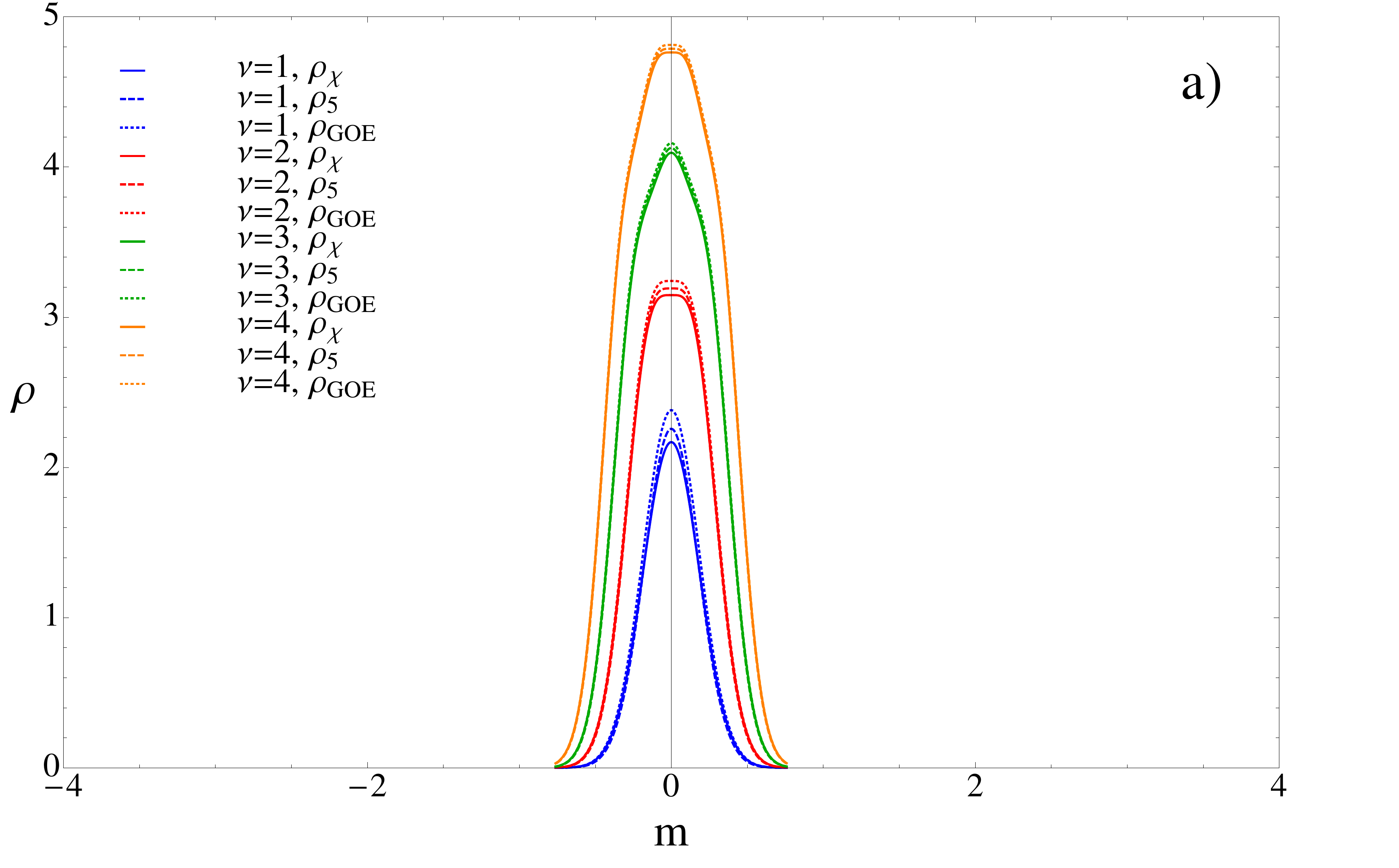}\hfill\includegraphics[width=0.47\textwidth]{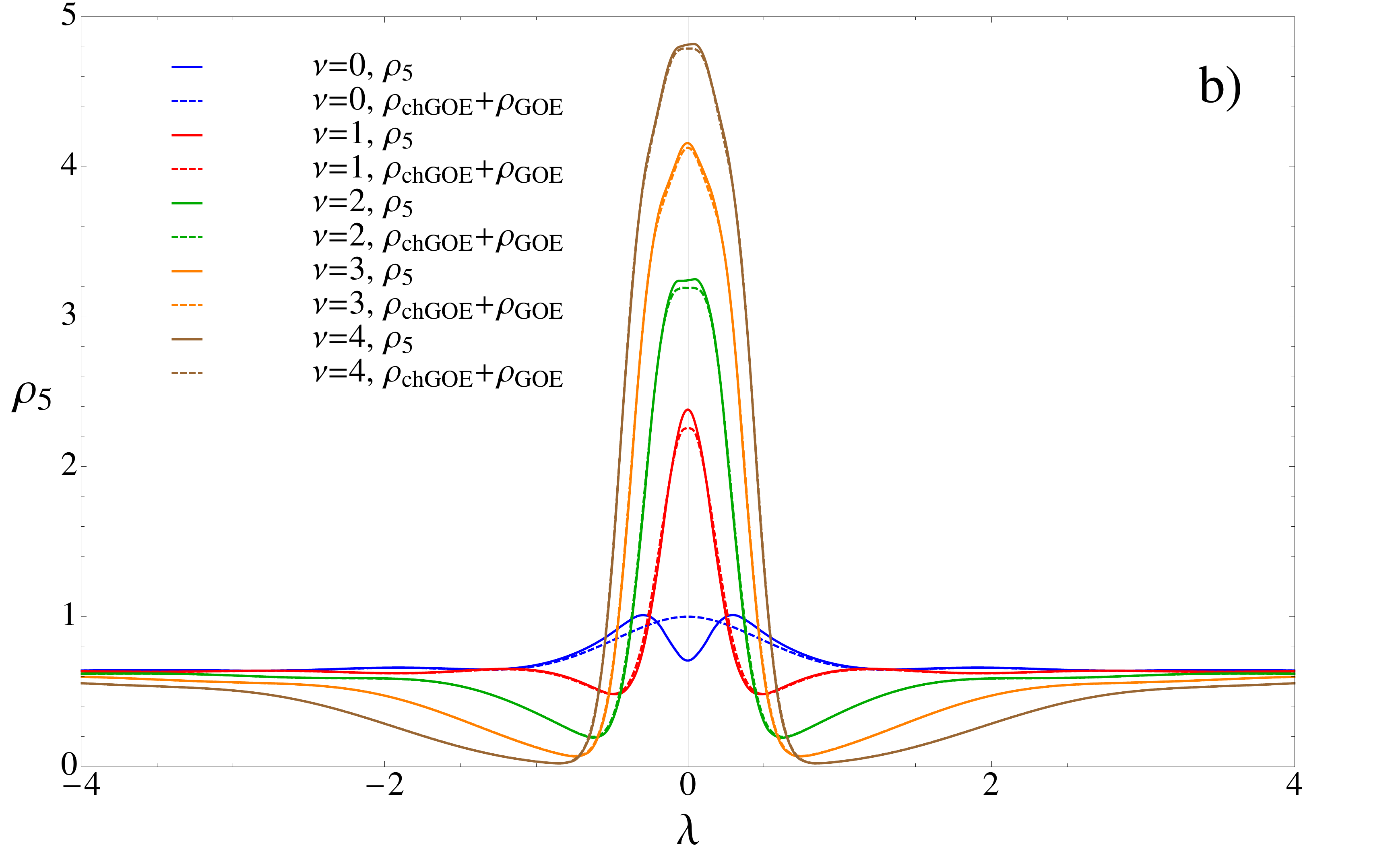}}
\centerline{\includegraphics[width=0.47\textwidth]{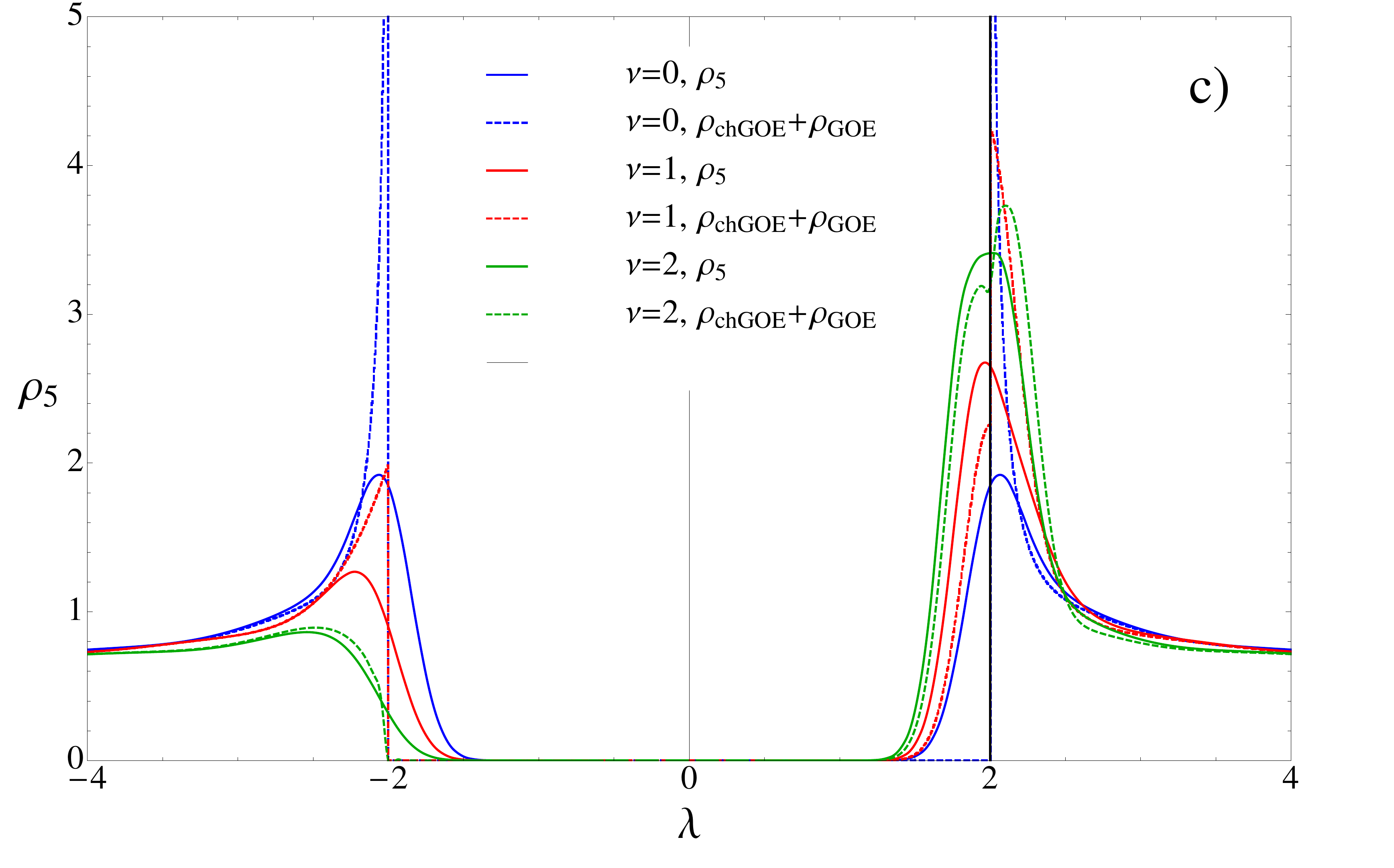}\hfill\includegraphics[width=0.47\textwidth]{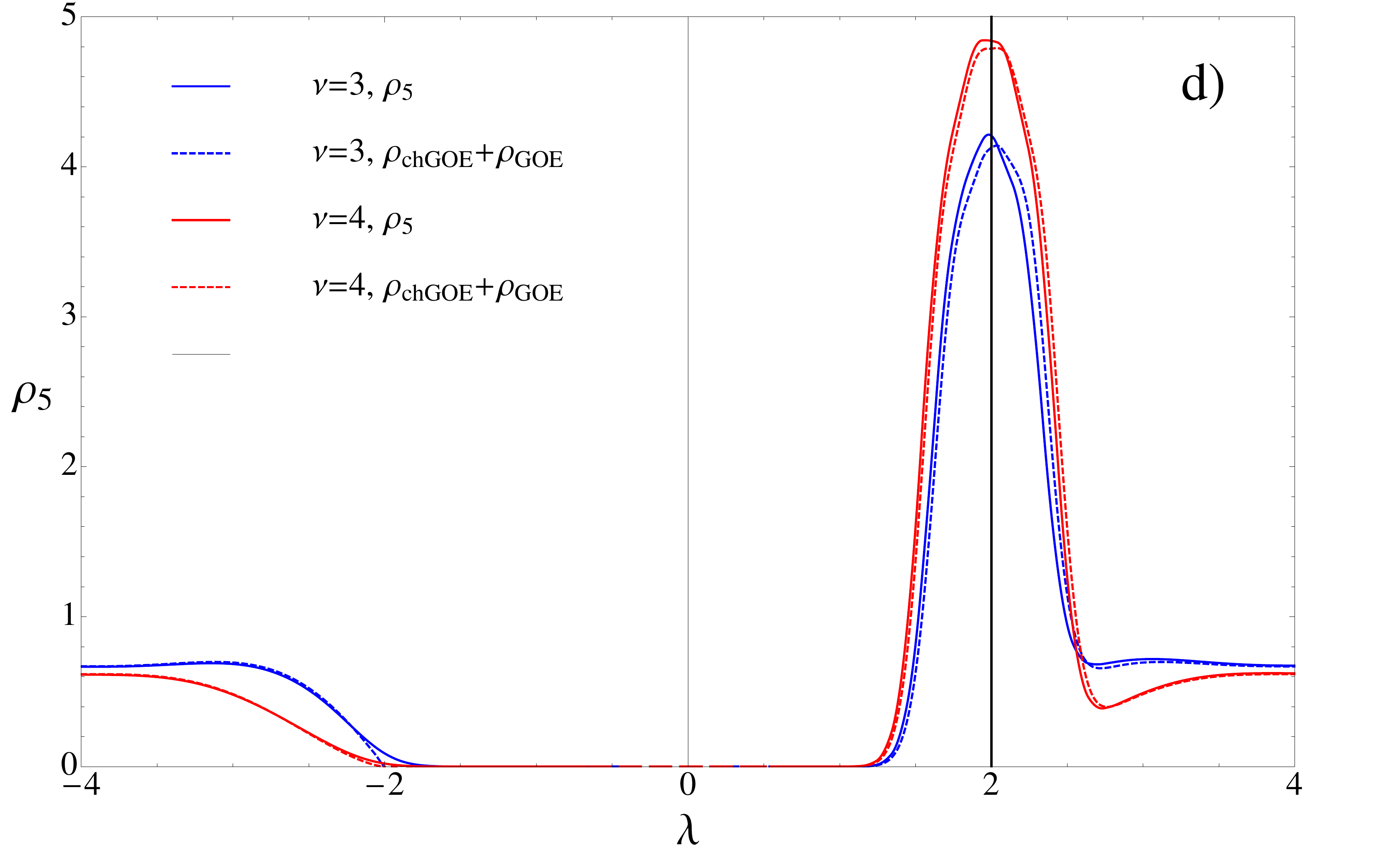}}
\centerline{\includegraphics[width=0.47\textwidth]{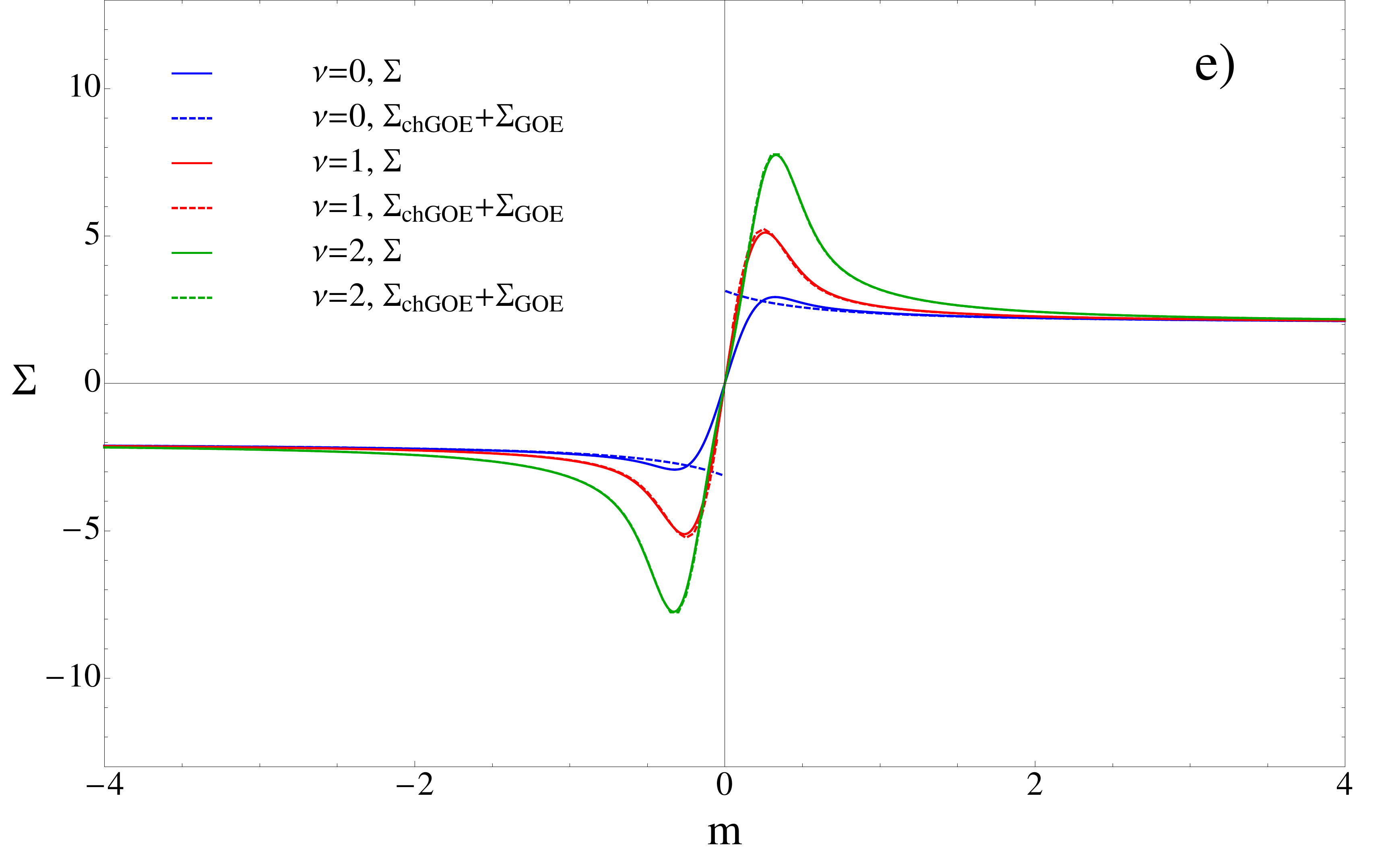}\hfill\includegraphics[width=0.47\textwidth]{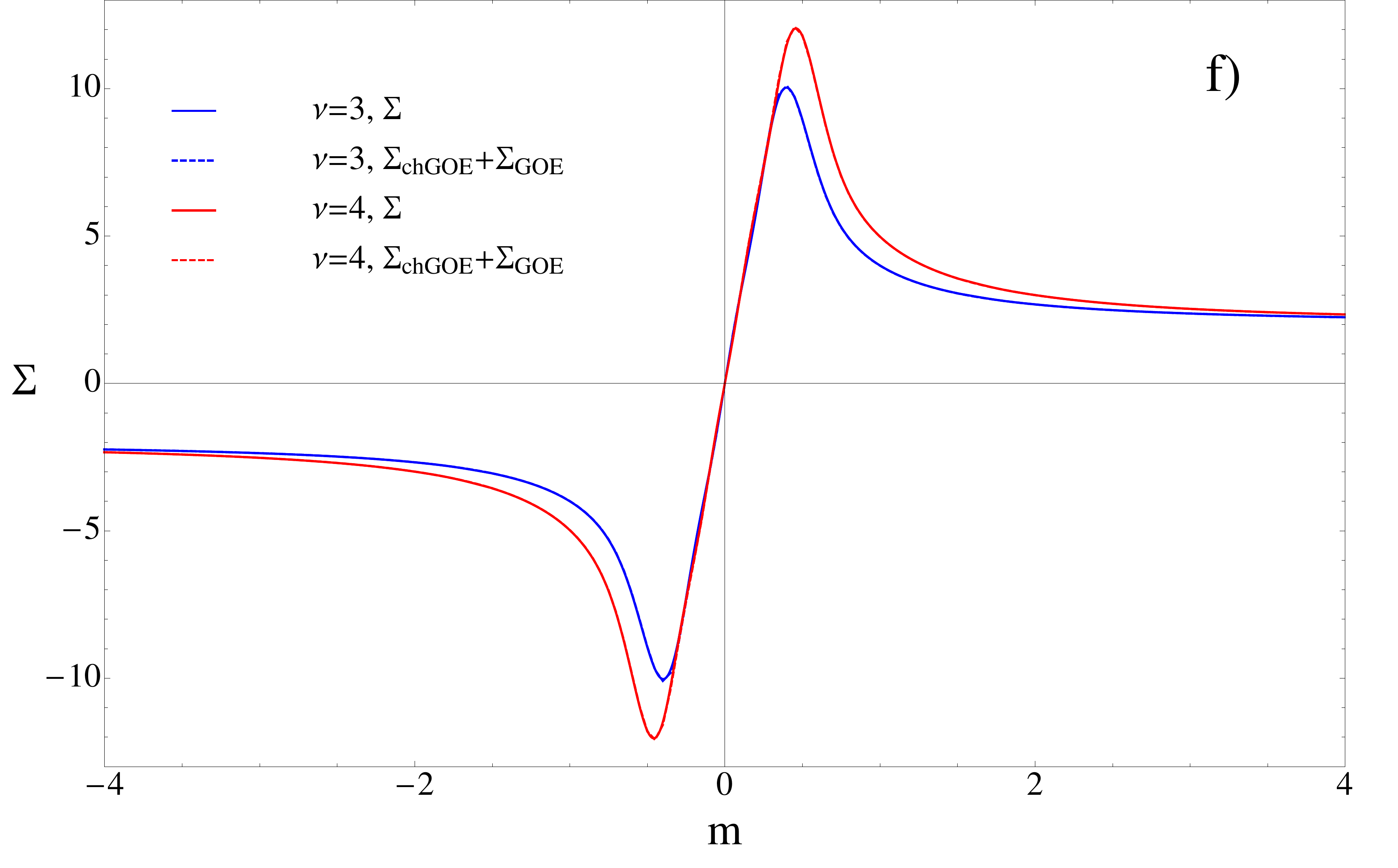}}
\caption{Comparison of some spectral observables of the Wilson Dirac operator  with lattice spacing $\sqrt{W_8 V}\widetilde{a}=|a|=0.0625$ and of the spectrum consisting of the sum of continuum QCD without zero modes and a $\nu\times\nu$ dimensional GOE for indices ($\nu=0,1,2,3,4$). 
Figure a) emphasizes that the distribution of chirality $\rho_\chi(m)$ 
over the real eigenvalues (solid curves), see Eq.~\eqref{rhochi-result}, 
and the level density $\rho_5(m,\lambda=0)$ of the Hermitian Wilson Dirac 
operator at the origin as a function of the quark mass (dashed curves), see Eq.~\eqref{rho5-result}, become the level density of a GOE.
 In subsection~\ref{sec:rhoreal} we argue that then also the level density of the real eigenvalues of the non-Hermitian Wilson Dirac operator $D_{\rm W}$ shares this distribution. The splitting of the spectrum of $D_5$ into 
the superposition of 
a  $\chiGOE$ and a $\GOE$ is shown in figure b) where the quark mass is set to $m=0$. The level density $\rho_5$ (solid curves) perfectly agrees 
with the sum $\rho_{\chiGOE}+\rho_{\GOE}$ (dashed curves) when 
the eigenvalue $\lambda$ stays away from the origin while the deviations remain close to the origin. Especially the case of vanishing index $\nu=0$ 
shows obvious deviations which are discussed in subsection~\ref{sec:rho5}. 
These deviations are particularly large for small index $\nu=0,1,2$ 
and become more prominent when the quark mass $m$ becomes non-zero, 
see figure c) for $m=2$ (black vertical line). 
The spectral discontinuities of the continuum Dirac operator are 
hardly suppressed by the $\GOE$ level density $\rho_{\GOE}$ shifted by the quark mass (dashed curves). 
However the level densities  $\rho_5(m=2,\lambda)$ (solid curves) are smooth at any finite values of the lattice spacing. Only for larger indices,
 here $\nu=3,4$, the agreement with a splitting into 
the two sub-spectra is almost striking, cf. figure d). 
The bad, see figure e), or good, see figure f), agreement carries over to the mass dependent chiral condensate $\Sigma(m)$, 
see Eq.~\eqref{sigma-def}. Also in these two plots the solid curves 
are the exact results~\eqref{sigma-result} at finite lattice spacing
 while the dashed curves are the approximation of the chiral condensate 
as the sum of the chiral condensate of continuum QCD without the contribution 
from the zero modes, $\Sigma_{\chiGOE}$, and of the ``chiral condensate" 
resulting from a finite dimensional GOE, $\Sigma_{\GOE}$, see Eqs.~\eqref{step2-a} and \eqref{Sigma-GOE}.}
\label{fig:comp}
\end{figure}
\FloatBarrier\noindent

\subsection{Limit to a very Coarse Lattice ($|a|\gg1$)}\label{sec:large-a}

We also consider the limit of  a coarse lattice $|a|\gg1$ to get a full understanding  of lattice 
artefacts in the spectrum of the Wilson Dirac operator. 
Especially this limit shows the change of scales in the spectrum  when increasing the lattice spacing. It may help to estimate the strength of the lattice artefacts.

In the limit of a very coarse lattice the quark mass $m$ and the axial mass $x$ 
are an order  smaller than $a^2$. Then the compact integral~\eqref{comp-int} is 
dominated by the term $4a^2\cos2\phi$. 
Expanding around the two saddle points $\varphi=\delta\varphi/a$ 
and $\varphi=\pi+\delta\varphi/a$ the function $\Phi_\mu$  becomes
\begin{eqnarray}
\Phi_\mu(m,x_0,a)&\overset{|a|\gg1}{\approx}&\frac{(-iL)^\mu e^{4a^2}}{2\pi a}\int_{-\infty}^\infty d\delta\varphi e^{-8\delta\varphi^2}\left(\exp\left[\frac{2Lm}{a}\delta\varphi-2iLx_0\right] +(-1)^\mu\exp\left[-\frac{2Lm}{a}\delta\varphi+2iLx_0\right]\right)\nn\\
&=&\frac{(-iL)^\mu }{\sqrt{8\pi} a}\exp\left[4a^2+\frac{m^2}{16a^2}\right]\frac{e^{-2iLx_0} +(-1)^\mu e^{-2iLx_0}}{2}.\label{comp-int-large-a}
\end{eqnarray}
A similar expansion can be done for the non-compact double integral~\eqref{non-comp-int}. 
However this time we have only one saddle point namely $s_{1/2}=\delta s_{1/2}/a$,
\begin{eqnarray}
S_{\mu,\alpha}(m',x_1,a)&\overset{|a|\gg1}{\approx}&\frac{(iL)^\mu}{8a^3}\exp[2iLx_1-4a^2]\int_{-\infty}^\infty d\delta s_1\int_{-\infty}^\infty d\delta s_2 \left|\delta s_1-\delta s_2\right |\exp\left[\frac{iLm'}{a}(\delta s_1+\delta s_2)-4(\delta s_1^2+\delta s_2^2)\right]\nn\\
&=&\frac{(iL)^\mu\sqrt{\pi}}{2^{11/2}a^3}\exp\left[2iLx_1-4a^2-\frac{m'^2}{16a^2}\right].\label{non-comp-int-small-a}
\end{eqnarray}
Since $m,x \ll a^2$   we can omit all terms in the sum~\eqref{final} which do not come with an $a^4$ factor. 
Using the approximations~\eqref{comp-int-large-a} and \eqref{non-comp-int-small-a} 
 the quenched partition function becomes
\begin{eqnarray}
Z_\nu(\widehat{M},\widehat{X},a)&\overset{a\gg1}{\approx}&\exp\left[2iL(x_1-x_0)+\frac{m^2-m'^2}{16a^2}\right].\label{chi-part-SUSY-large-a}
\end{eqnarray}
The second term in Eq.~\eqref{comp-int-large-a} depending on the sign $(-1)^\mu$ cancels in the sum. Therefore the dependence on the topological charge is completely lost.

We wish  to emphasize that this scaling is only valid around the origin and applies for the eigenvalues which are of order $\mathcal{O}(a)$, cf. Figs.~\ref{fig:thermo}. A more physical scale for the level density of the real eigenvalues and the ``mesoscopic" spectrum around the origin of the Hermitian Wilson Dirac operator is discussed in the next subsection.

\subsection{Thermodynamic Limit}\label{sec:thermo-dyn}

In this section we  discuss the thermodynamic limit sometimes also referred to as the mean
 field limit where the quark mass $m$, the axial mass $\lambda$, and the squared lattice spacing
 are of the same order, and satisfy  $m=\Sigma V \widetilde{m}, \lambda=\Sigma V\widetilde{ \lambda}, a^2=W_8 V\widetilde{a}^2 \gg 1$.
This limit is best performed in an eigenvalues representation of 
the supersymmetric integral~\eqref{zgen}. Choosing $V= -iLU$ as a new variable, the
 exponent in the partition function~\eqref{zgen} is given by
\begin{equation}\label{partition-thermo}
\exp\left[ -\frac{1}{2} \Str\widehat{M} (V + V^{-1}) -\frac{1}{2} \Str \widehat{X}(V- V^{-1}) -a^2 \Str (V^2 +V^{-2})\right].
\end{equation}
The corresponding saddle point equation reads
\begin{equation}\label{saddlepoint-equation}
p(V)=\frac{m}{2} (V-V^{-1}) +\frac{\lambda}{2}(V+V^{-1}) -2a^2 (V^2 - V^{-2})=0.
\end{equation}
The solution of this equation is computed in appendix~\ref{app:saddlepointsol}. For $\lambda=0$ it is
\begin{equation}\label{saddle-lambda0-ch}
V=\left\{\begin{array}{cl} \displaystyle \left(\frac{m}{8a^2}-iL\sqrt{1-\left(\frac{m}{8a^2}\right)^2}\right)\eins_4, & |m|<8a^2, \\ \sign\, m\eins_4, & |m|>8a^2. \end{array}\right.
\end{equation}
while for $\lambda\neq0$ it takes the quite complicated form
\begin{equation}\label{sol-a-ch}
V=  e^{\vartheta+L i \varphi}\eins_4\ {\rm with}\ \varphi=\arccos\left[\frac{m\cosh\vartheta+\lambda\sinh\vartheta}{8a^2\cosh2\vartheta}\right]
\end{equation}
and
\begin{eqnarray}\label{saddle-sol-ch}
\hspace*{-0.2cm}\sinh 2\vartheta=-\left(\frac{m\lambda}{(8a^2)^2}+\sqrt{\frac{m^2\lambda^2}{(8a^2)^4}-\frac{1}{27}\left(\frac{m^2-\lambda^2}{(8a^2)^2}-1\right)^3}\right)^{1/3}\hspace*{-0.3cm}-\left(\frac{m\lambda}{(8a^2)^2}-\sqrt{\frac{m^2\lambda^2}{(8a^2)^4}-\frac{1}{27}\left(\frac{m^2-\lambda^2}{(8a^2)^2}-1\right)^3}\right)^{1/3}\hspace*{-0.3cm}.\hspace*{0.2cm}
\end{eqnarray} 

\begin{figure}[t!]
\centerline{\includegraphics[width=0.47\textwidth]{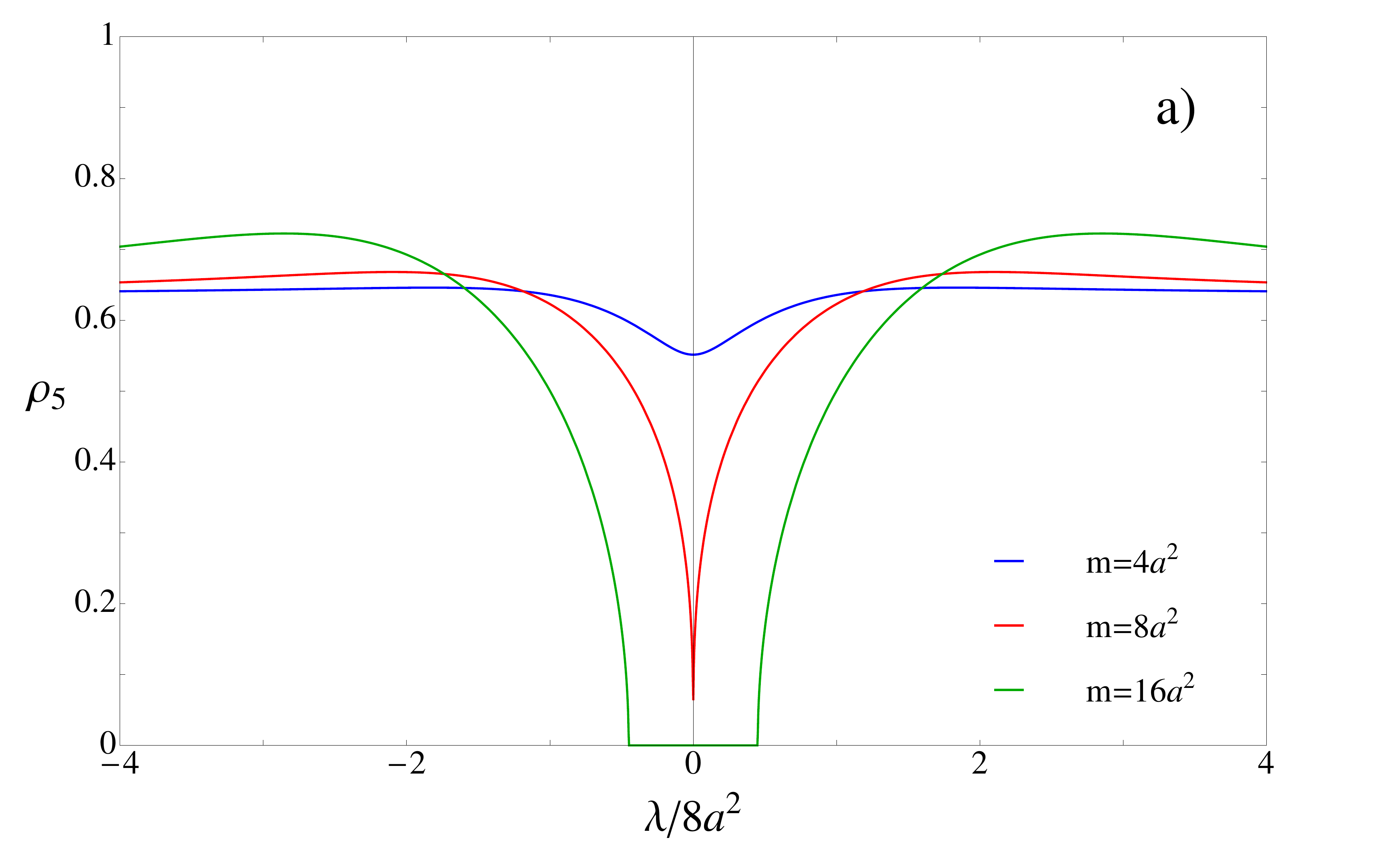}\hfill\includegraphics[width=0.47\textwidth]{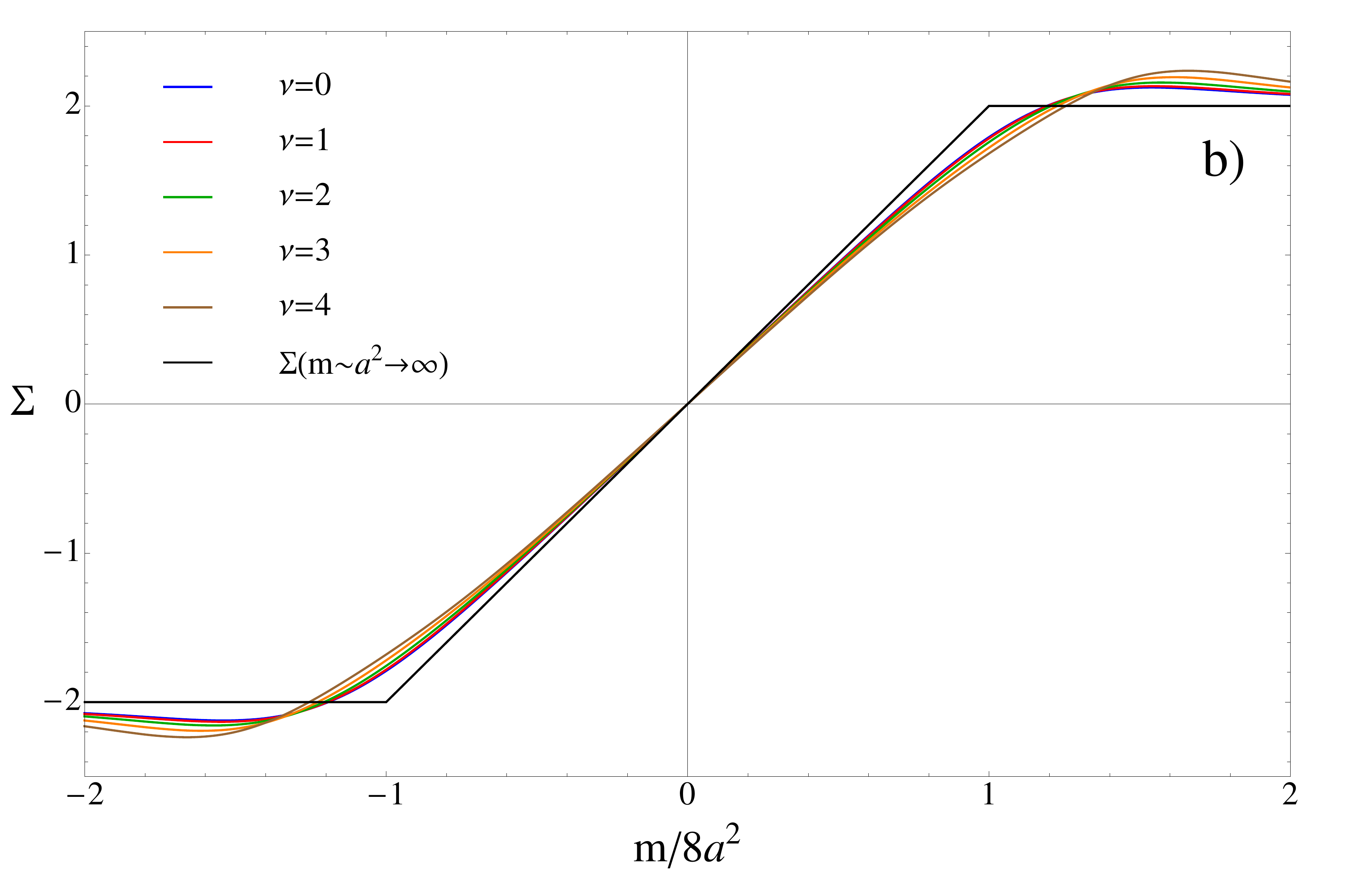}}
\centerline{\includegraphics[width=0.47\textwidth]{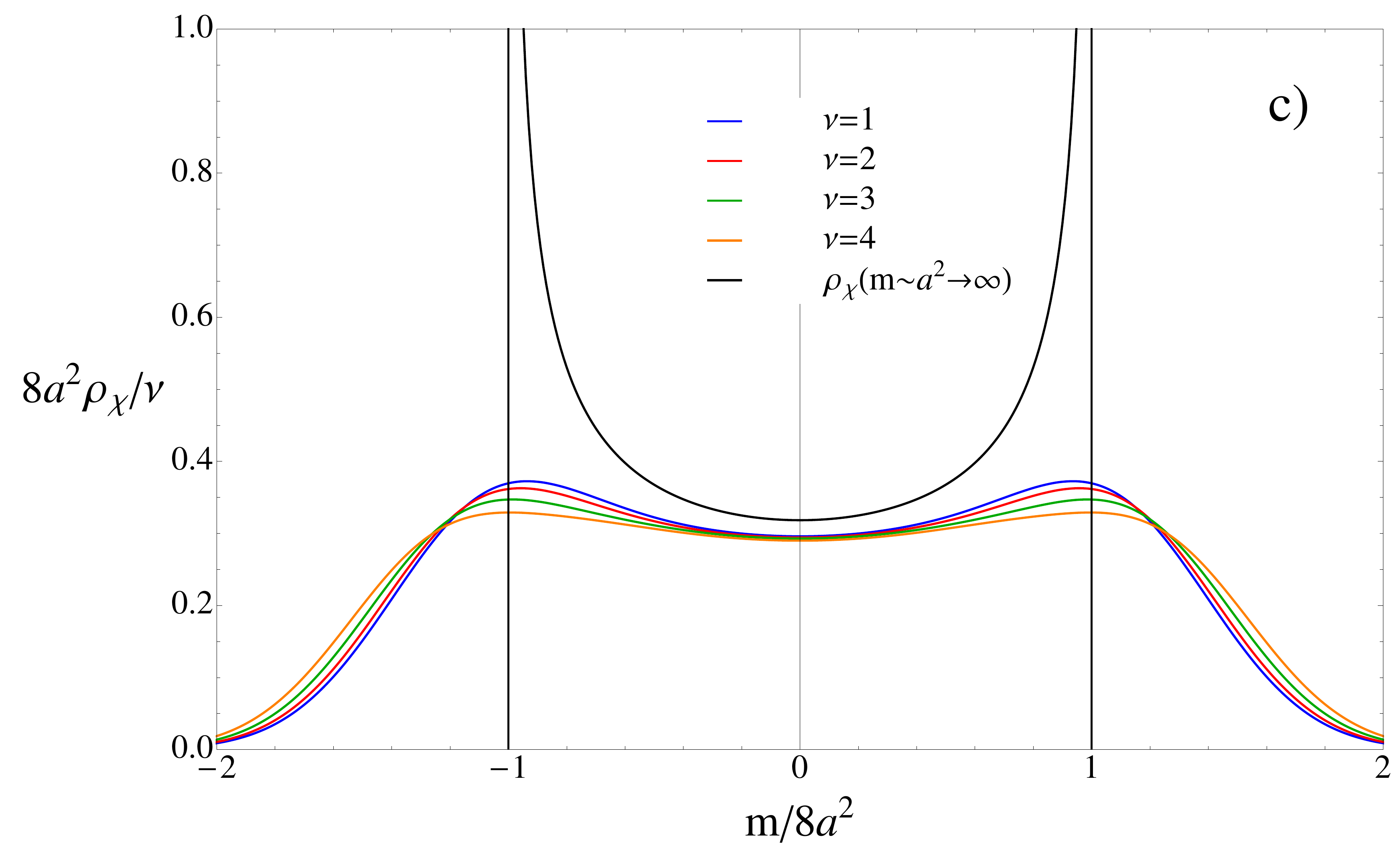}\hfill\includegraphics[width=0.47\textwidth]{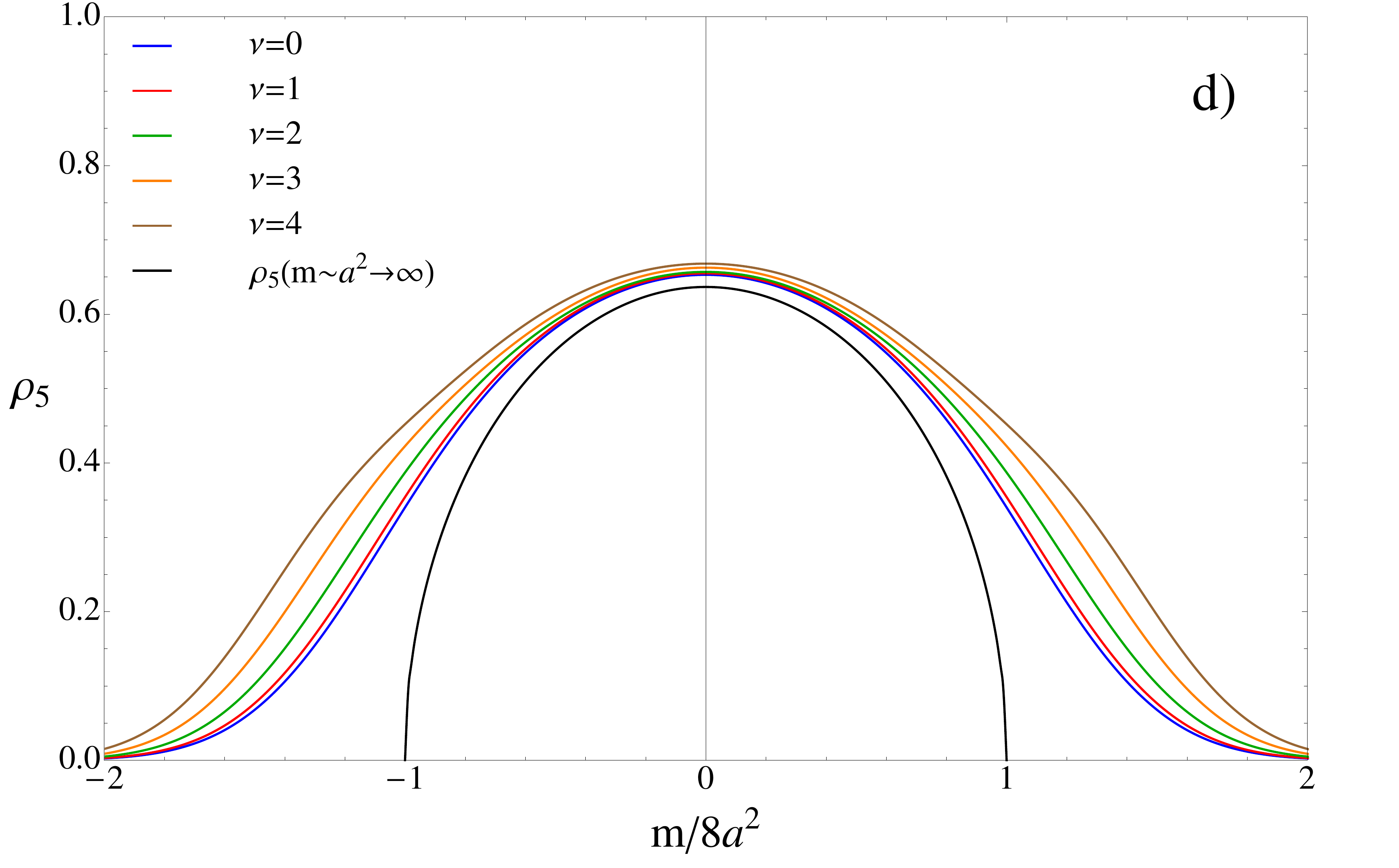}}
\caption{In plot a) we show three different cases for the thermodynamic limit of the level density  $\rho_5$. The Hermitian Wilson Dirac operator $D_5$ exhibits a spectral gap around the origin when the quark mass $|m|$ is larger than $8a^2$. This gap closes at $m=8a^2$ and the system enters the Aoki phase. Hence the value of the level density $\rho_5$ at the origin is an order parameter for the Aoki phase. In plots b), c) and d) we compare the thermodynamic limit (black curves) of the chiral condensate, the distribution of the chiralities over the real eigenvalues, and the mass dependence of the level density $\rho_5$ at the origin $\lambda=0$, respectively, with the behavior at finite lattice spacing $\sqrt{VW_8}\widetilde{a}=a=1$. Although the finite $a$ result has still large deviations from the thermodynamic limit, the phase transition building up at $m=\pm8a^2$ is clearly visible. Also the dependence of the observables on the index $\nu$ has almost disappeared. }
\label{fig:thermo}
\end{figure}

 The level density of the Hermitian Wilson Dirac operator can be calculated from these saddlepoint solutions via
\begin{equation}\label{saddle-rho5}
\rho_5(m,\lambda,a)=\left[\frac{\nu}{2\pi}{\rm Im}\,\Str V^{-1}\partial_{x_1}V+\frac{L}{2}\Str\diag(0,0,1,1){\rm Im}\,(V-V^{-1})\right]_{\substack{m=m'\\x_1=x_2=\lambda}}.
\end{equation} 
The first term is sub-leading such that it can be omitted in the discussion
The density $\rho_5$ is plotted in Fig.~\ref{fig:thermo}.a) and d). It exhibits two different scenarios, either $\rho_5$ has a spectral gap at the origin or the gap is closed. The critical points in the case of a gap can be read off the solution~\eqref{saddle-sol-ch} which are
\be\label{critical}
\lambda= \pm m \left ( 1 -\left (\frac {8a^2}m\right )^{2/3}\right )^{3/2}.
\ee
Instead of the involved derivation of these critical points via the saddle point solution presented in appendix~\ref{app:saddlepointsol} one can also simply find them by a substitution $W={\rm sign}\,m(V-V^{-1})/2$ in Eq.~\eqref{partition-thermo} such that we have to minimize the function
\begin{equation}
q(W)=|m|\sqrt{\eins_4+W^2}+\lambda W-4a^2 W^2.
\end{equation}
At the critical points the first two derivatives of $q(W)$ vanish, i.e.
\be
q'(W)=\frac{mW}{\sqrt{\eins_4+W^2}}+\lambda-8a^2 W=0\ {\rm and}\ q''(W)=\frac{m}{(\eins_4+W^2)^{3/2}}-8a^2=0.
\ee
This yields $W=\pm\sqrt{(m/8a^2)^{2/3}-1}$ implying the critical point~\eqref{critical} from the first derivative. Hence the spectrum of $D_5$ has a gap for $m>8a^2$ in the interval
\be
\lambda\in\left [ -|m| \left ( 1 -\left (\frac {8a^2}m\right )^{2/3}\right )^{3/2},
|m| \left ( 1 -\left (\frac {8a^2}m\right )^{2/3}\right )^{3/2} \right ].
\ee
For $m<8a^2$ and $\lambda=0$ the gap is closed and the system is in the Aoki phase.

The distribution of chirality over the real eigenvalues and the mass dependent chiral condensate are given by the saddlepoints solution $V$ via the relations
\begin{eqnarray}
\rho_\chi(m,a)&=&\left[-\frac{\nu}{2\pi}{\rm Im}\,\Str V^{-1}\partial_{m'}V-\frac{1}{2}\Str\diag(0,0,1,1){\rm Im}\,(V+V^{-1})\right]_{\substack{m=m'\\x_1=x_2=0}},\nn\\
\Sigma(m,a)&=&\left[-\frac{\nu}{2\pi}{\rm Re}\,\Str V^{-1}\partial_{m'}V-\frac{1}{2}\Str\diag(0,0,1,1){\rm Re}\,(V+V^{-1})\right]_{\substack{m=m'\\x_1=x_2=0}}.\label{observe-thermo}
\end{eqnarray}
Since the axial mass is set to zero we only need to consider the solution~\eqref{saddle-lambda0-ch}.  The first term is again sub-leading for $\Sigma(m,a)$ while it is leading for $\rho_\chi(m,a)$. This different behaviour hints to a separation of scales which indeed happens, see Fig.~\ref{fig:thermo}.

We discuss the thermodynamic limit of the spectral observables in more detail in the next section.

  \section{Spectral Statistics of the Wilson Dirac Operator of two color QCD}\label{sec:spectrum}

In this section we discuss the spectral observables in more detail 
and derive their analytical expressions. In particular, we summarize the continuum and thermodynamic limit. 
We study the following spectral observables: The unquenched partition function $Z_\nu^{N_{\rm f}}$
 and its chiral condensate $\Sigma^{N_{\rm f}}$ in subsection~\ref{sec:partition}, 
the level density $\rho_5$ of the Hermitian Wilson Dirac operator $D_5$ in subsection~\ref{sec:rho5}, 
the chiral condensate $\Sigma(m)$ of the quenched theory in subsection~\ref{sec:sigma}, 
the distribution of chirality $\rho_{\chi}$ over the real eigenvalues of the non-Hermitian Wilson Dirac operator $D_{\rm W}$ 
in subsection~\ref{sec:rhochi}, and the level density of the real eigenvalues and the number of 
the additional real eigenvalues in subsection~\ref{sec:rhoreal}.

\begin{figure}[t!]
\centerline{\includegraphics[width=0.47\textwidth]{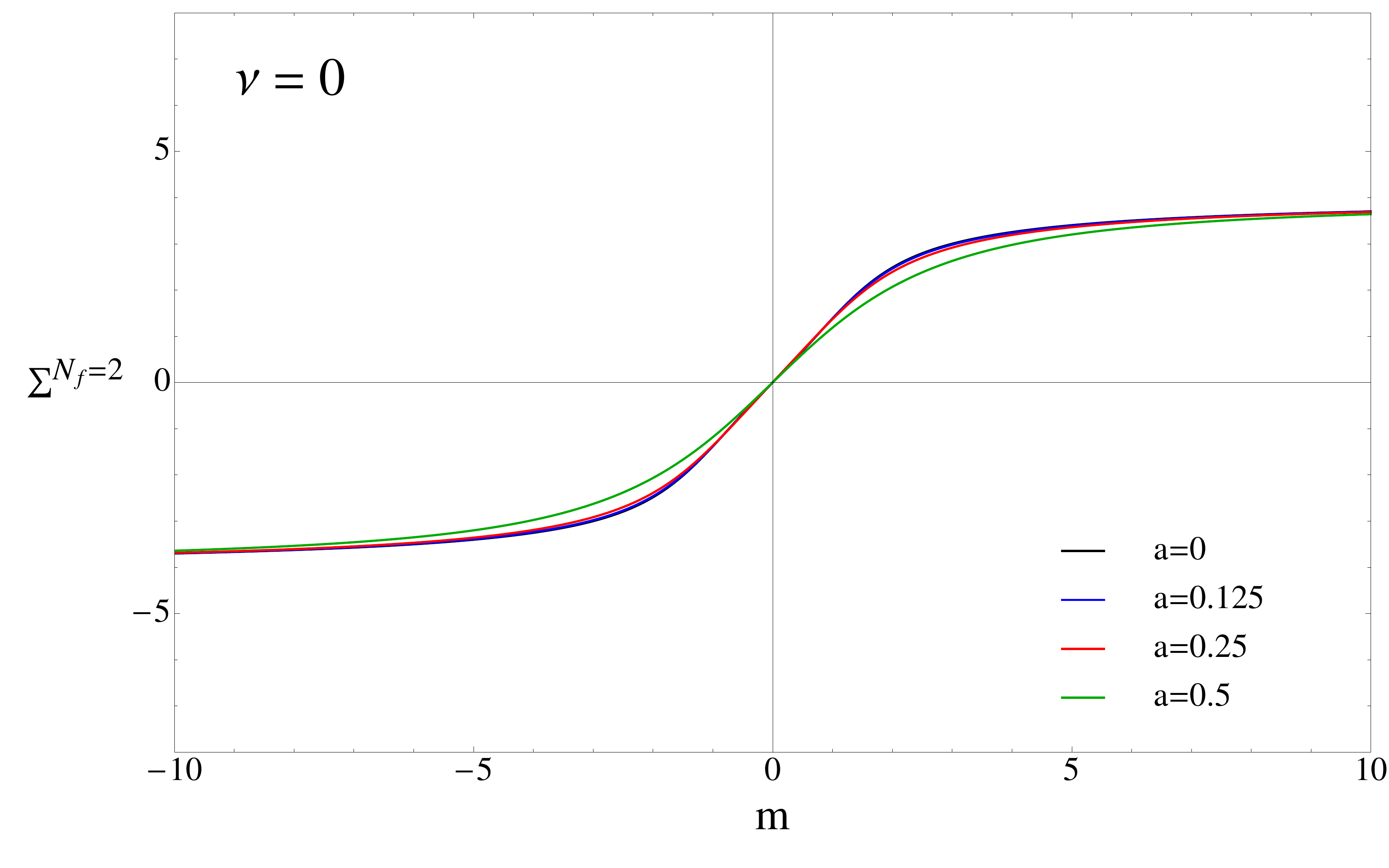}\hfill\includegraphics[width=0.47\textwidth]{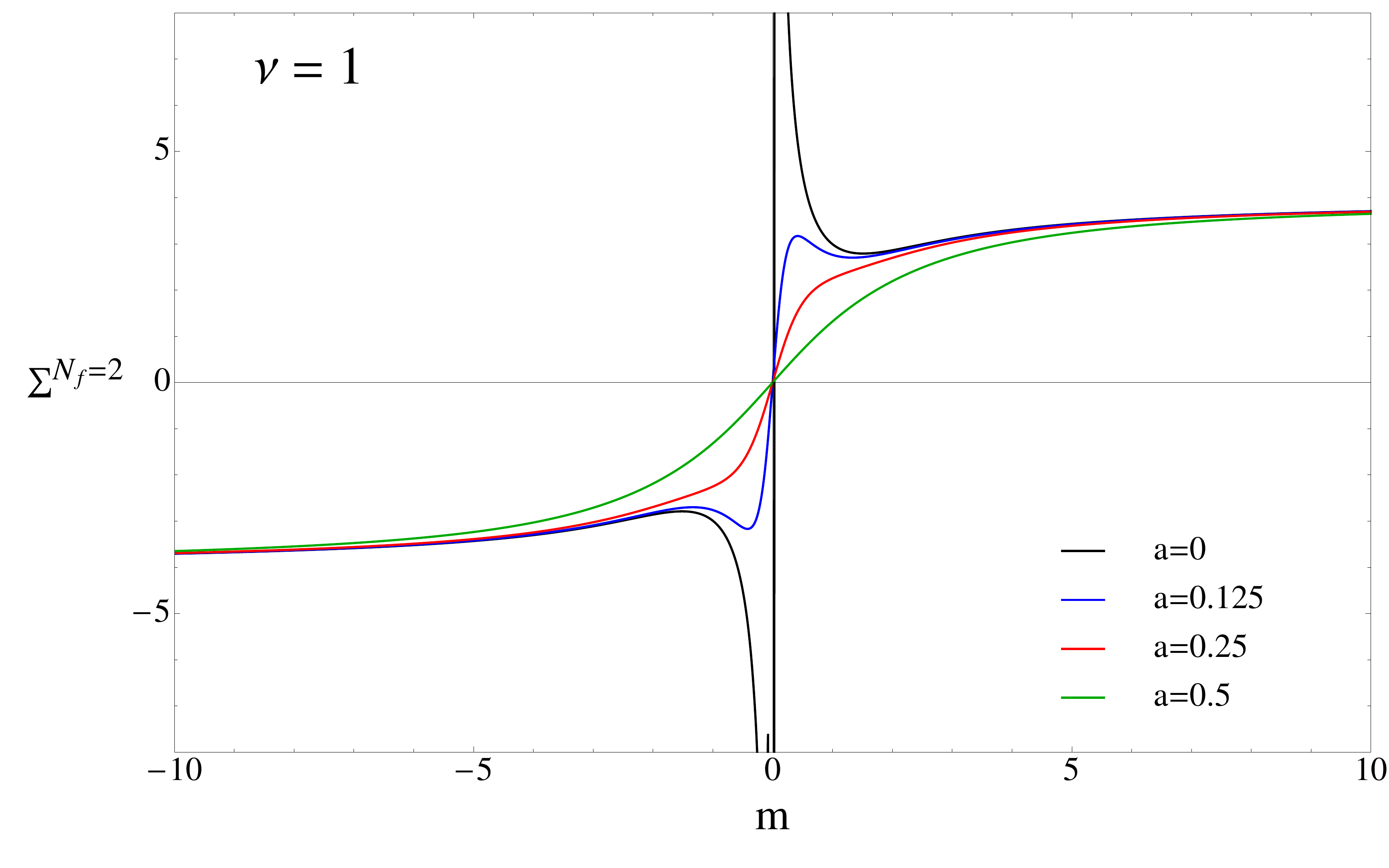}}
\centerline{\includegraphics[width=0.47\textwidth]{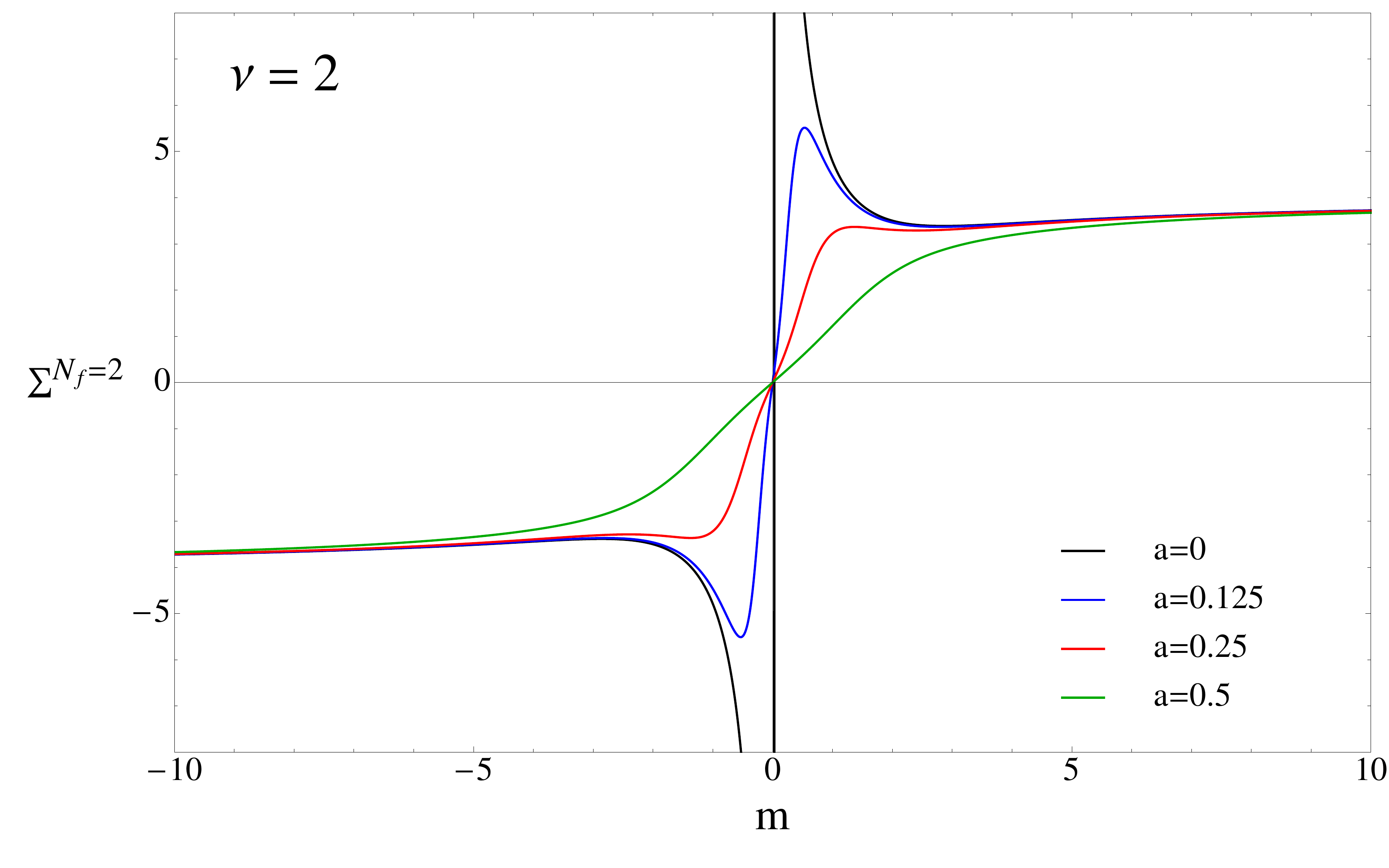}\hfill\includegraphics[width=0.47\textwidth]{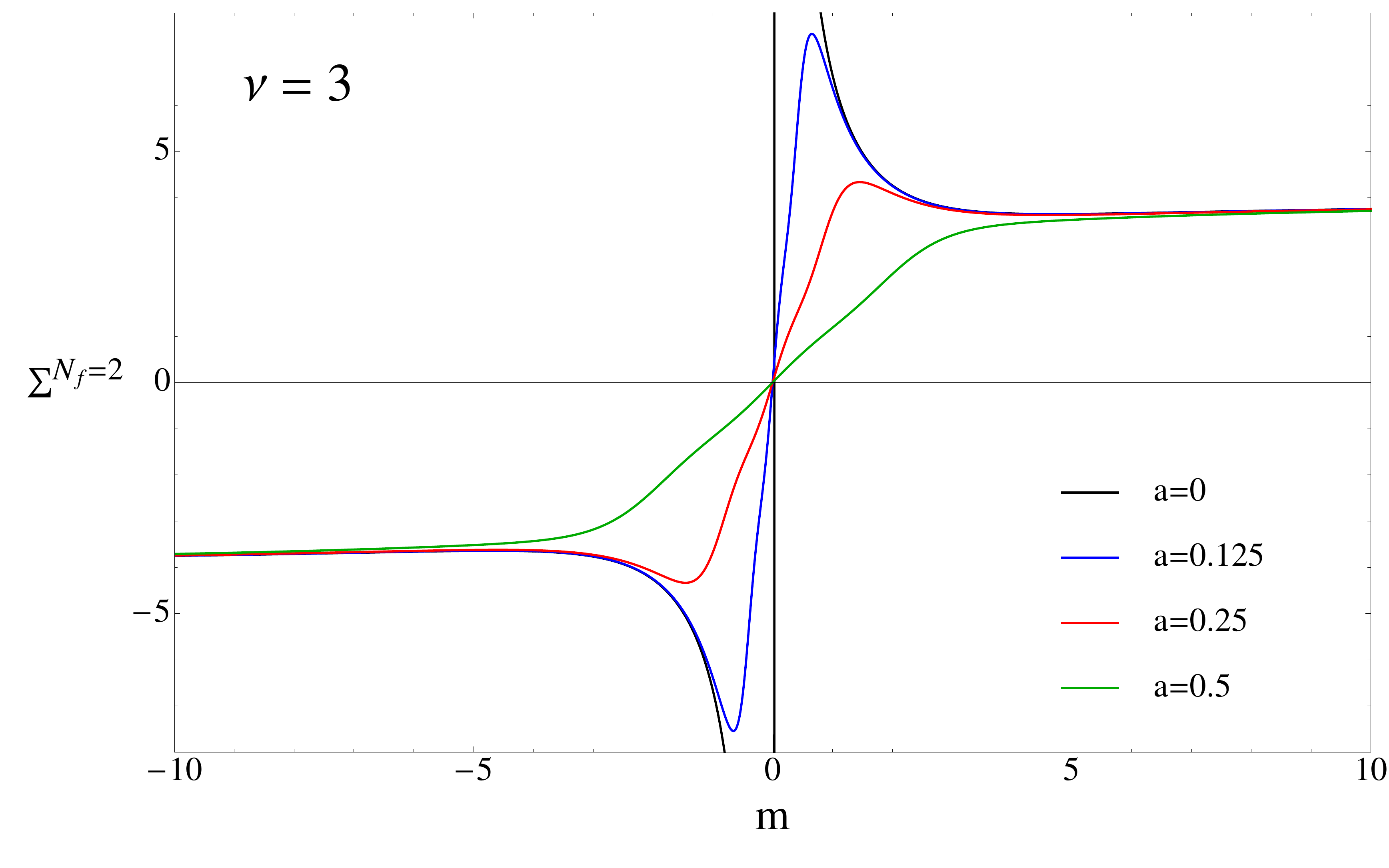}}
\caption{Chiral condensate for two dynamical quarks for index $\nu=0,1,2,3$ and lattice spacing 
$\sqrt{VW_8}\widetilde{a}=a=0,0.125,0.25,0.5$. 
The continuum limit (black curves) diverges as $\nu/m$ for $m \to 0$
 due to the zero modes. For larger index $\nu$, 
the peaks reminiscent of this singularity are 
more persistent at finite lattice spacing. 
The reason is the localization of real eigenvalues
 around the origin which are smoothed  out only very slowly.}
\label{fig:SigmaNf}
\end{figure}

\subsection{Partition Function of Dynamical Quarks}\label{sec:partition}

For $N_{\rm f} =1$ there is no spontaneous symmetry breaking and the QCD 
partition function for QCD with two colors is the same as for QCD
with three or more colors. Indeed for $N_{\rm f} =1$ the partition function~\eqref{zchi} can be written as
\be\label{part-nf1}
Z_{\nu}^{N_{\rm f}=1}(m,a)=\int_{-\pi}^\pi \frac{d\varphi}{2\pi} e^{i\nu\phi} \exp[2m\cos\varphi -4a^2 \cos 2\varphi],
\ee 
which coincides with the one-flavor partition function for $\beta =2$, see \cite{ADSV10}.

For two or more flavors  the partition function is the $N_{\rm f}(2N_{\rm f}-1)$ dimensional integral~\eqref{zchi}.  For simplicity we assume that all quark masses are equal to $m$. Then the partition function
\be\label{part-nf}
Z_{\nu}^{N_{\rm f}}(m,a)=\int_{\CSE(2N_{\rm f})} d\mu(U) {\det}^{\nu/2}U \exp\left[\frac{m}{2}\tr(U+U^{-1})-a^2\tr(U^2+U^{-2})\right]
\ee
can be calculated by diagonalizing $U$ and applying de Bruijn's integration theorem~\cite{DeBruijn}. Then we end up with a Pfaffian (essentially an exact square root of a determinant of an antisymmetric matrix) \cite{Smilga-V}
\be\label{part-nf-b}
Z_{\nu}^{N_{\rm f}}(m,a)\propto \Pf[A]\quad {\rm with}\quad A_{kl}=(k-l)\Phi^0_{\nu-2N_{\rm f}+k+l-1}(m,x_0=0,a), 1\le k,l \le 2N_f.
\ee
where $\Phi_0^\mu$ is the integral given in Eq.~(\ref{comp-int}) evaluated of $x_0 =0$,
\be
\Phi^0_\mu(m,x_0=0,a)&=&(-i)^\mu\int_{-\pi}^\pi\frac{d\varphi}{2\pi}e^{i\mu\varphi}\exp\left[2m\sin\varphi+4a^2\cos2\varphi\right].
\ee
 The chiral condensate is easily calculated by taking the derivative of the free energy ${\rm ln}\,Z_{\nu}^{N_{\rm f}}(m,a)$ with respect to the quark mass resulting in
\be\label{sigma-nf}
\Sigma_{\nu}^{N_{\rm f}}(m,a)=\frac 1{2}\partial_m {\rm ln}\,Z_{\nu}^{N_{\rm f}}(m,a)= \frac{1}{4}\tr A^{-1}\partial_m A.
\ee
The derivative of $A$ is given by
\be\label{A-der}
\partial_m A_{kl}=(k-l)[\Phi^0_{\nu-2N_{\rm f}+k+l}(m,x_0=0,a)+\Phi^0_{\nu-2N_{\rm f}+k+l-2}(m,x_0=0,a)].
\ee
The chiral condensate is normalized to the asymptotics 
$\displaystyle{ \lim_{m\to+\infty}\Sigma_{\nu}^{N_{\rm f}}(m,a)=N_f}$. 
In Fig.~\ref{fig:SigmaNf} we illustrate the behavior for particular indices $\nu=0, 1, 2, 3$.

As already mentioned in subsection~\ref{sec:ferm}, the unquenched partition function~\eqref{part-nf} 
factorizes in the partition function of continuum QCD without the zero modes and in the partition 
function of a $\nu\times\nu$ dimensional GOE in the continuum limit $|a|\ll1$. 
The spectrum of the finite dimensional GOE  is of the order $a$ which shrinks to the origin in the continuum limit.
This creates a singularity of the chiral condensate at $a=0$ for configurations with $\nu\neq0$, cf. Fig.~\ref{fig:SigmaNf}.

\begin{figure}[t!]
\centerline{\includegraphics[width=0.47\textwidth]{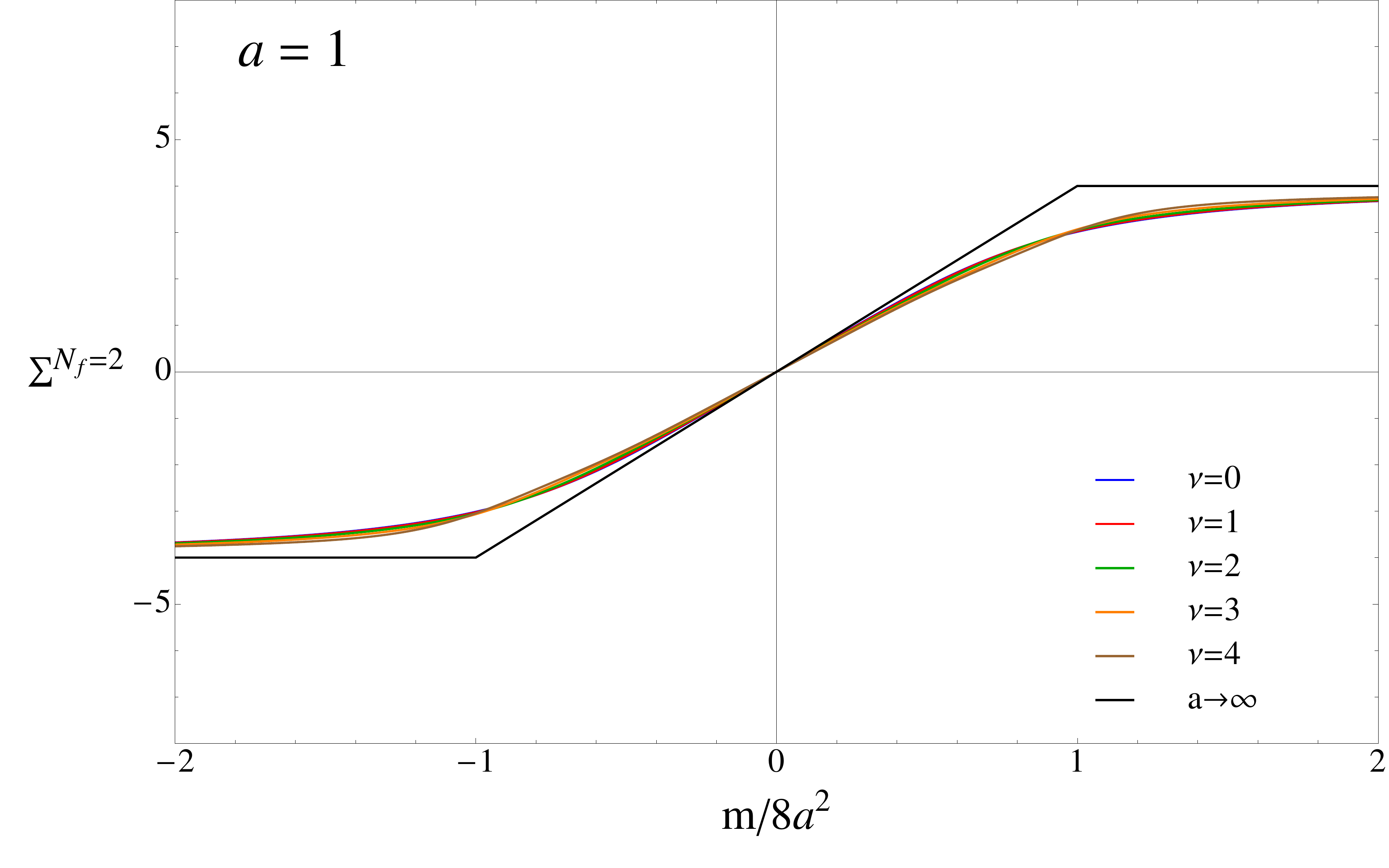}\hfill\includegraphics[width=0.47\textwidth]{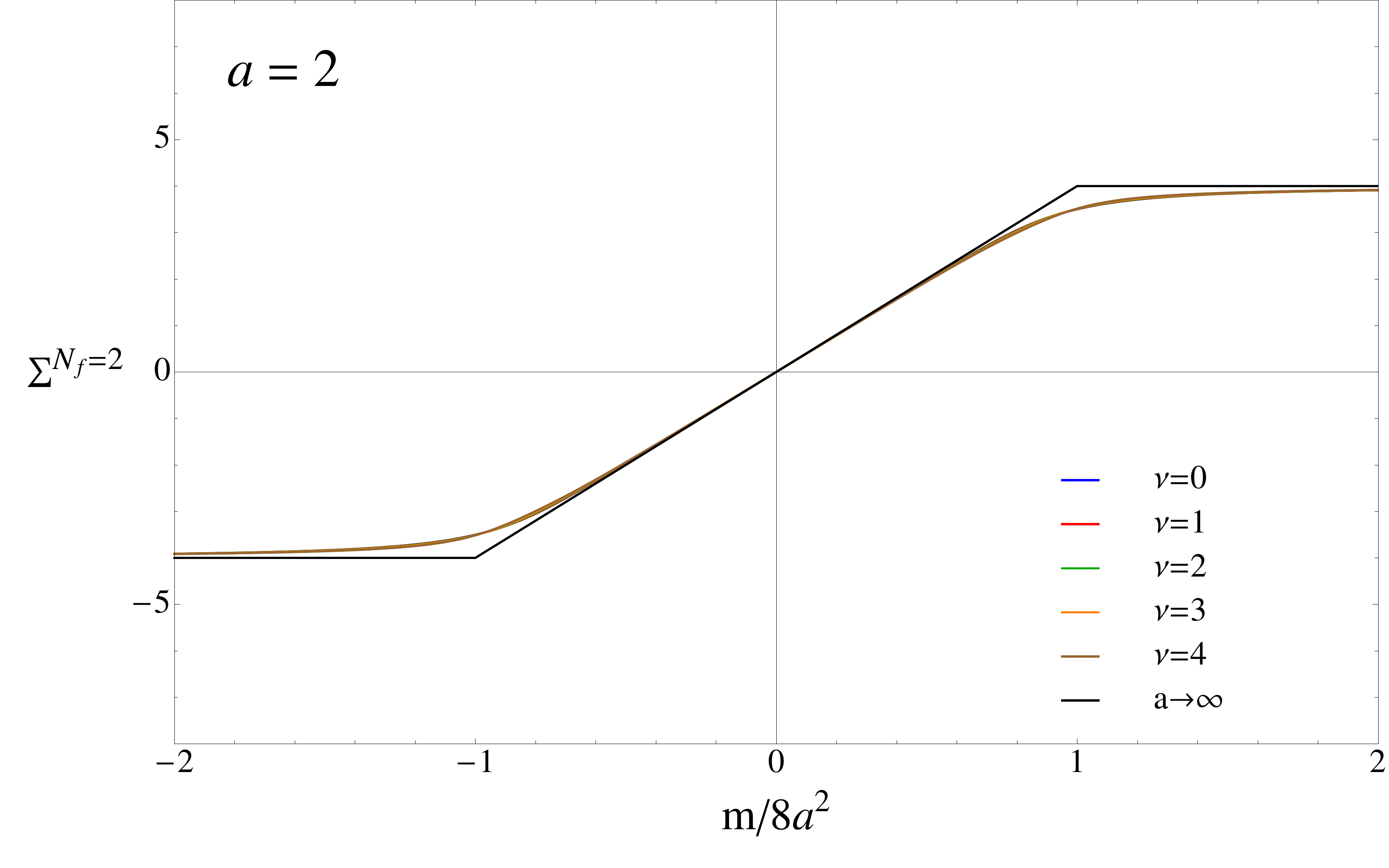}}
\caption{Comparison of the chiral condensate with two dynamical quarks
 for index $\nu=0,1,2,3,4$ with the thermodynamic limit (black curve).
 Although around the origin  the thermodynamic limit 
is a good approximation 
for  lattice spacing $\sqrt{VW_8}\widetilde{a}=a=1$,
 we have still large deviations in the phase transition region 
around the quark mass $m=8a^2$. 
At $\sqrt{VW_8}\widetilde{a}=a=2$ the thermodynamic limit is almost approached. Nonetheless the $\nu$ dependence is barely visible already at smaller lattice spacings.}
\label{fig:SigmaNFabig}
\end{figure}

In the thermodynamic limit the partition function is either dominated by the saddlepoint 
$\U_0=\sign\, m$ for $|m|\geq 8a^2$ or by the saddlepoints satisfying $U_0+U_0^{-1}=m/4a^2$ for $|m|\leq 8a^2$. 
Then the resulting chiral condensate is given by
\begin{equation}\label{sigmanf-thermo}
 \langle\bar\psi\psi\rangle_{N_{\rm f}}=\frac 1{N_f}\Sigma_{\nu}^{N_{\rm f}}(m,a)\approx\left\{\begin{array}{cl} \displaystyle \frac{m}{4a^2}, & |m|<8a^2, \\  \sign\, m, & |m|>8a^2. \end{array}\right.
\end{equation}
This result does not depend on $N_{\rm f}$ and we will show
in subsection~\ref{sec:sigma} that it is also valid for  the quenched theory. 
The thermodynamic limit is already well approximated for a lattice spacing $\sqrt{VW_8}\widetilde{a}=a\approx2$, see Fig.~\ref{fig:SigmaNFabig}.

Finally, let us show that the sign of $W_8$ has to be positive. We start from the observation that the partition function~\eqref{part-nf} has to be the same as the average over the eigenvalues over $D_5+m\gamma_5$ which are real,
\be
Z_\nu^{N_{\rm f}}(m,a)&=&(-1)^{\nu N_{\rm f}} \left\langle \prod_j(\lambda_j^{5}(m))^{N_{\rm f}}\right\rangle_{D_5} ,\label{part-nf-c}
\ee 
Hence the partition function has to be positive definite for  $N_{\rm f}$ even. The sign $(-1)^{\nu N_{\rm f}}$ in front of the average results from multiplying $\gamma_5$ with the non-Hermitian Wilson Dirac operator $D_{\rm W}$.
As is the case for $\beta =2$ the partition function satisfies the general relation (see Eq. (\ref{zchi}))

\be
Z_\nu^{N_{\rm f}}(m=0,a^2)=(i)^{N_{\rm f}\nu}Z_\nu^{N_{\rm f}}(m=0,-a^2).
\label{zrel}
\ee
Therefore, the partition function for $N_{\rm f}\in4\mathbb{N}_0+2$ cannot be positive definite for both values of the sign of $W_8$.
This can also be seen from the explicit expression for the two-flavor partition function, which,
 at $ m= 0$
is given by the two dimensional integral
\be\label{part-nf2}
Z_{\nu}^{N_{\rm f}=2}(m=0,a)=\frac{1}{2}\int_{-\pi}^\pi \frac{d\varphi_1}{2\pi}\int_{-\pi}^\pi \frac{d\varphi_2}{2\pi} 
|e^{i\varphi_1}-e^{i\varphi_2}|^4 e^{i\nu(\varphi_1+\varphi_2)} \exp[-4a^2 \cos 2\varphi_1-4a^2 \cos 2\varphi_2].
\ee 
All odd powers of the phases $ e^{i\varphi_1}$ and $ e^{i\varphi_2}$ vanish since the $a$ dependent term has double the frequency. Therefore,
for odd $\nu$,  
only two  terms remain  in the expansion of the Vandermonde determinant. They can be combined as 
\be\label{part-nf2-b}
Z_{\nu}^{N_{\rm f}=2}(m=0,a)=-4I_{(\nu-1)/2}(-4a^2)I_{(\nu+1)/2}(-4a^2)=4I_{(\nu-1)/2}(4a^2)I_{(\nu+1)/2}(4a^2).
\ee 
This term is always positive definite if $a$ is real and thus $W_8$ is 
positive while it is negative when $VW_8\widetilde{a}^2=a^2<0$. 
Hence a negative value of $W_8$
  contradicts to the positivity of the partition function.
  
  The unquenched partition function~\eqref{part-nf2-b} is shown in Fig.~\ref{fig:partition} for $\nu=1$ and $N_{\rm f}=2, 4$. 
Indeed the partition function for $N_f=4$ does not depend on the sign of $a^2$ at $ m = 0$ (see Eq. (\ref{zrel})), but the figure shows that the
partition function  is not positive definite for a negative value of $W_8$.
\begin{figure}[t!]
\centerline{\includegraphics[width=0.47\textwidth]{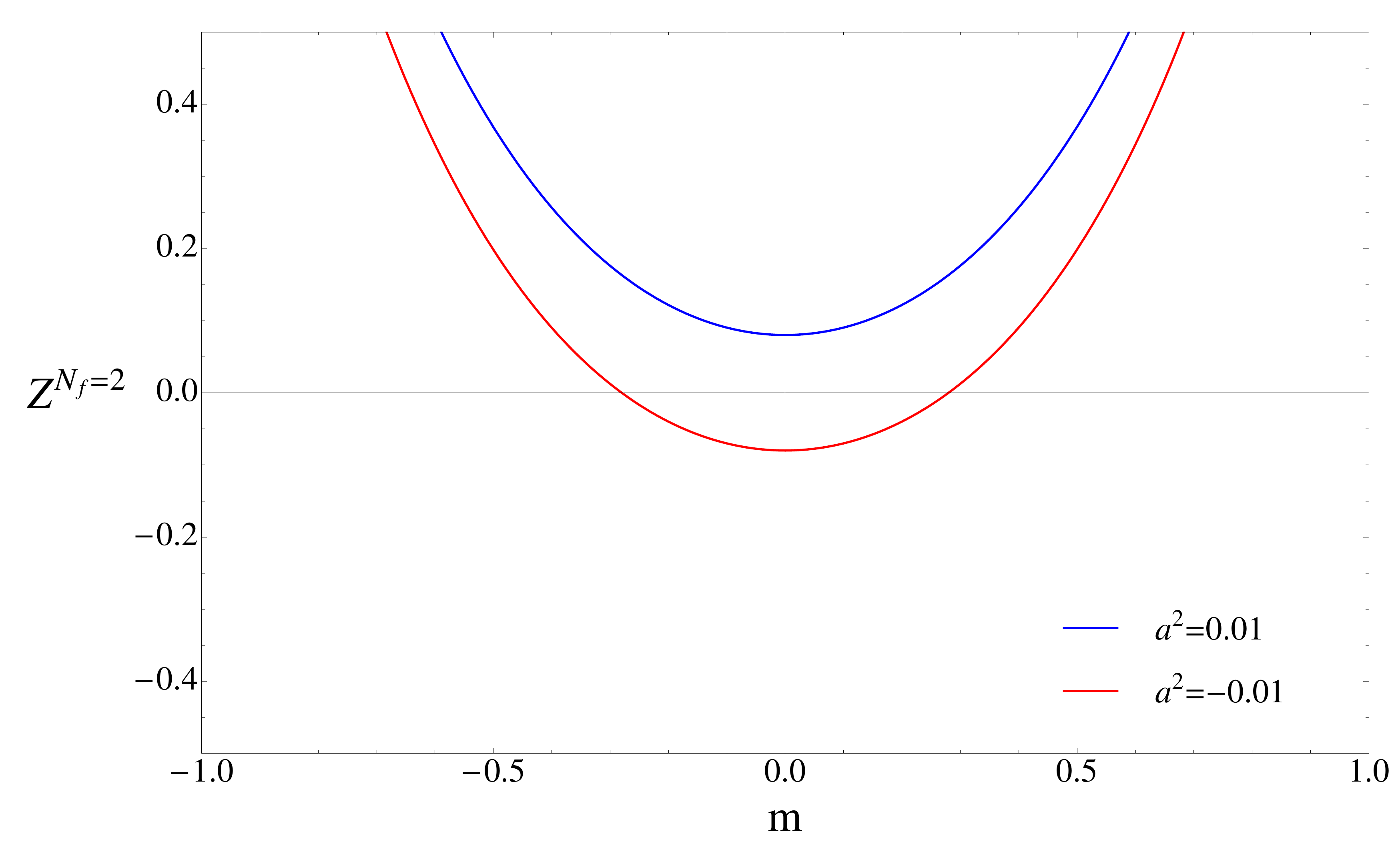}\hfill\includegraphics[width=0.47\textwidth]{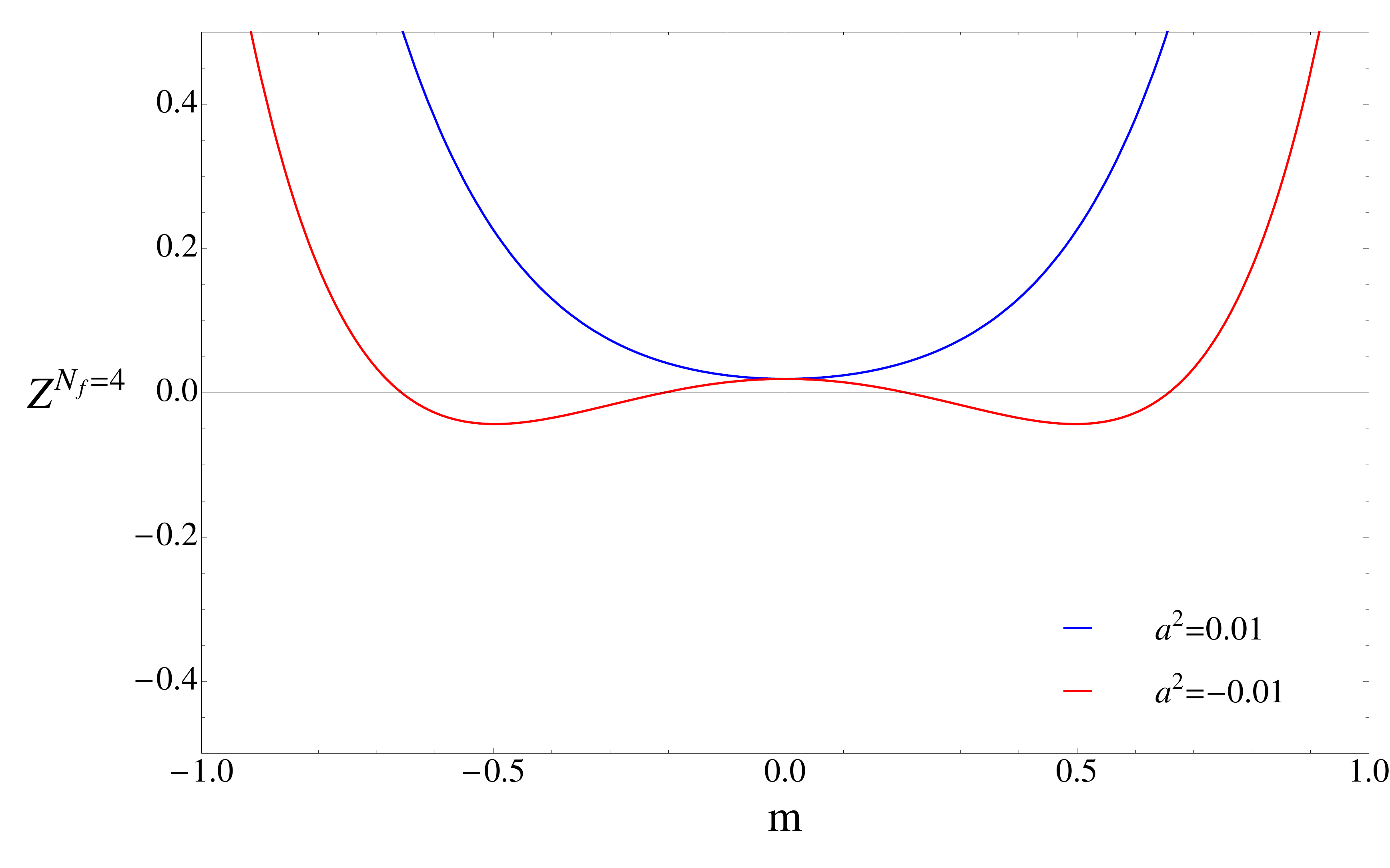}}
\caption{The mass dependence of the two flavor (left) and the four flavor partition function (right)
for $VW_8\widetilde{a}^2=a^2 =\pm0.01 $ and $\nu =1$. In the case where the low energy constant is negative we have always
  regions where the chiral partition function is negative although it is positive when considering the full theory.}
\label{fig:partition}
\end{figure}

\subsection{Level Density of $D_5$}\label{sec:rho5}

The level density $\rho_5$ given by
\be
\rho_5(m,\lambda,a)&=&\left.\partial_{x_1} {\rm Im}\,Z_\nu(\widehat{M},\widehat{X},a)\right|_{\substack{\widehat{M}=m\eins_4\\ \widehat{X}=(\lambda+i\varepsilon)\eins_4\to\lambda\eins_4}}\nn
\ee
immediately results from combining Eqs.~\eqref{G5-def}, \eqref{rho5-def}, \eqref{final}, and \eqref{derivatives-S}. We obtain a quite complicated expression 
(see Eq.~(\ref{rho5-result}) of  appendix~\ref{app:explicit})
which can be evaluated numerically. In Fig.~\ref{fig:MCcomprho5} we compare this result with Monte Carlo simulations of the random matrix theory~\eqref{RMT}.

\begin{figure}[t!]
\centerline{\includegraphics[width=0.47\textwidth]{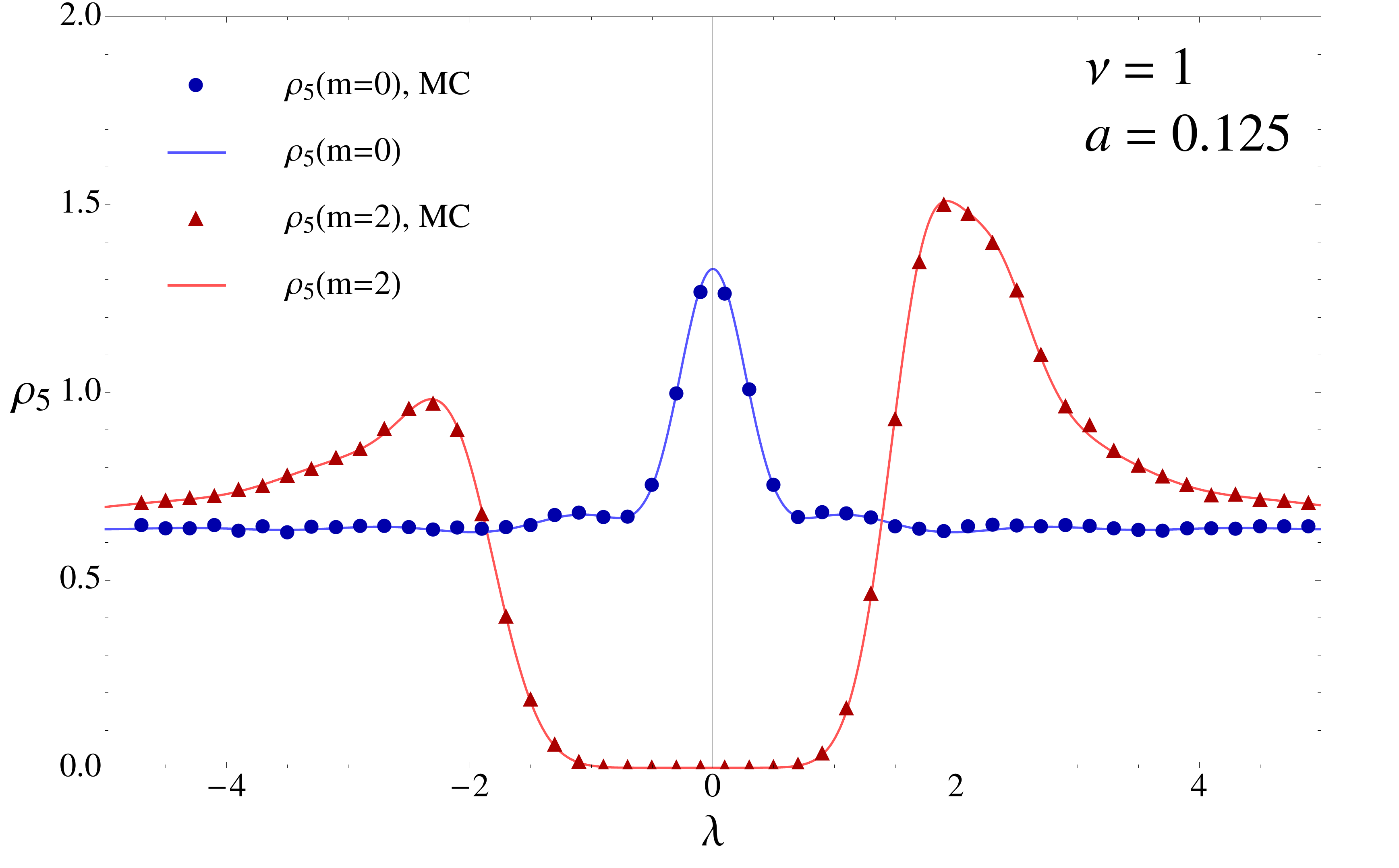}\hfill\includegraphics[width=0.47\textwidth]{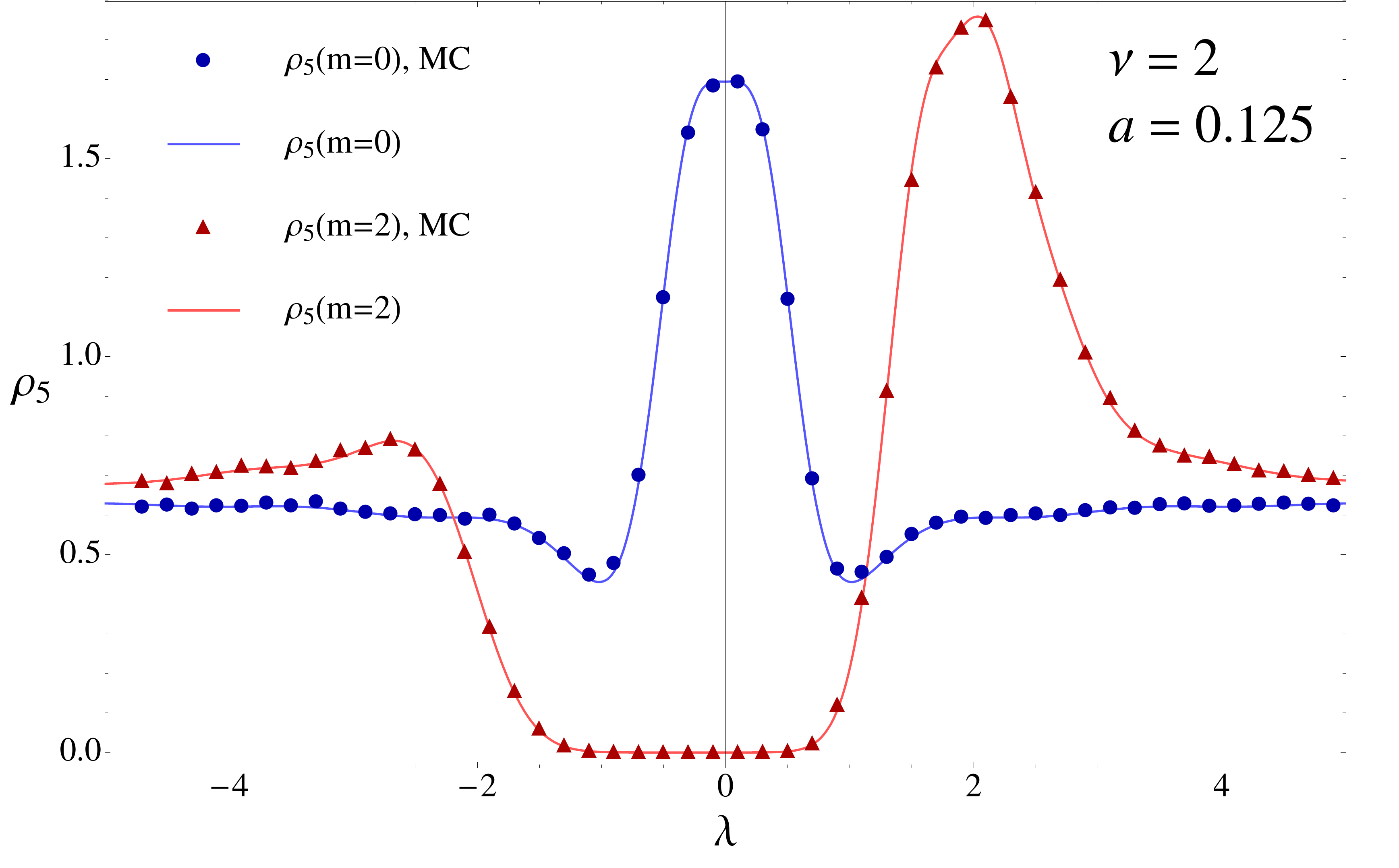}}
\centerline{\includegraphics[width=0.47\textwidth]{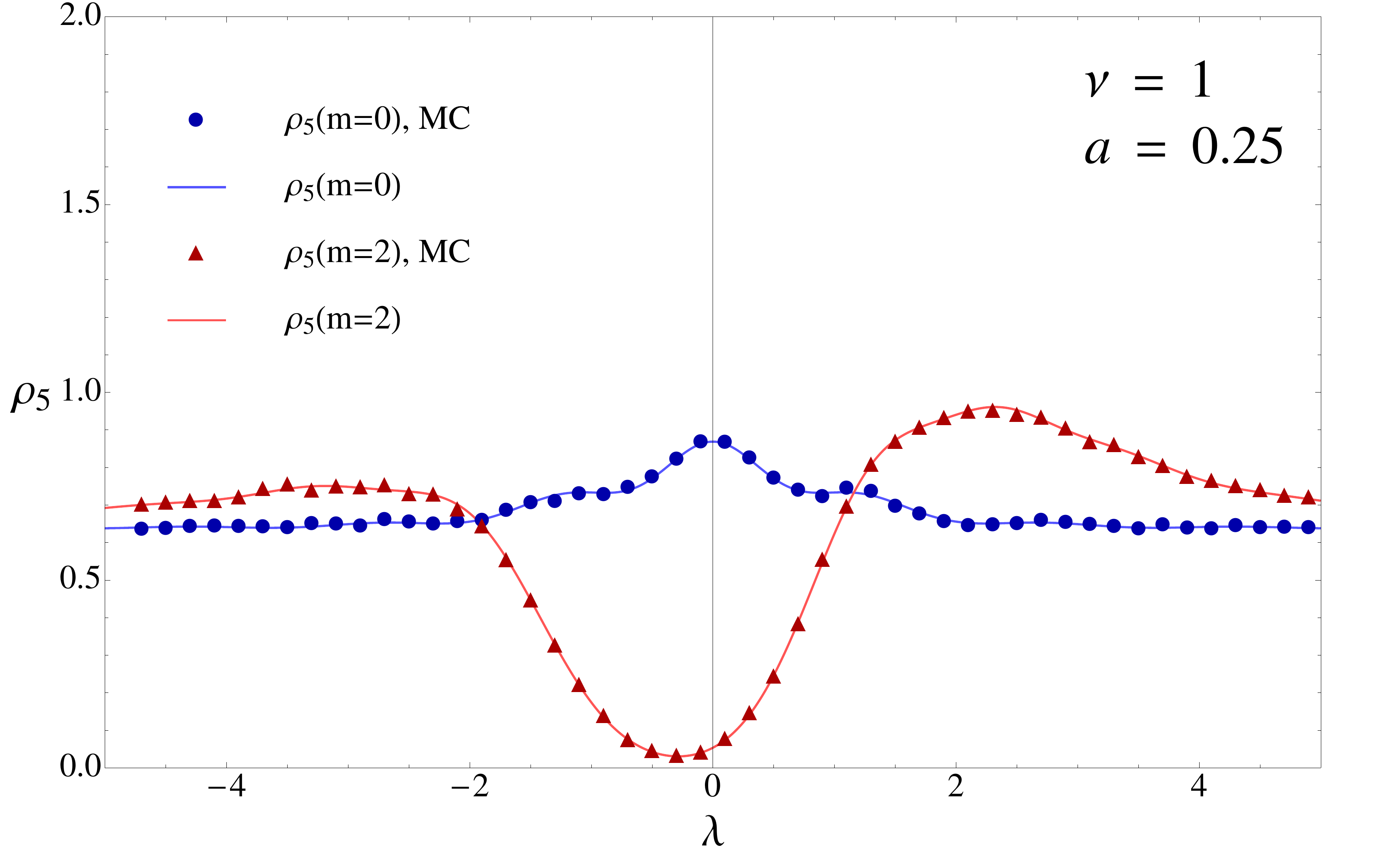}\hfill\includegraphics[width=0.47\textwidth]{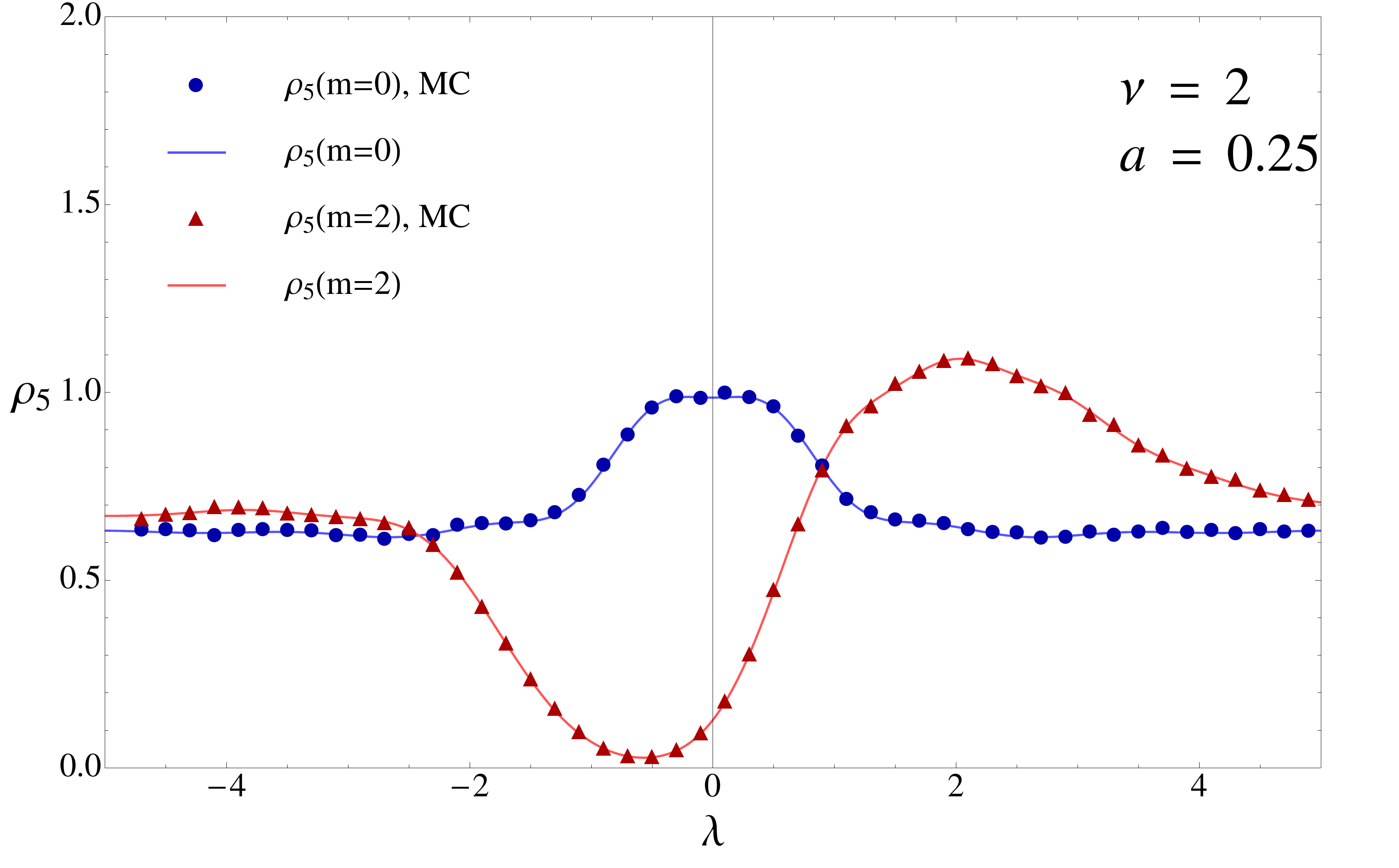}}
\centerline{\includegraphics[width=0.47\textwidth]{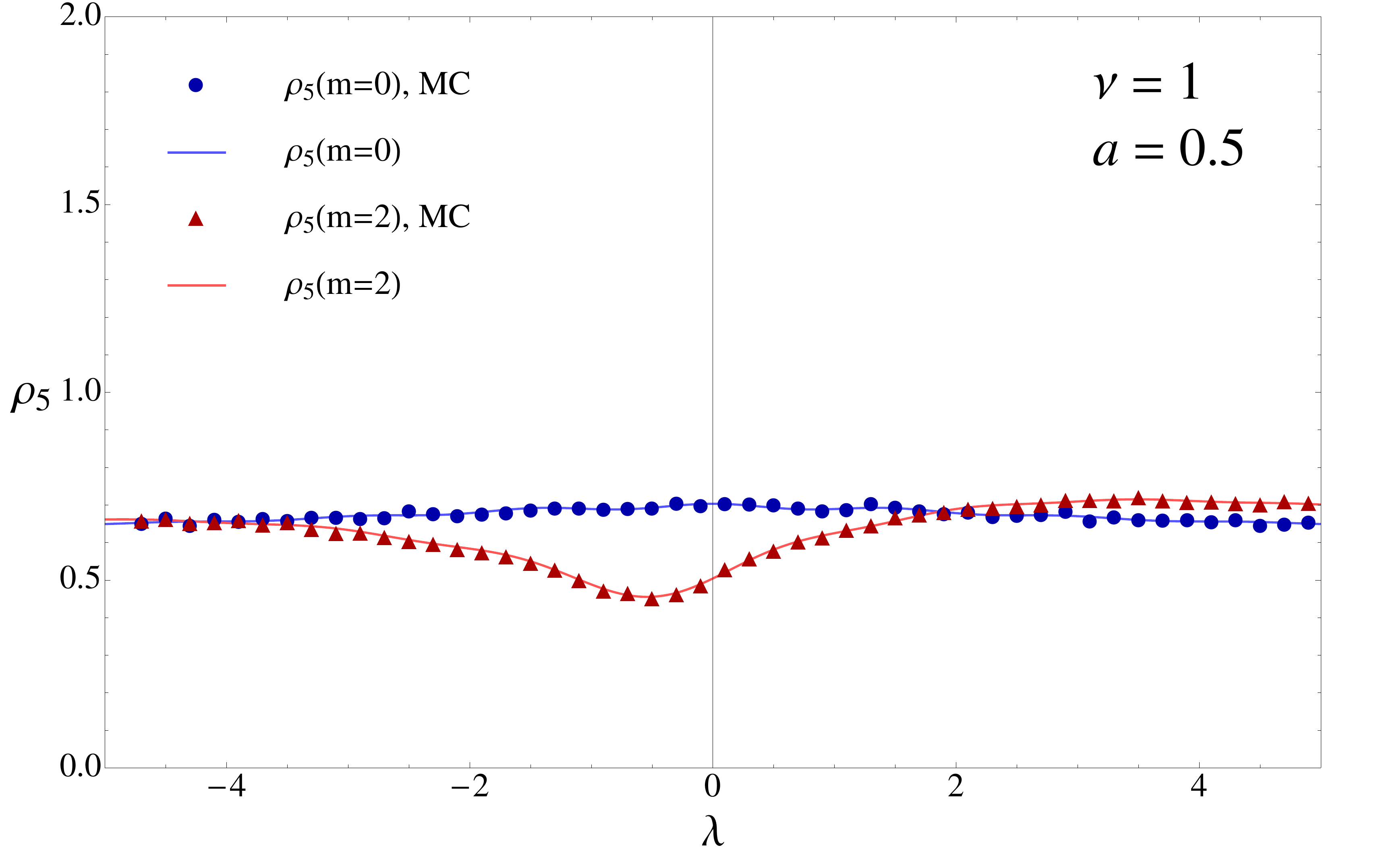}\hfill\includegraphics[width=0.47\textwidth]{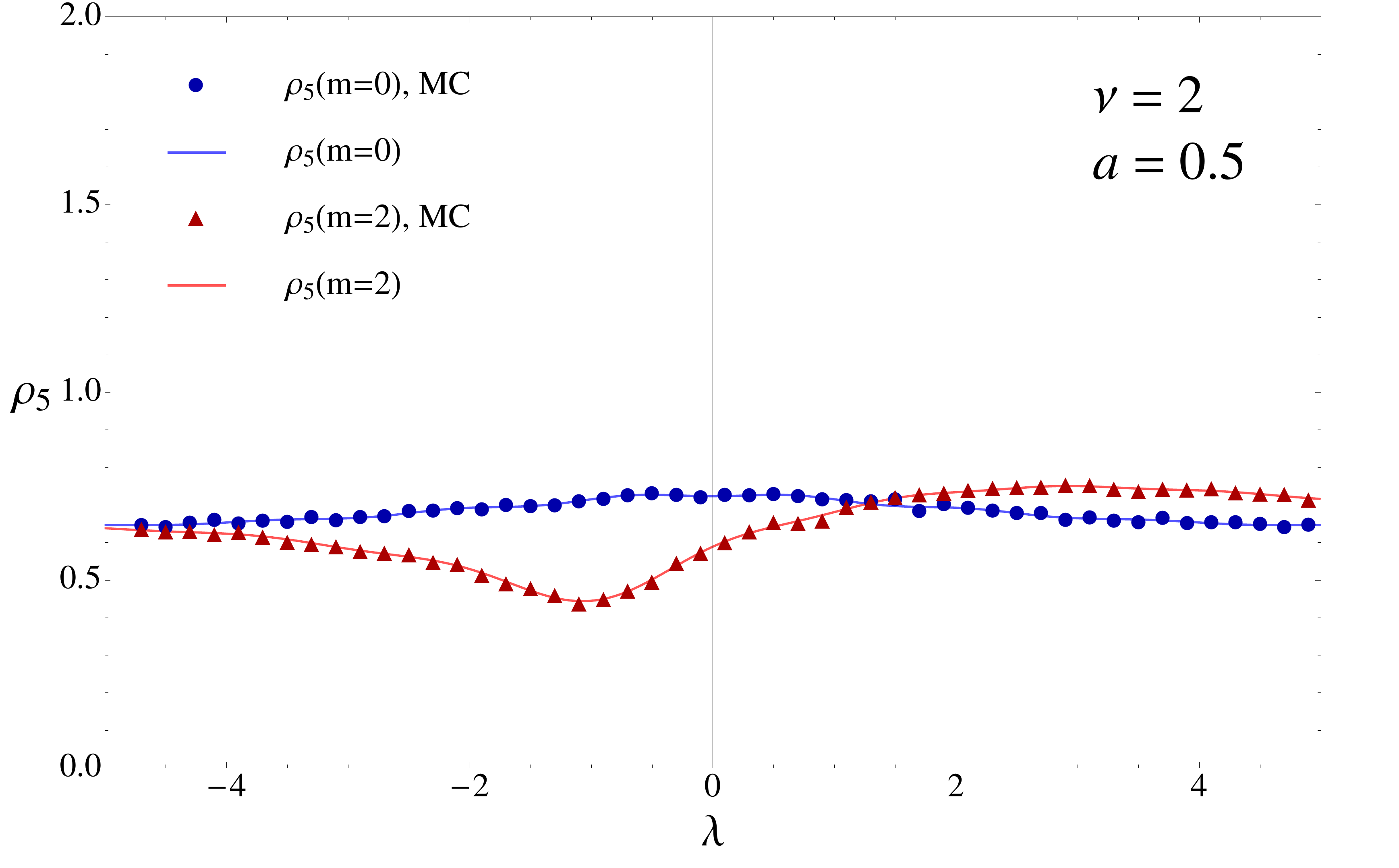}}
\caption{Comparison of the analytical result~\eqref{rho5-result} for 
the level density $\rho_5$ (solid curves) with Monte Carlo simulations (MC, symbols).
 We have simulated for two different values for  the valence quark mass,
$m=0$ (blue curves) and  $V\Sigma\widetilde{m}=m=2$ (red curves). The index has been chosen $\nu=1$ (left plots) and $\nu=2$ (right plots) and the lattice spacing is $\sqrt{VW_8}\widetilde{a}=a=0.125,0.25,0.5$.
 For each plot we have generated $10^5$ matrices distributed according to 
 Eq.~\eqref{RMT-Prob}. The prominent peaks at $m=0$ and $m=2$ result from the broadened former zero modes. They will become Dirac delta functions in the continuum limit.}
\label{fig:MCcomprho5}
\end{figure}

When taking the continuum limit $a\to0$ the peak around the quark mass $m$  shrinks to a single Dirac delta function with area $\nu$. 
At non-zero $a$, the peaks are no longer degenerate and broaden to a width $\sim a$.
Due to the weak level repulsion which is linear for real matrices we do not observe separate peaks 
for  $\nu>1$.
 
In the limit $ a\to 0$  the density of the  cluster of $\nu$ peaks around $m$ 
combine into  the level density $\rho_{\GOE}$, 
see Eq.~\eqref{level-GOE}, of a $\nu\times\nu$ dimensional Gaussian, cf. Figs.~\ref{fig:comp} b), c)  and d). In this limit we 
expect that $\rho_5$ can be approximated by
 \begin{equation}\label{rho5-approx}
 \rho_5(m,\lambda,a)\overset{|a|\ll1}{\approx}\frac{|\lambda|}{\sqrt{\lambda^2-m^2}}
\rho_{\chiGOE}^{(\nu)}\left(\sqrt{\lambda^2-m^2}\right)\Theta(\lambda^2-m^2)+\rho_\GOE^{(\nu)}\left(\frac{m-\lambda}{4a}\right)
 \end{equation}
 with $\rho^{\nu}_\GOE$ the level densities~\eqref{rho-GOE-nu0}, \eqref{rho-GOE-nu1}, and \eqref{level-GOE} depending on $\nu$ and 
 \begin{equation}\label{rho-chiGOE}
 \rho_{\chiGOE}^{(\nu)}(x)=J_\nu(2|x|)\left(1-\int_0^{2|x|}dyJ_\nu(y)\right)+2|x|\left(J_\nu^2(2|x|)-J_{\nu-1}(2|x|)J_{\nu+1}(2|x|)\right)
 \end{equation}
 the microscopic level density of continuum QCD and, thus, of chiral GOE without zero modes \cite{GOE-level}. This approximation is shown in Figs.~\ref{fig:comp}.b), c) and d).
 
 As we already pointed out in subsection~\ref{sec:small-a} the continuum limit via the approximation~\eqref{rho5-approx} is not uniform at $\lambda=m$ for
small values of  $\nu$. In particular,
as can be seen by a brief computation (see appendix~\ref{app:a=0}),  for $m=\lambda=0$ and $\nu=0$ we never reach the correct value at the origin in the continuum limit, see Fig.~\ref{fig:rhonu0}. 
Thus we obtain the limit
\begin{equation}
\lim_{a\to0}\rho_5(m=0,\lambda=0,a)=\frac{1}{ \sqrt{2}}\neq1=\lim_{\lambda\to0}\rho_5(m=0,\lambda,a=0).\label{rho5-result-d}
\end{equation}
The non-commutativity of the  two limits has  also been confirmed by Monte Carlo simulations, see Fig.~\ref{fig:rhonu0}. 
Therefore the continuum limit is not uniform in $m$ so that one has to be careful with quantities which depend essentially  on  eigenvalues close to the origin.
 
\begin{figure}[t!]
\centerline{\includegraphics[width=0.75\textwidth]{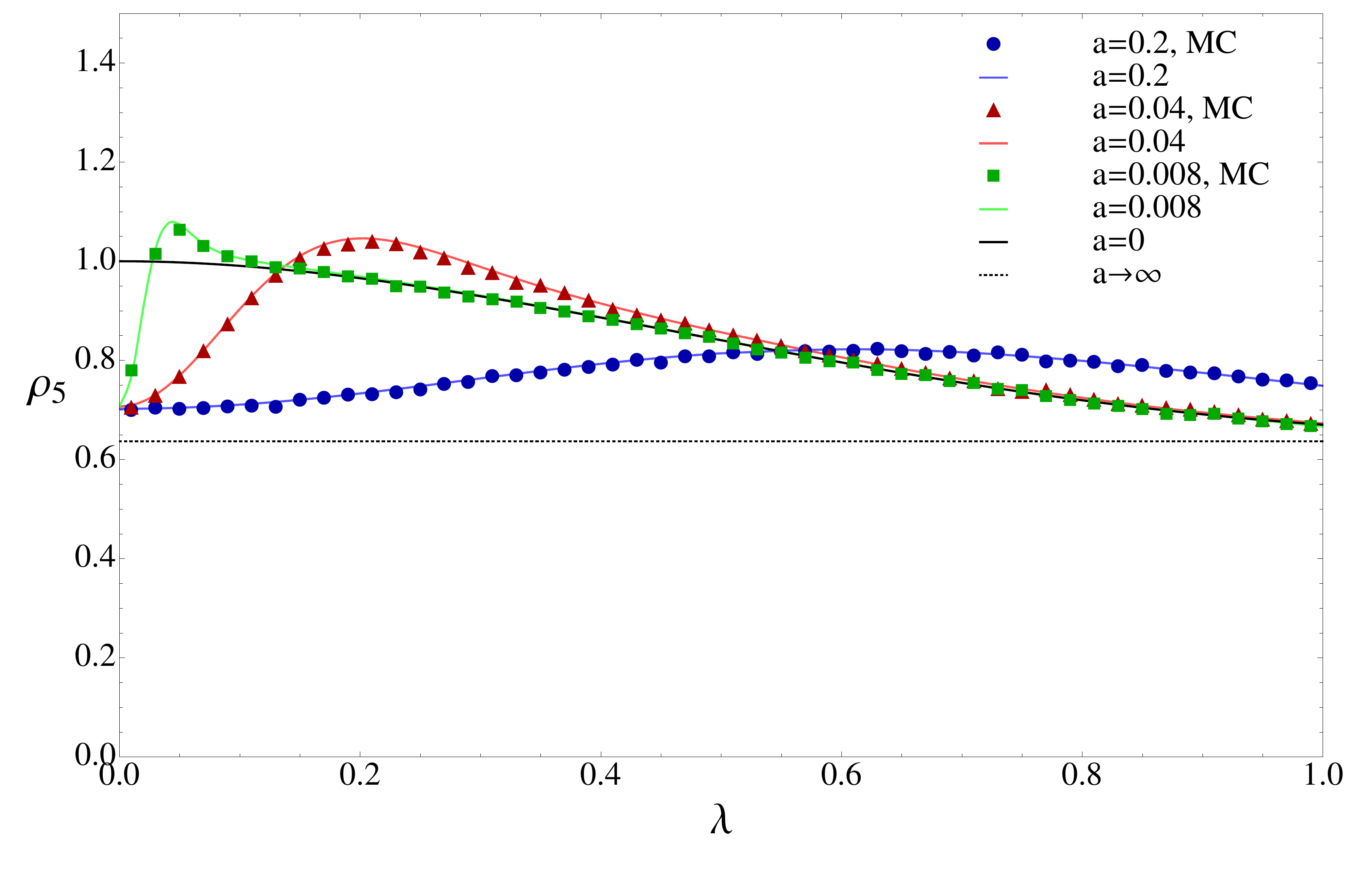}}
\caption{Continuum limit of the level density $\rho_5$ of the Hermitian Wilson Dirac operator. The index and the quark mass are set to zero, $\nu=m=0$.
 The continuum limit (black solid curve) is not uniformly approached by the analytical result~\eqref{rho5-result}.
 We have confirmed our analytical result by Monte Carlo simulations of the random matrix model~\eqref{RMT} and \eqref{RMT-Prob}. The ensemble of 
 $2.5\times10^6$ matrices gives an accuracy of about $1\%$.  The dotted black curve is the limit for a very coarse lattice where $\rho_5=2/\pi$ is a constant.}
\label{fig:rhonu0}
\end{figure}

As was discussed in Eq. \eqref{rho5-b}, the mass dependence 
of $\rho_5(m,\lambda=0,a) $ is given by the 
density of the real eigenvalues of $D_{\rm W}$ weighted by the absolute value of the inverse chirality of the states. 
In the continuum limit the density of the real eigenvalues  is  well approximated by the level density $\rho_{\GOE}$ , 
see Eq.~\eqref{level-GOE}, of exactly the same finite dimensional GOE which also describes the former zero modes, 
cf. Fig.~\ref{fig:MCcomprho} with height that is of order $1/a$ 
and a  width that is of order $a$. Since the chirality of the states with real eigenvalues 
is  $|\langle k|\gamma_5|k\rangle|\approx1$, in the continuum limit, we also find that the mass dependence of 
$\rho_5(m,\lambda=0,a)$ is given by a GOE in this limit.

In the thermodynamic limit the level density $\rho_5$ takes the form
\begin{equation}\label{rho5-thermo}
\rho_5\left(\frac{m}{8a^2},\frac{\lambda}{8a^2}\right)\approx\frac{2}{\pi}\cosh\left[\vartheta\left(\frac{m}{8a^2},\frac{\lambda}{8a^2}\right)\right]\sin\left[\varphi\left(\frac{m}{8a^2},\frac{\lambda}{8a^2}\right)\right]\Theta\left[\lambda^2-(|m|^{2/3}-(8a^2)^{2/3})^3\right]
\end{equation}
where $\varphi$ and $\vartheta$ are given by Eqs.~\eqref{sol-a-ch} and~\eqref{saddle-sol-ch}.
The Heaviside theta function implies a spectral gap  if $|m|>8a^2$. If the gap is closed the system is in the Aoki phase.
 The order parameter is the pion condensate which is proportional 
to $\del_x\log Z |_{x=0}$  and, hence, to $\rho_5(m,\lambda=0,a)$. 
In the thermodynamic limit this quantity becomes a semi-circle
\begin{equation}\label{rho5-lambda0-thermo}
\langle\bar\psi\gamma\tau_3\psi\rangle_{N_{\rm f}=0}\propto\rho_5\left(\frac{m}{8a^2},\frac{\lambda}{8a^2}=0\right)\approx\frac{2}{\pi}\sqrt{1-\frac{m^2}{(8a^2)^2}}\Theta\left[8a^2-|m|\right],
\end{equation}
(see Fig.~\ref{fig:thermo}.d). It immediately shows that the phase transition at $|m|=8a^2$ is of second order as in the case of three color QCD with Wilson fermions. Furthermore we can say that the height of the limit~\eqref{rho5-lambda0-thermo} is of order $\mathcal{O}(1)$ and its integral is of order $\mathcal{O}(a^2)$. These two pieces of information become important in the discussion of the real eigenvalues in subsection~\ref{sec:rhoreal}.

\subsection{The Chiral Condensate}\label{sec:sigma}

 The analytical result for the chiral condensate 
\begin{eqnarray}
\Sigma(m,a)&=&\left.\partial_{m'} {\rm Re}\,Z_\nu(\widehat{M},\widehat{X},a)\right|_{\substack{\widehat{M}=m\eins_4\\ \widehat{X}=i\varepsilon\eins_4\to0}}
\ee
for the quenched theory is explicitly shown in Eq.~(\ref{sigma-result}) of appendix~\ref{app:explicit}.     
 Its behavior is shown in Fig.~\ref{fig:SigmaNF0}.

\begin{figure}[t!]
\centerline{\includegraphics[width=0.47\textwidth]{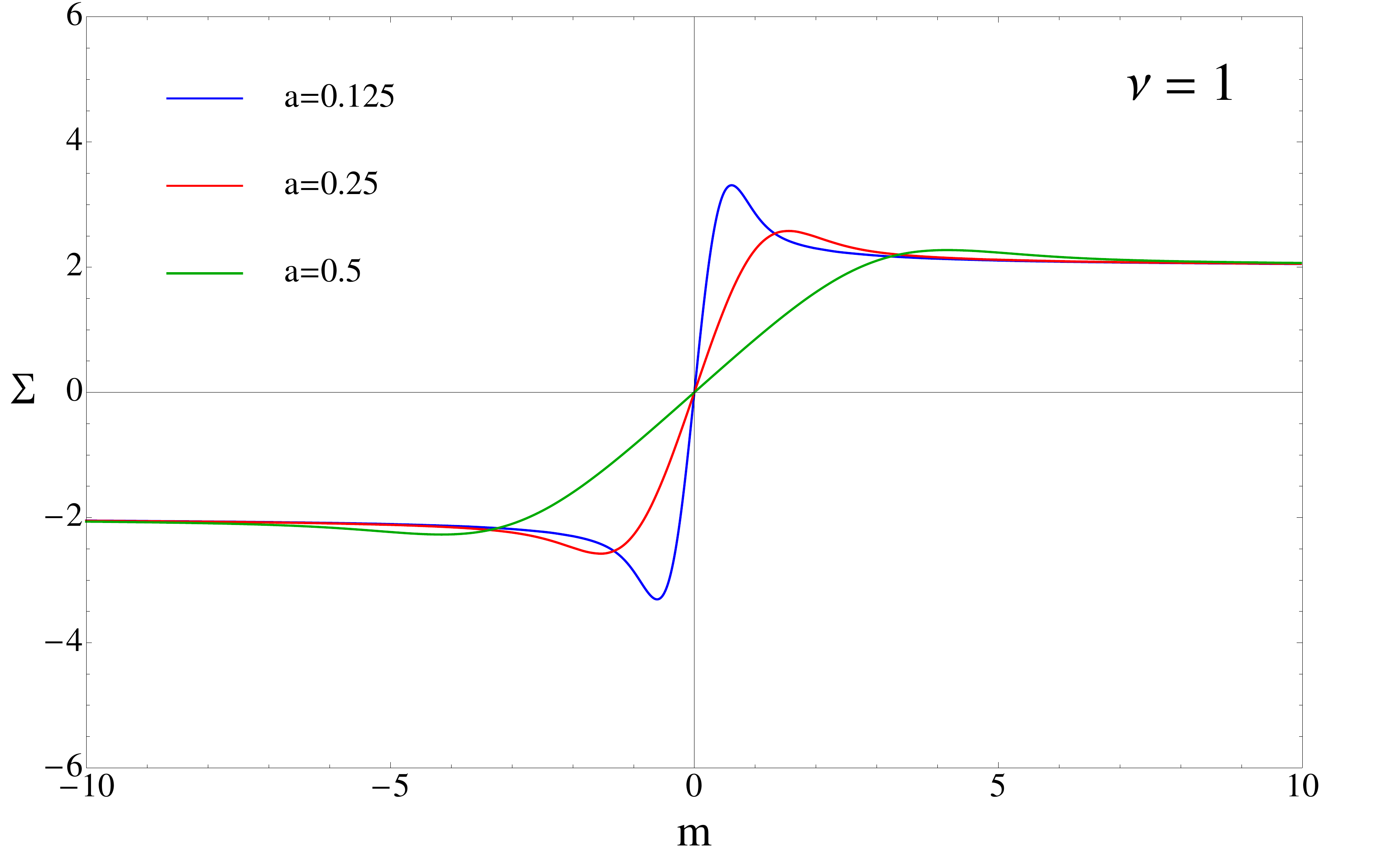}\hfill\includegraphics[width=0.47\textwidth]{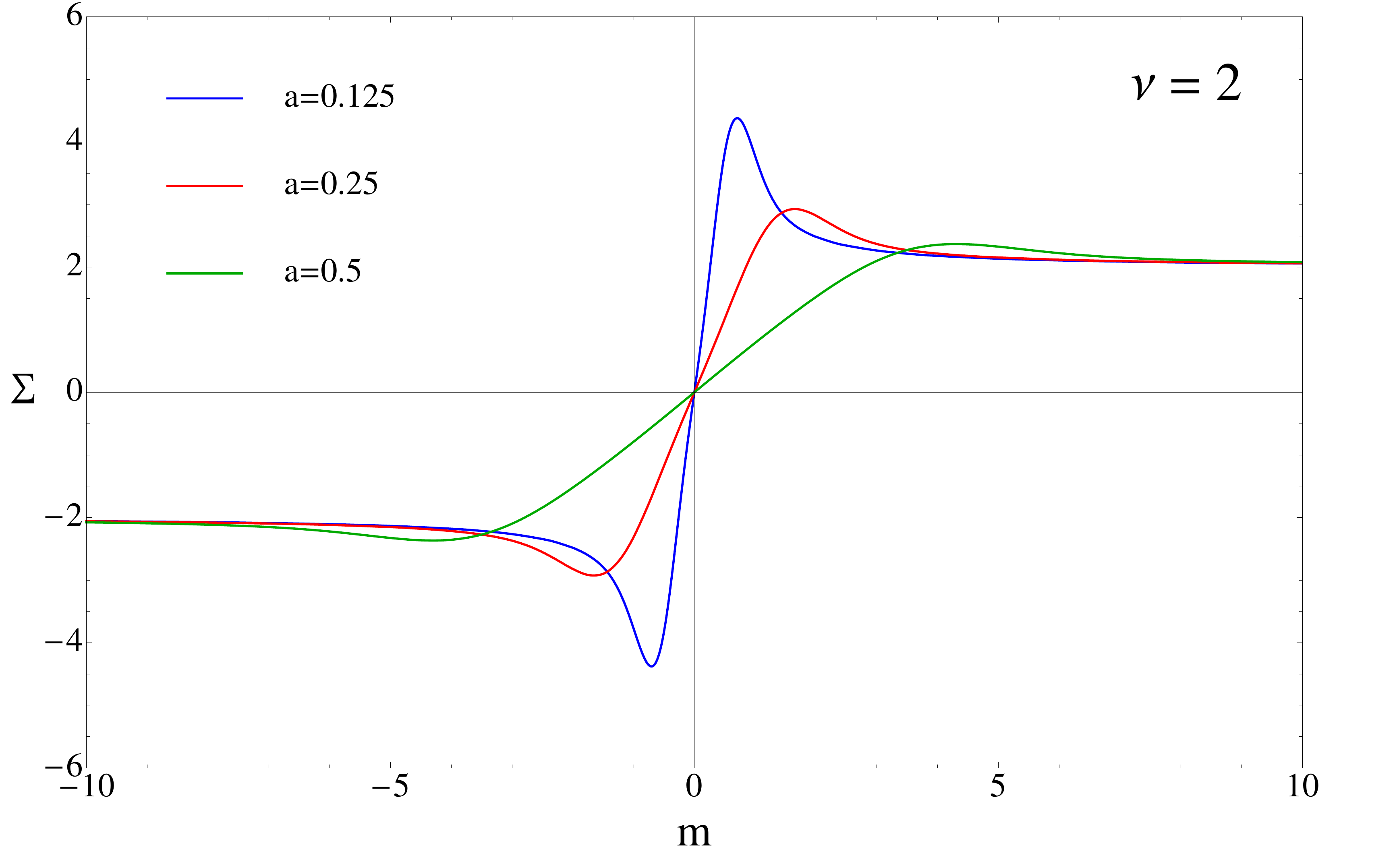}}
\caption{The chiral condensate for parameters used for the 
simulation in Figs.~\ref{fig:MCcomprho5} and \ref{fig:MCcomprho}.
 We underline that the chiral condensate is for the quenched theory 
while the one shown in Fig.~\ref{fig:SigmaNf} is  for the theory with dynamical quarks. Interestingly, the chiral condensate in the quenched theory approaches the asymptotic value $2$ from above while it is approached from below 
for the theory with dynamical quarks. }
\label{fig:SigmaNF0}
\end{figure}

As in the theory with dynamical quarks we have a $1/m$ singularity for 
$m\to 0$ due to the zero modes at non-zero index $\nu\neq0$ in the continuum theory. These singularities are washed out by the finite dimensional GOE according to the approximation
\begin{equation}\label{sigma-approx}
\Sigma(m,a)\overset{|a|\ll1}{\approx}\Sigma_{\chiGOE}^{(\nu)}(m)+\Sigma_\GOE^{(\nu)}\left(m\right),
\end{equation}
where the continuum limit of the chiral condensate without the contribution of the zero modes, $\Sigma_{\chiGOE}^{(\nu)}(m)$, is given in Eq.~\eqref{sigma-chiGOE}. The continuum result is in agreement with earlier
work by Damgaard et al. \cite{damgaard-chgoe}.
Indeed the contribution of $\Sigma_\GOE^{(\nu)}$, see Eq.~\eqref{Sigma-GOE}, works quite well for small but finite lattice spacing and $|\nu|>2$, cf. Figs.~\ref{fig:comp} e) and f). 
The large  deviations for smaller indices result from the non-uniform continuum limit of the smallest eigenvalues. 
The former zero modes help to push the spectrum away from the origin. This is the reason why the approximation by the GOE and continuum QCD without zero modes is very accurate for larger indices.

The thermodynamic limit of the chiral condensate~\eqref{sigma-result} can be easily taken via the Eqs.~\eqref{observe-thermo} and \eqref{saddle-lambda0-ch} yielding
\begin{equation}\label{Sigma-thermo}
\langle\bar\psi\psi\rangle_{N_{\rm f}=0}=\Sigma\left(\frac{m}{8a^2}\right)\approx\left\{\begin{array}{cl} \displaystyle \frac{m}{4a^2}, & |m|<8a^2 \\ 2\sign\, m\eins_4, & |m|>8a^2. \end{array}\right.
\end{equation}
This limit is already visible for a lattice spacing $\sqrt{VW_8}\widetilde{a}=a\approx1$. Moreover, it is exactly the same result~\eqref{sigmanf-thermo} as 
obtained for dynamical quarks.

\subsection{The Distribution of Chirality over the Real Eigenvalues}\label{sec:rhochi}

Also the distribution of the chiralities over the real eigenvalues 
\begin{eqnarray}
\rho_\chi(m,a)&=&-\left.\frac{1}{\pi}\partial_{m} {\rm Im}\,Z_\nu(\widehat{M},\widehat{X},a)\right|_{\substack{\widehat{M}=m\eins_4\\ \widehat{X}=(\lambda+i\varepsilon)\eins_4\to\lambda\eins_4}}
\ee
can  be 
written  as a sum of products of a  compact integral and a non-compact twofold integral ( see Eq. (\ref{rhochi-result}) of \ref{app:explicit}).
 Comparisons with Monte Carlo simulations of the random matrix model~\eqref{RMT} are shown in Fig.~\ref{fig:MCcomprho}.

\begin{figure}[h!]
\centerline{\includegraphics[width=0.47\textwidth]{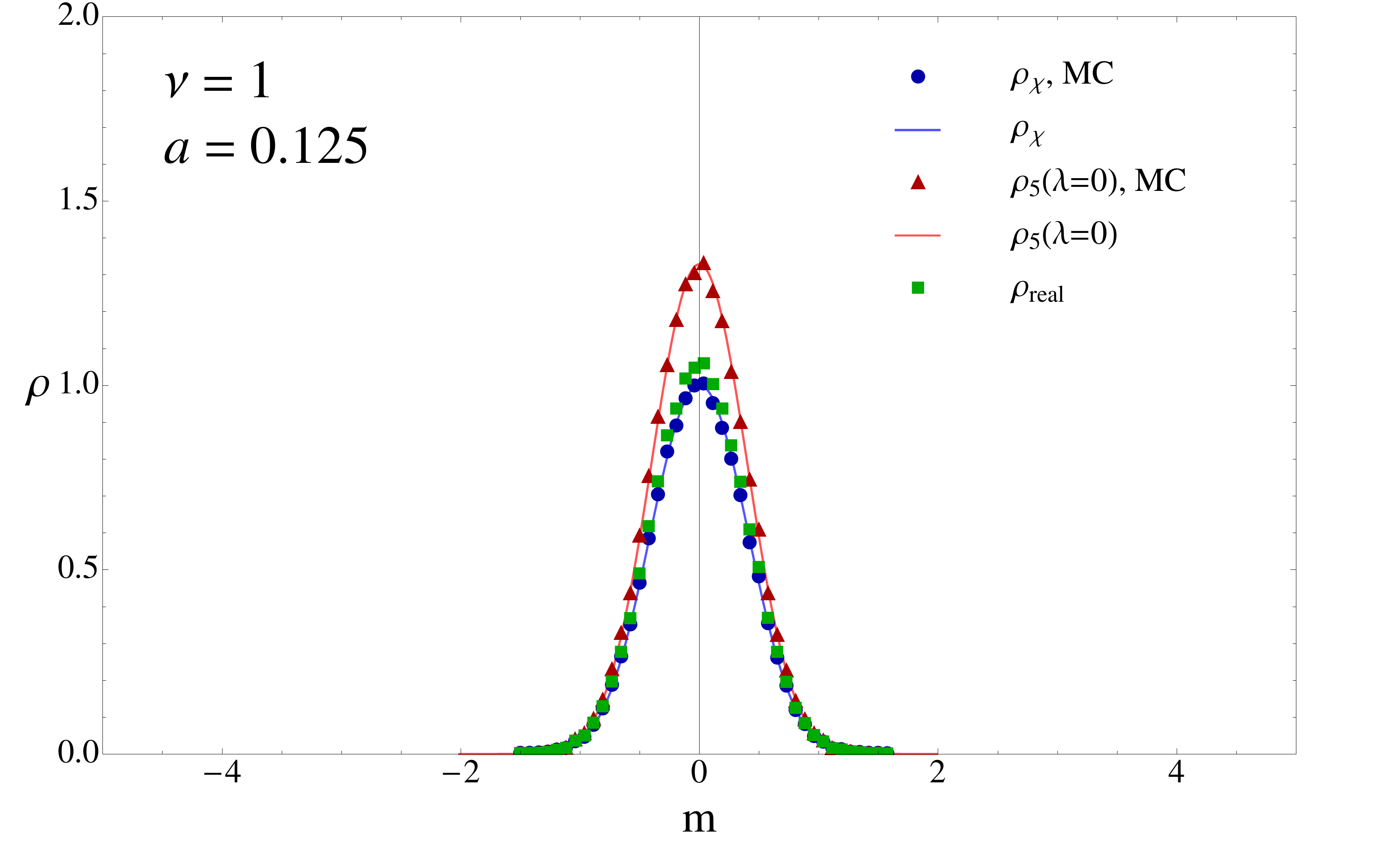}\hfill\includegraphics[width=0.47\textwidth]{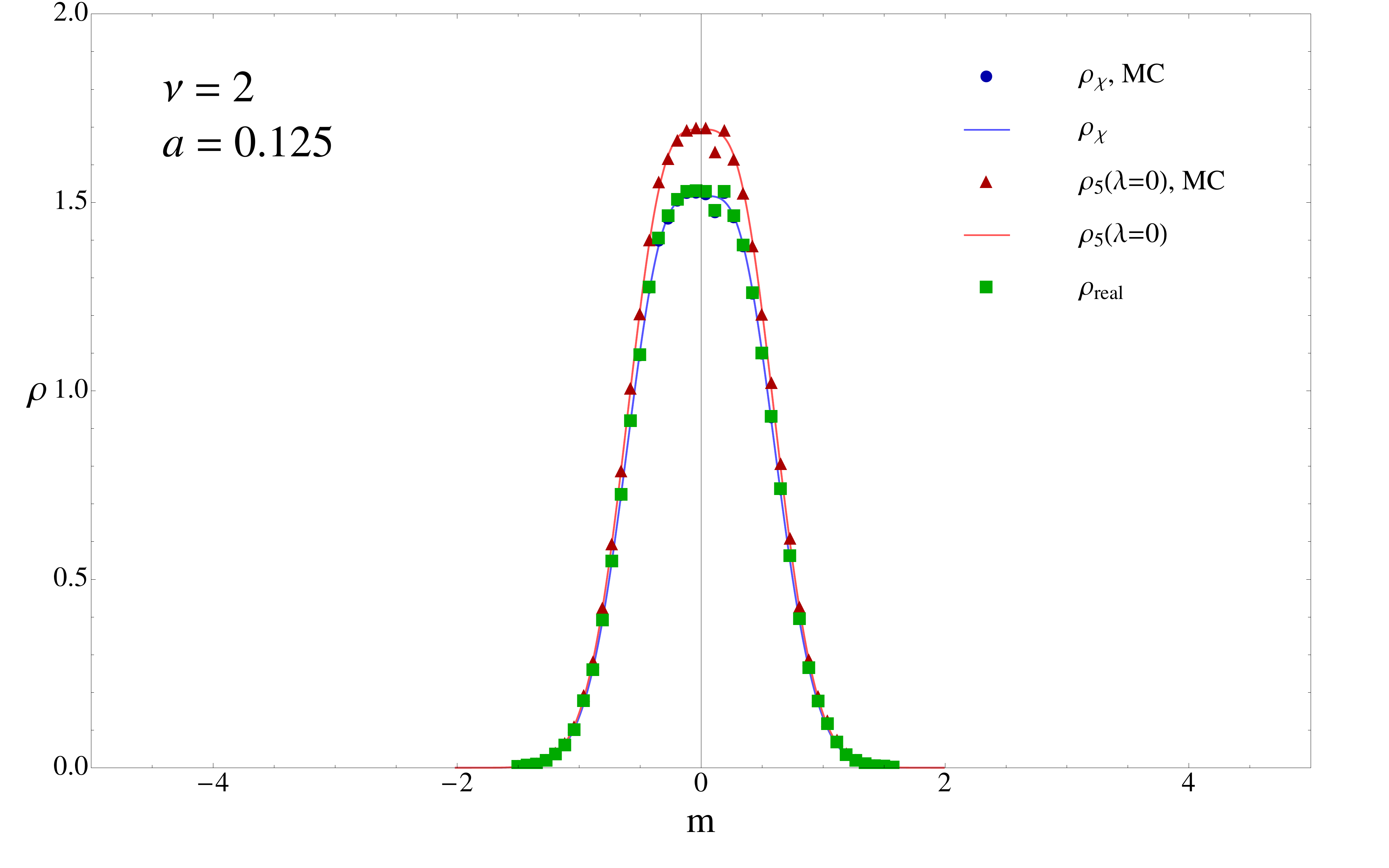}}
\centerline{\includegraphics[width=0.47\textwidth]{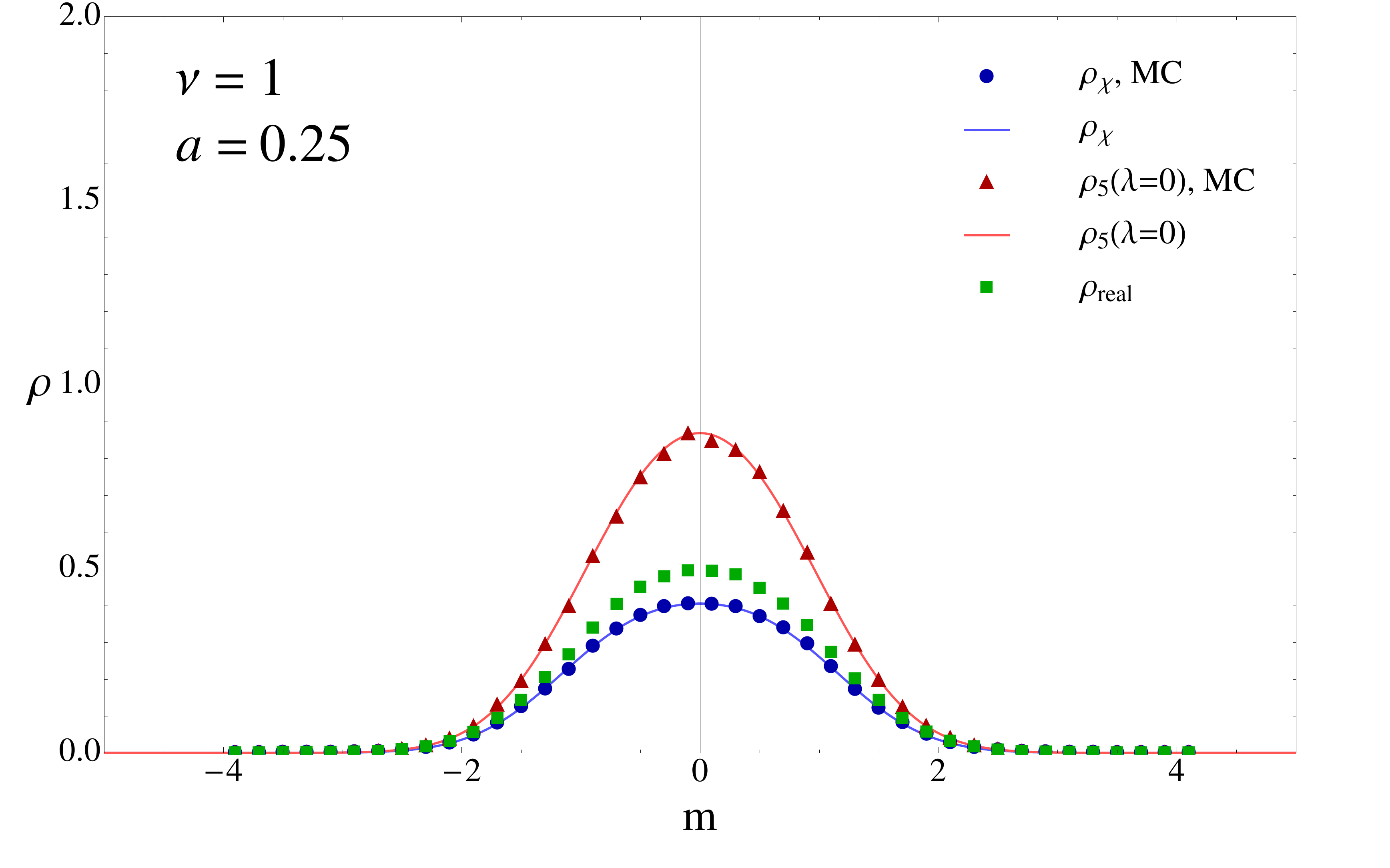}\hfill\includegraphics[width=0.47\textwidth]{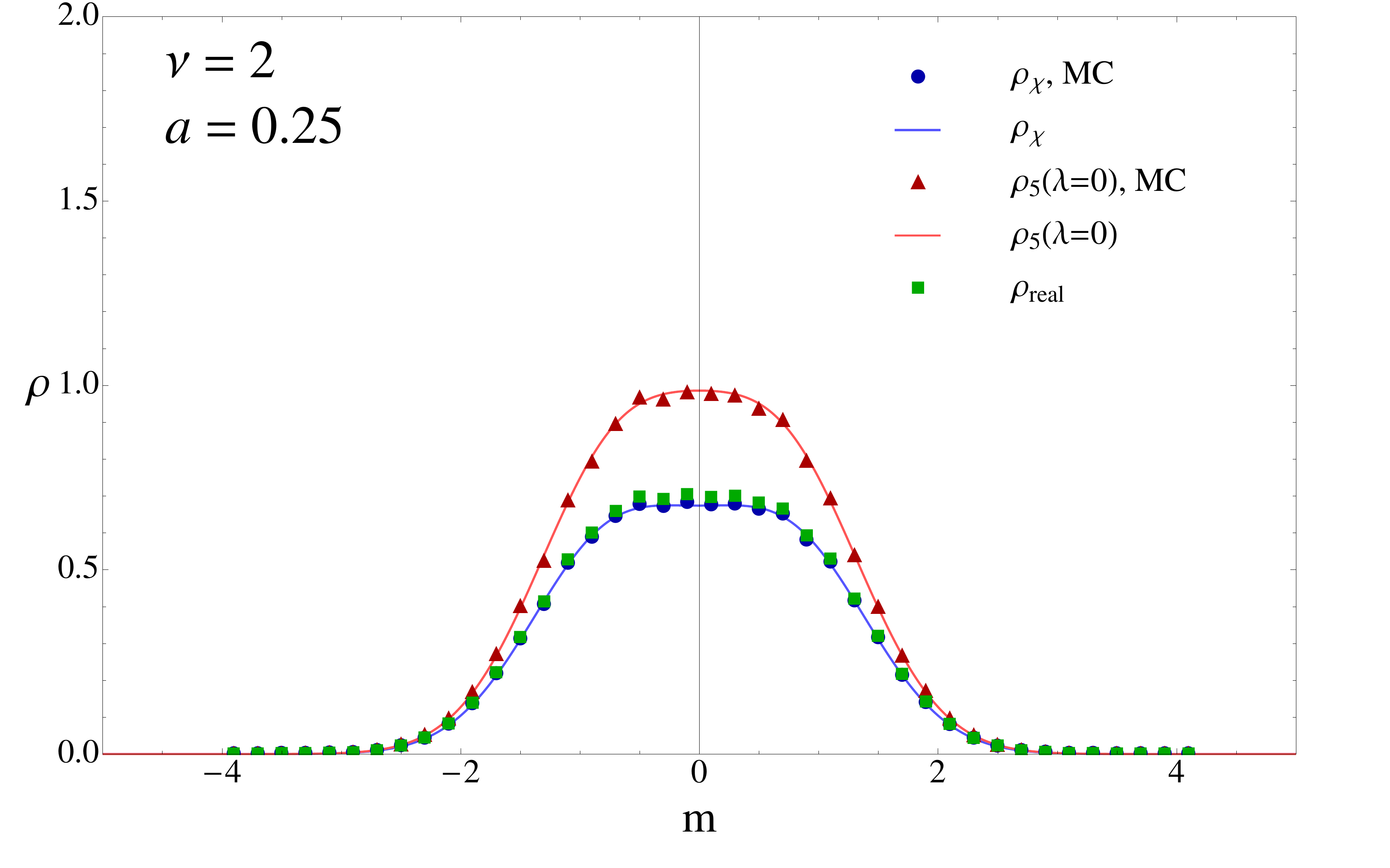}}
\centerline{\includegraphics[width=0.47\textwidth]{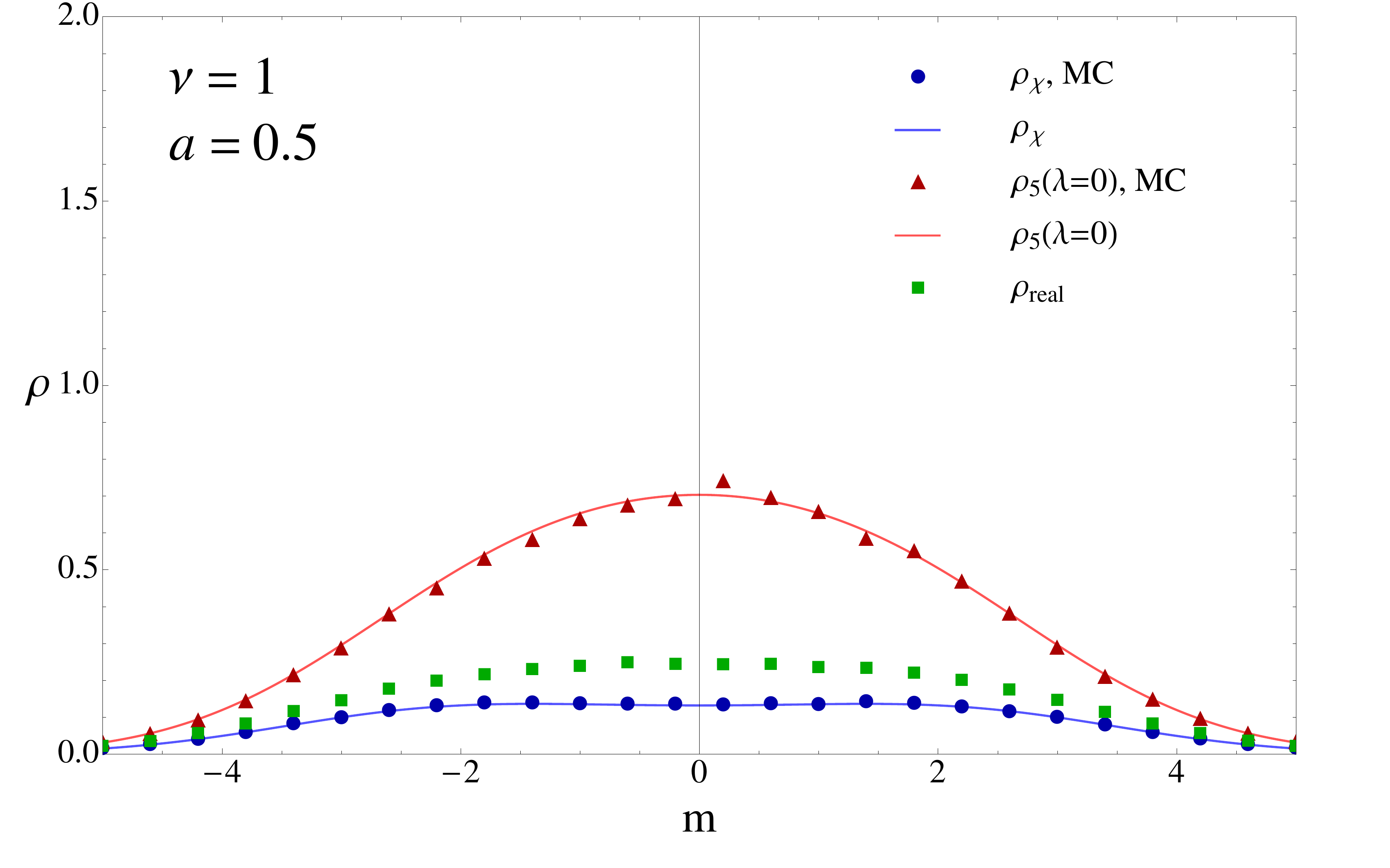}\hfill\includegraphics[width=0.47\textwidth]{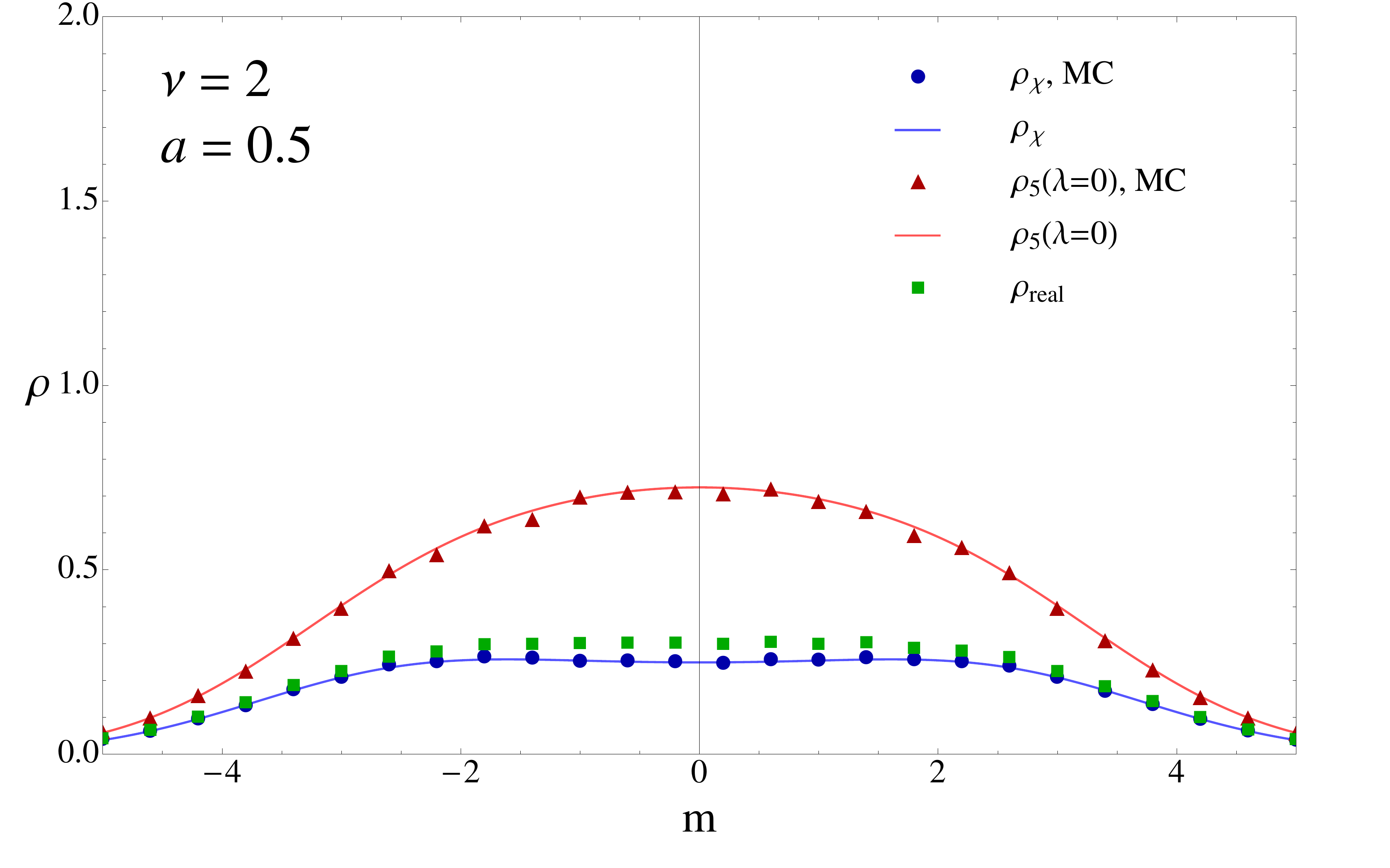}}
\caption{Comparison of the analytical results~\eqref{rho5-result} (red solid curves) and \eqref{rhochi-result} (blue solid curves) of the mass dependence of the level density $\rho_5(\lambda=0)$ at the origin and the distribution of chirality over the real eigenvalues, respectively, with Monte Carlo simulations of the random matrix model~\eqref{RMT} (MC, symbols).
 We have generated an ensemble of $10^5$ matrices both for index 
$\nu=1$ (left plots) and index $\nu=2$ (right plots) at lattice  spacings $\sqrt{VW_8}\widetilde{a}=a=0.125, 0.25, 0.5$. We have also calculated 
the distribution of the real eigenvalues $\rho_{\rm real}$ 
of the non-Hermitian Wilson Dirac operator $D_{\rm W}$ (green symbols). As correctly predicted by the inequality~\eqref{inequ-c} the level density $\rho_{\rm real}$ lies always in between $\rho_\chi$ and $\rho_5(\lambda=0)$. Interestingly,
for small lattice spacings,  the level density is better approximated by $\rho_\chi$ than by $\rho_5(\lambda=0)$ for small lattice spacings. 
Moreover, we notice a significant difference between the integral of the three distributions for large lattice spacings while their support remains the same.}
\label{fig:MCcomprho}
\end{figure}

In the limit $ a \to 0$ the chirality distribution, $\rho_\chi(\lambda)$ becomes the level density $\rho_{\GOE}$ of a $\nu$-dimensional GOE  which, for $\nu =1$, is a Gaussian, see Eqs.~\eqref{level-GOE}, \eqref{rho-GOE-nu0} and \eqref{rho-GOE-nu1}. We compared this behavior of $\rho_\chi$ with $\rho_{\GOE}$ and $\rho_5(\lambda=0)$ in Fig.~\ref{fig:comp}.a) at a lattice spacing $\sqrt{VW_8}\widetilde{a}=a=0.0625$. The agreement is almost perfect.
An interesting distinction from the case  of three color QCD, where  the distribution of the chiralities over the Dirac spectrum  in the continuum limit
also agrees with a finite dimensional 
Gaussian random matrix model, is that  the peaks corresponding to the zero modes are only barely visible humps. 
As in the case of three color QCD the number of these humps is equal to $\nu$. However they merge with the main peak at the origin due to the very weak level repulsion.
 This becomes apparent when comparing Figs.~\ref{fig:comp}.a) and  \ref{fig:MCcomprho} with Fig. 1 of \cite{ADSV10}.

The thermodynamic limit of $\rho_\chi$ follows from the first equality of Eq.~\eqref{observe-thermo}.  The supermatrix $V+V^{-1}$ is real at the saddle point~\eqref{saddle-lambda0-ch} such that the second term in  Eq.~\eqref{observe-thermo} vanishes. Hence the first term is the leading one. The derivative in the mass yields an inverse square root behavior,
\be
\label{rhochi-thermo}
\rho_\chi(m)\approx\frac{\nu}{\pi}
\frac{\Theta(8a^2-|m|)}{\sqrt{(8a^2)^2-m^2}}.
\ee
Therefore, $\rho_\chi$ of two color QCD exhibits the same square root singularities at the boundary of the support 
as in the case of three color QCD, see \cite{KVZ-prl}. We observe this 
asymptotic behavior in Monte Carlo simulations where it is starting to build up, cf. Fig.~\ref{fig:MCcomprho}, and in the numerical evaluation for even larger lattice spacings $a=1$ shown in Fig.~\ref{fig:thermo}.c).

\subsection{Level Density of the Real Modes of $D_{\rm W}$}\label{sec:rhoreal}

Since the analytical derivation of $\rho_{\rm real}$  for $\beta =2$ was a {\it tour
de force} \cite{Kie,KVZ-prl},
 it is not surprising that we did not succeed to find an
analytical result for the  density of the real eigenvalues of the non-Hermitian 
Wilson Dirac operator $D_{\rm W}$ for two color QCD with $\beta=1$.
Nonetheless we can extract some information via the inequality~\eqref{inequ-c} and thus via the distributions $\rho_5(m,\lambda=0)$ and $\rho_\chi(\lambda)$.

For small level spacing $|a|\ll1$ the distributions 
$\rho_5(m,\lambda=0)$ and $\rho_\chi(\lambda)$ are well approximated by the  
density $\rho_\GOE$ of the finite dimensional GOE, see Fig.~\ref{fig:comp}.a).
 Therefore also the level density of the real eigenvalues has to be 
$\rho_{\rm real}\approx \rho_\GOE$ for $|a|\ll1$. 
This is indeed the case already for the lattice spacing 
$\sqrt{VW_8}\widetilde{a}=a\approx0.1$, see Fig.~\ref{fig:MCcomprho}. 
Surprisingly the distribution of the chiralities over the Dirac spectrum  agrees quite well with $\rho_{\rm real}$ 
even for $\sqrt{VW_8}\widetilde{a}=a\approx0.25$ 
when the index is $\nu>2$. Since the integral of  the difference 
$\rho_{\rm real}-\rho_\chi$ is equal to the average number of additional 
real modes $N_{\rm add}$ we deduce that this number is  
highly suppressed for configurations with a larger index $\nu$ at small 
lattice spacing. For the Wilson Dirac operator of three color QCD  
the average number $N_{\rm add}$ is of order $\mathcal{O}(a^{2\nu+1})$, 
see \cite{KVZ-prl}. The reason is that the probability of finding an
additional real eigenvalue is proportional to
\be
\prod_{\lambda \in {\mathbb R}}(\lambda - \lambda_k)^{\beta_{\rm D}},
\ee
where the real eigenvalues $\lambda_k$ are of order $a$. 
In the present two color case we expect 
that $N_{\rm add}$ is of order $\mathcal{O}(a^{\nu+1})$ because of 
the smaller level repulsion.  Our expectation is confirmed by Monte-Carlo 
simulations of the random matrix model~\eqref{RMT}, see Fig.~\ref{fig:Nadd}.

\begin{figure}[t!]
\centerline{\includegraphics[width=0.7\textwidth]{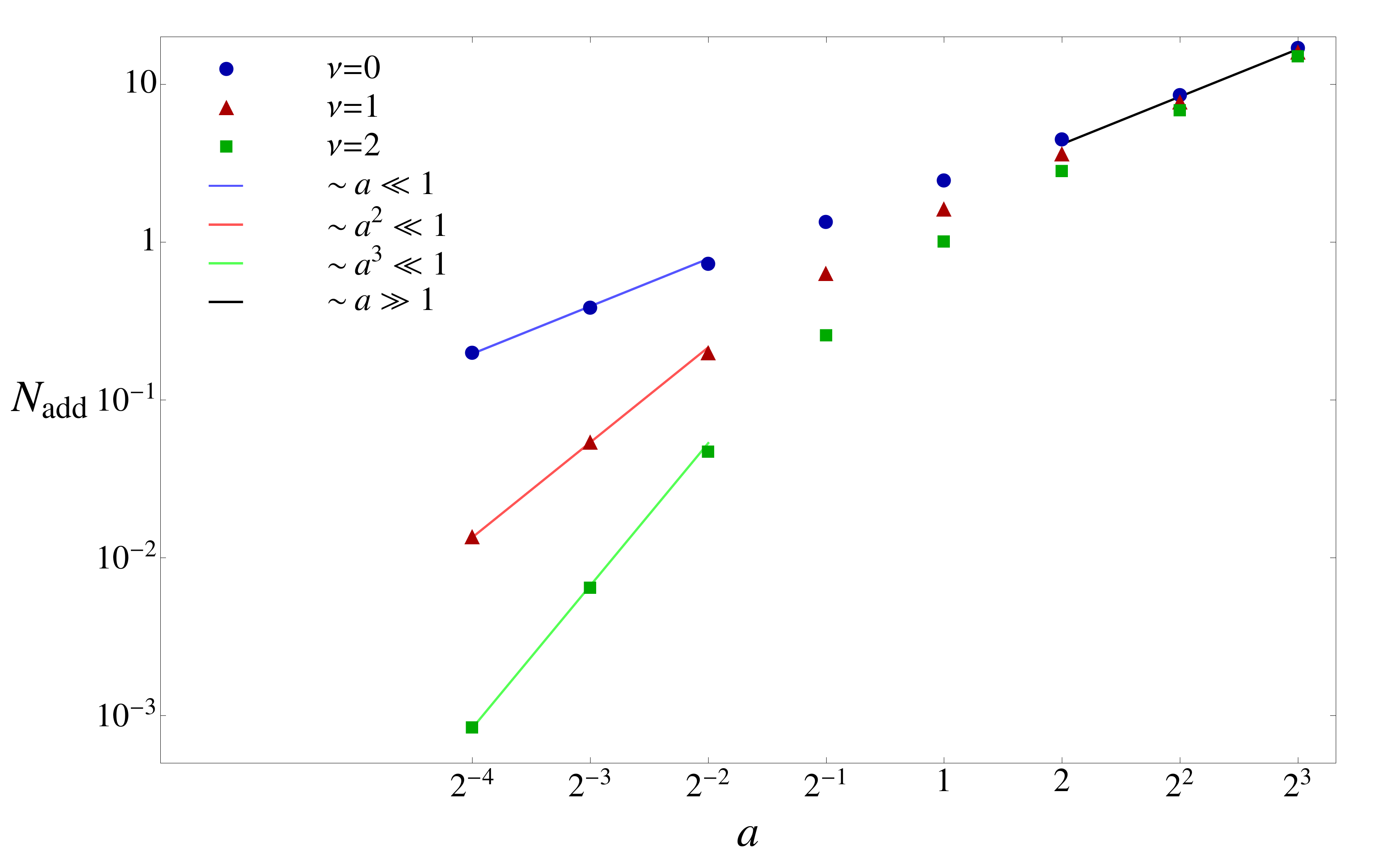}}
\caption{Log-log plot of the dependence of the average number 
of additional real modes on the lattice spacing $a$. The Monte Carlo simulations (symbols) of the random matrix model~\eqref{RMT} for 
index $\nu=0,1,2$ were fitted by the behavior $\propto a^\nu$ for small lattice spacing (colored lines) and $\propto a$ for large lattice spacing (black line), separately. 
We generated  random matrices for various matrix dimensions and ensemble sizes to keep the statistical and systematic error below 2\%. 
The transition region, $\sqrt{VW_8}\widetilde{a}=a=0.25-2$, for  the
scaling behavior is  excluded from the fits.}
\label{fig:Nadd}
\end{figure}

In the thermodynamic limit the height of $\rho_{\rm real}$ 
has to lie between the square root singularity of $\rho_\chi$ on the scale $\mathcal{O}(a^{-2})$ 
and the semi-circle of $\rho_{5}(\lambda=0)$ on the scale $\mathcal{O}(1)$. 
From the simulations shown in Fig.~\ref{fig:MCcomprho} we notice some kind of flattening of $\rho_{\rm real}$ and a departure from $\rho_\chi$
for larger values of $a$. Moreover, the average number of additional real modes extracted from Monte Carlo simulations, see Fig.~\ref{fig:Nadd}, 
suggest a $\nu$ independent behavior where $N_{\rm add}\propto a$. Since also the distribution of the real eigenvalues lives on the support $[-8a^2,8a^2]$, we conjecture that $\rho_{\rm real}$ becomes a constant flat plateau on this interval with a height of order $\mathcal{O}(a^{-1})$ in the thermodynamic limit. Interestingly this is exactly same behavior as in three color QCD, cf. \cite{KVZ-prl}.

\section{Conclusions}\label{sec:conclusio}

We have computed and analyzed the microscopic spectrum of the Wilson 
Dirac operator for QCD with two colors and the quarks in the fundamental 
representation.
The discretization effects are very similar
to the three color case. Especially in the thermodynamic limit, 
$|VW_8 \widetilde{a}^2|=|a|^2\propto|V\Sigma \widetilde{m}| =|m|\propto|V\Sigma \lambda|\gg1$, they are  essentially the same.
 When the quark mass is large enough, $|m|>8a^2$, the microscopic spectrum of the Hermitian Wilson Dirac operator 
$D_5$ develops a symmetric gap around the origin of a width $[|m|^{2/3}-(8a^2)^{2/3}]^{3/2}$. 
The system enters the Aoki phase when the gap closes, or in terms of the eigenvalues of the non-Hermitian Dirac operator $D_{\rm W}$,
when the mass $|m|$ hits the strip of eigenvalues. In this phase, parity is broken spontaneously with a non-vanishing  
pion condensate that  is proportional to the eigenvalue density of $D_5$ at the origin. 
In the thermodynamic limit, the support of  the  distribution of chirality over the real eigenvalues of $D_{\rm W}$ as well as the support of the 
level density of the real eigenvalues is of the order  $a^2$ while  the number of the additional real modes increases linearly in $a$.

In the continuum limit $|a|\ll1$ the scaling behavior is crucially different.
 Then the spectrum can be approximated by the microscopic spectrum of the continuum Dirac operator 
without zero modes plus the spectrum of a finite dimensional GOE.
 The GOE describes the broadening of the former zero modes which scale like $a$ in the continuum limit 
and stands in contrast to the thermodynamic limit.
 Moreover, we have on average $a^{\nu+1}$ additional real modes. Hence additional real modes are strongly suppressed for 
larger topological indices $\nu$ as  is the case for QCD with three colors. However the suppression is much weaker than for three color QCD. 
The reason is the weaker level repulsion which is linear for the two color case. 
This repulsion is also the reason why no separate peaks are visible.

Surprisingly the continuum limit is not uniform for the microscopic 
level density $\rho_5$ of $D_5$ at the origin. In the continuum we would expect $\rho_5(m=0,\lambda\to0,a=0)=1$.  
However the continuum limit yields $\rho_5(m=0,\lambda=0,a\to0)=1/\sqrt{2}$. This shows that one has to be careful when 
considering observables which essentially depend on the eigenvalue nearest to the origin 
for configurations with $\nu=0$ and also $\nu=1,2$ if the quark mass is non-zero. 
The mass dependent chiral condensate of the quenched theory confirms this statement. 
It can be quite accurately approximated with continuum QCD without zero modes and the finite dimensional GOE for larger
values of the index, while it has the strongest deviations 
for small topological indices.
 The $\nu$ zero modes broadened along the real axis push the spectrum away from the quark mass $m$ which is the point where the limit is non-uniform.

As is the case for QCD with three colors, we have also found that  
the low energy constant $W_8$ has to be positive. 
The reason is   that the chiral partition function with an even number of dynamical quarks is not positive definite if $W_8<0$.
Effects from the low energy constants $W_6$ and $W_7$ were not considered in the present work. 
They may weaken this conclusion, though we would not expect this due to the analogy with the case of three color QCD.

Technically the two color case is much more complicated than the three color QCD. Particularly, there are no analytical results for the spectral density  
of $ D_{\rm W}$. At least for small lattice spacings the density of the real eigenvalues
 is tightly constrained by the distribution of the chirality over the 
real eigenvalues of $D_{\rm W}$ and by the mass dependence of $\rho_5$ at the origin. 
It is not clear if it is at all possible to derive analytical expressions for  the microscopic spectral density of $D_{\rm W}$. 
For the three color case our derivation relied heavily on
the existence of an Itzykson-Zuber integral over a non-compact group. Such an integral is not available for
two color QCD. But there is another way to deduce some results for the real eigenvalues of $D_{\rm W}$ in two color QCD. We expect that the results of three color QCD should also apply to the two color case. Mean field results generally do not depend on the Dyson index $\beta_{\rm D}$. Indeed we have seen that apart from factors of 2 for continuum QCD. This is sometimes referred to as orbifolding \cite{hanada}.

Let us summarize. Our results provide an analytical control of the smallest eigenvalues of the Dirac operator 
which can potentially compromise  numerical simulations. It is noteworthy that the absence of level repulsion from the origin for the case of $N_{\rm c}=2$ QCD makes this effect much more pronounced than in the case of $N_{\rm c}=3$ QCD. 

\section*{Acknowledgments}

We thank Gernot Akemann and Kim Splittorff for helpful comments. MK and SZ are financially supported by the Alexander-von-Humboldt Foundation. JV and SZ are supported by U.S. DOE Grant No. DE-FG-88ER40388 and MK is also partially financially supported by the CRC 701: \textit{Spectral Structures and Topological Methods in Mathematics} of the Deutsche Forschungsgemeinschaft.

\appendix

 \section{Some Properties of Bessel Functions}\label{app:Bessel}
 
In this section we  recall some properties of Bessel functions which play a crucial role in the spectral statistics 
of chiral perturbation theory of QCD. More information on Bessel functions can 
be found in \cite{Abramowitz}. 

Especially, we need the Bessel function of the first kind,
 \begin{equation}\label{BesselJ}
  J_{\mu}(x)=\int_{-\pi}^{\pi}\frac{d\varphi}{2\pi} \exp[-i x\sin\varphi]e^{i\mu\varphi},\ {\rm for}\ \mu\in\mathbb{Z}\ {\rm and}\ x\in\mathbb{C},
 \end{equation}
 the modified Bessel function of the first kind
 \begin{equation}\label{BesselI}
  I_{\mu}(x)=\int_{-\pi}^{\pi}\frac{d\varphi}{2\pi} \exp[x\cos\varphi]e^{i\mu\varphi},\ {\rm for}\ \mu\in\mathbb{Z}\ {\rm and}\ x\in\mathbb{C},
 \end{equation}
 and the modified Bessel function of the second kind
 \begin{equation}\label{BesselK}
  K_{\mu}(x)=\frac{1}{2}\int_{-\infty}^{\infty}dt\exp[-x\cosh t]e^{-\mu t},\ {\rm for}\ \mu,x\in\mathbb{C}\ {\rm and}\ {\rm Re}\,x>0.
 \end{equation}
 The Bessel function $K_\mu$ is particularly important because of the integral
 \begin{equation}\label{BesselK-b}
  \frac{1}{2}\int_{-\infty}^{\infty}dt\exp[-2(\varepsilon-i L x)\cosh t+2i L y\sinh t]e^{-\mu t}=\left|\frac{x+y}{x-y}\right|^{\mu/2}e^{i\mu\phi_-} K_\mu(\sqrt{|y^2-x^2|}e^{i\phi_+})
 \end{equation}
 with the two angles
 \begin{equation}
 \phi_{\pm}=-L\frac{\pi}{4}\left(\sign(x+y)\pm\sign(x-y)\right).
 \end{equation}
 We have assumed that $\varepsilon>0$ is infinitesimally small, $L=\pm1$  and $x$ and $y$ are real. 
The right hand side of Eq.~\eqref{BesselK-b} can be obtained by the complex shift $t\to t+{\rm ln}|(x-t)/(x+t)|/2-i\phi_-$.  The angles satisfy the relations
 \begin{equation}
 \phi_+=-L\frac{\pi}{2}\sign(x)\Theta(x^2-y^2)\quad {\rm and}\quad 
e^{i(\phi_--\phi_+)}=iL\sign(x-y).
 \end{equation}
 This results in the very useful identity
 \begin{equation}\label{BesselK-c}
  \frac{(iL)^\mu}{2}\int_{-\infty}^{\infty}dt\exp[-2(\varepsilon-i L x)\cosh t+2i L y\sinh t]e^{-\mu t}=\frac{1}{(y-x-iL\varepsilon)^\mu} \left(\sqrt{|y^2-x^2|}e^{i\phi_+}\right)^{\mu}K_\mu(\sqrt{|y^2-x^2|}e^{i\phi_+}).
 \end{equation}
 
All Bessel functions satisfy a symmetry regarding the transformation $\mu\to-\mu$,
\begin{equation}\label{index-rel}
J_{-\mu}(x)=J_{\mu}(-x)=(-1)^\mu J_{\mu}(x),\ I_{-\mu}(x)=I_{\mu}(x),\ {\rm and}\ K_{\mu}(x)=K_{-\mu}(x).
\end{equation}
Except for the Bessel function $K_\mu$ this relation is only valid for integer $\mu$.

The functions $J_\mu$ and $I_\mu$ have a particular simple representation as an absolutely convergent series which extends its index to general $\mu\in\mathbb{C}$,
\begin{equation}\label{Besselseries}
J_{\mu}(x)=\sum_{j=0}^\infty\frac{(-1)^j}{j!\Gamma(\mu+j+1)}\left(\frac{x}{2}\right)^{2j+\mu}
\quad {\rm and}\quad 
I_{\mu}(x)=\sum_{j=0}^\infty\frac{1}{j!\Gamma(\mu+j+1)}\left(\frac{x}{2}\right)^{2j+\mu}.
\end{equation}
These series representations make it obvious that the two Bessel functions $J_\mu$ and $I_\mu$ are related by
\begin{equation}\label{rel-J-I}
 J_{\mu}(x)=e^{-i\mu\pi/2}I_\mu(e^{i\pi/2}x).
\end{equation}
Also the Bessel function $K_\mu$ is related to the other two Bessel functions. 
This relation is based on a representation of the Bessel functions as Meijer G-functions~\cite{Abramowitz} which are given by contour integrals,
\begin{equation}\label{BesselJ-con}
 x^\mu J_{\mu}(2x)=\int_{\mathcal{C}}\frac{ds}{2\pi i}\frac{\Gamma[\mu-s]}{\Gamma[1+s]}x^{2s}
\end{equation} 
and
\begin{equation}\label{BesselK-con}
x^\mu K_{\mu}(2x)=\frac{1}{2}\int_{\mathcal{C}}\frac{ds}{2\pi i}\Gamma[\mu-s]\Gamma[-s]x^{2s}.
\end{equation} 
The contour $\mathcal{C}$ encircles the positive real axis clockwise and thus all poles of the Gamma functions. When choosing $x=e^{i\pi n/2 }x'$ imaginary ($n=\pm1$ and $x'>0$, cf. Eq.~\eqref{BesselK-c}) we can take the imaginary part of Eq.~\eqref{BesselK-con} and obtain
\begin{eqnarray}
{\rm Im}\left[e^{i \pi n\mu/2}x'^\mu K_{\mu}(2e^{i \pi n/2}x')\right]&=&\frac{1}{2}{\rm Im}\,\int_{\mathcal{C}}\frac{ds}{2\pi i}\Gamma[\mu-s]\Gamma[-s] e^{i\pi ns} x'^{2s}\nn\\
&=&\frac{1}{2}{\rm Im}\,\int_{\mathcal{C}}\frac{ds}{2\pi i}\frac{\Gamma[\mu-s]}{\Gamma[1+s]}\frac{\pi}{\sin(-\pi s)}e^{i\pi ns} x'^{2s}\nn\\
&=&-n\frac{\pi}{2}\int_{\mathcal{C}}\frac{ds}{2\pi i}\frac{\Gamma[\mu-s]}{\Gamma[1+s]} x'^{2s}\nn\\
&=&-n\frac{\pi}{2}x'^\mu J_{\mu}(2x').\label{K-J-rel}
\end{eqnarray} 
We have used the reflection formula of the Gamma function $\Gamma[z]\Gamma[1-z]=\pi/\sin\pi z$.
Since the integral is equal to a sum over the residues at the poles of the Gamma functions which are real, the imaginary part is obtained by replacing
$ \exp[ i\pi ns] \to \sin\pi ns=n\sin\pi s$.

One can also understand the relation~\eqref{K-J-rel} from the logarithmic cut along the imaginary axis of $K_\mu$ which follows from the series expansion
\begin{eqnarray}
 K_\mu(z)&=&\frac{1}{2}\left(\frac{2}{z}\right)^\mu\,\sum_{k=0}^{\mu-1}\frac{(\mu-k-1)!}{k!}\left(-\frac{z^2}{4}\right)^k+(-1)^{\mu+1}{\rm ln}\left(\frac{z}{2}\right)I_\mu(z)\nn\\
 &&+\frac{(-1)^\mu}{2}\left(\frac{z}{2}\right)^\mu\sum_{k=0}^\infty\frac{\psi(k+1)+\psi(\mu+k+1)}{k!(\mu+k)!}\left(\frac{z^2}{4}\right)^k,\label{BesselK-series}
\end{eqnarray}
where $\psi(k)=\partial_k{\rm ln}\,\Gamma(k)$ is the Digamma function, meaning the logarithmic derivative of the Gamma function. This series can indeed be
calculated by taking the residues of the contour integral~\eqref{BesselK-con}.

Another relation which is less known but essential to derive  
results for two color QCD is based on an integral over the Bessel 
functions $K_\mu$ and $J_\mu$. In particular we consider the imaginary 
part of the integral
\begin{equation}\label{integral-def}
 S'_{\mu,\alpha}(x)=x^\mu\int_1^\infty dy y^\alpha K_\mu(2xy)
\end{equation}
with $\mu,\ \alpha\in\mathbb{R}_+$, $x\in\mathbb{C}$, and ${\rm Re}\,x>0$. By introducing an auxiliary Gaussian $e^{-\epsilon y^2}$ in the integrand we can extend the definition to purely imaginary $x=e^{i \pi n/2} x'$ ($n=\pm1$ and $x'>0$) and ensure that the integral over $y$ is always absolutely convergent.  Then we can perform the following calculation
\begin{eqnarray}
 {\rm Im}\,S'_{\mu,\alpha}\left(e^{i \pi n/2}x'\right)\hspace*{-0.5cm} &=&\lim_{\epsilon\to0}\int_1^\infty dy y^{\alpha-\mu}e^{-\epsilon y^2} {\rm Im}\left[\left(e^{i\pi n/2}x'y\right)^\mu K_\mu\left(2e^{i\pi n/2}x'y\right)\right]\label{calc-K-J-a}\\
&\overset{{\rm Eq.}~\eqref{K-J-rel}}{=}&-n\frac{\pi}{2}x'^\mu\lim_{\epsilon\to0}\left(\int_0^\infty-\int_0^1\right)dy y^\alpha e^{-\epsilon y^2} J_\mu\left(2x'y\right)\nn\\
 &=&\hspace*{-0.5cm}n\frac{\pi}{2}x'^\mu\int_0^1dy y^\alpha J_\mu\left(2x'y\right)-n\frac{\pi}{4}\frac{\Gamma[(\mu+\alpha+1)/2]}{\Gamma[\mu+1]}\lim_{\epsilon\to0}\frac{x'^{2\mu}}{\epsilon^{(\mu+\alpha+1)/2}} \ _1F_1\left(\frac{\mu+\alpha+1}{2};\mu+1;-\frac{x'^2}{\epsilon}\right).\nn
\end{eqnarray}
The function $ _1F_1$ is the hypergeometric function,
\begin{equation}\label{hypergeo}
 _1F_1\left(\frac{\mu+\alpha+1}{2};\mu+1;-\frac{x'^2}{\epsilon}\right)=\sum_{j=0}^\infty\frac{\Gamma[j+(\mu+\alpha+1)/2]\Gamma[\mu+1]}{j!\Gamma[(\mu+\alpha+1)/2]\Gamma[\mu+j+1]}\left(-\frac{x'^2}{\epsilon}\right)^{j}.
\end{equation}
The limit $\epsilon\to0$ can be performed by rewriting the Gamma functions as an integral where the limit is trivial,
\begin{eqnarray}
&&\frac{\pi}{4}\frac{\Gamma[(\mu+\alpha+1)/2]}{\Gamma[\mu+1]}\lim_{\epsilon\to0}\frac{x'^{2\mu}}{\epsilon^{(\mu+\alpha+1)/2}} \ _1F_1\left(\frac{\mu+\alpha+1}{2};\mu+1;-\frac{x'^2}{\epsilon}\right)\nn\\
&=&\frac{\pi}{4}\frac{1}{\Gamma[(\mu-\alpha+1)/2]}\lim_{\epsilon\to0}\frac{x'^{2\mu}}{\epsilon^{(\mu+\alpha+1)/2}}\int_0^1dt t^{(\mu+\alpha-1)/2}(1-t)^{(\mu-\alpha-1)/2}\exp\left[-\frac{x'^2}{\epsilon}t\right]\nn\\
&=&\frac{\pi}{4}\frac{\Gamma[(\mu+\alpha+1)/2]}{\Gamma[(\mu-\alpha+1)/2]}x'^{\mu-\alpha-1}.
\end{eqnarray}
The intermediate step is only true for $\mu-\alpha>-1$. However one can analytically extend this identity to arbitrary $\mu$ and $\alpha$ since both expressions are analytical in  the combination $\mu-\alpha$. They are also bounded by the analytic function $x'^{\mu-\alpha}/\Gamma[(\mu-\alpha)/2]$ on the positive complex half-plane ${\rm Re}\,(\mu-\alpha)>1$. Hence Carlson's theorem, see \cite{mehta}, can be applied which allows such a unique analytic continuation.

 Summarizing this calculation we have
\begin{equation}
 {\rm Im}\left[\left(e^{i\pi n/2}x'\right)^\mu\int_1^\infty dy y^\alpha K_\mu\left(2e^{i\pi n/2}x'y\right)\right]=n\frac{\pi}{2}x'^\mu\int_0^1dy y^\alpha J_\mu\left(2x'y\right)-n\frac{\pi}{4}\frac{\Gamma[(\mu+\alpha+1)/2]}{\Gamma[(\mu-\alpha+1)/2]}x'^{\mu-\alpha-1}\label{calc-K-J-b}.
\end{equation}
It becomes important to find the well-known results of continuum QCD from our calculations.

\section{Evaluation of the Partition Function}\label{app:eval}

The quenched partition function~\eqref{zgen}, in particular its imaginary part, plays a crucial role in the spectral statistics of the Wilson Dirac operator. Since its evaluation is cumbersome we split the calculation in the integration over the four Grassmann variables $(\alpha,\alpha^*,\beta,\beta^*)$, see subsection~\ref{app:Grass}, and in the explicit calculation of the imaginary part in subsection~\ref{app:Im}. In subsection~\ref{app:sphi} we present some properties of the integrals involved. The explicit expressions for the spectral observables are summarized in subsection~\ref{app:explicit}.

\subsection{Integration over the Grassmann Variables}\label{app:Grass}

We evaluate the Grassmann integrals of the supersymmetric
partition function~\eqref{zgen} using the representation of $U$ given 
in Eq.~\eqref{U-mat}.
To calculate the inverse of $U$ and its traces thereof we
 split $U$ into a numerical part $U_0$ which is a diagonal matrix and a nilpotent part $G$,
\be
U = \left(\begin{array}{cccc} e^{i\varphi} & 0 & \alpha^* & \beta^* \\ 0 & e^{i\varphi} & -\alpha & -\beta \\ \alpha & \alpha^*  & e^{s_1} & 0 \\ \beta & \beta^* & 0 & e^{s_2} \end{array}\right)=U_0 +  G.
\ee
For this purpose it is suitable to know what the square of $G$ is,
\begin{equation}
G^2=\left(\begin{array}{ccc} (\alpha^*\alpha+\beta^*\beta)\eins_2 & 0 & 0 \\ 0 & 2\alpha\alpha^* & \alpha\beta^*+\beta\alpha^* \\ 0 & \alpha\beta^*+\beta\alpha^* & 2\beta\beta^* \end{array}\right).
\end{equation}
Additionally we need
\begin{eqnarray}
(U_0^{-1}G)^2&=&\left(\begin{array}{ccc} (e^{-i\varphi-s_1}\alpha^*\alpha+e^{-i\varphi-s_2}\beta^*\beta)\eins_2 & 0 & 0 \\ 0 & 2e^{-i\varphi-s_1}\alpha\alpha^* & e^{-i\varphi-s_1}(\alpha\beta^*+\beta\alpha^*) \\ 0 & e^{-i\varphi-s_2}(\alpha\beta^*+\beta\alpha^*) & 2e^{-i\varphi-s_2}\beta\beta^* \end{array}\right),\nn\\
(U_0^{-1}GU_0^{-1})^2&=&\left(\begin{array}{ccc} (e^{-2i\varphi-2s_1}\alpha^*\alpha+e^{-2i\varphi-2s_2}\beta^*\beta)\eins_2 & 0 & 0 \\ 0 & 2e^{-2i\varphi-2s_1}\alpha\alpha^* & e^{-2i\varphi-s_1-s_2}(\alpha\beta^*+\beta\alpha^*) \\ 0 & e^{-2i\varphi-s_1-s_2}(\alpha\beta^*+\beta\alpha^*) & 2e^{-2i\varphi-2s_2}\beta\beta^* \end{array}\right),\nn\\
(U_0^{-1}G)^4&=&2e^{-2i\varphi-s_1-s_2}\,\diag(\eins_2,-\eins_2)\alpha^*\alpha\beta^*\beta,\nn\\
(U_0^{-1}GU_0^{-1})^2(U_0^{-1}G)^2&=&e^{-3i\varphi-s_1-s_2}\,\diag([e^{-s_1}+e^{-s_2}]\eins_2,-2e^{-s_1},-2e^{-s_2})\alpha^*\alpha\beta^*\beta,\nn\\
{[(U_0^{-1}G)^2U_0^{-1}]}^2&=&2e^{-2i\varphi-s_1-s_2}\,\diag(e^{-2i\varphi}\eins_2,-e^{-s_1-s_2}\eins_2)\alpha^*\alpha\beta^*\beta.
\end{eqnarray}
The inverse of $U$ is given by a finite geometric sum
\begin{eqnarray}
U^{-1} &=&\left [1- U_0^{-1}G +(U_0^{-1}G)^2 -(U_0^{-1}G)^3 +(U_0^{-1}G)^4 
\right ]   U_0^{-1}.
\end{eqnarray}
Now we are ready to compute the traces.

Only even powers of the Grassmann variables and, thus, even powers of $G$ contribute to the traces, 
\begin{eqnarray}
\Str(\widehat{M}+\widehat{X})U& =& \Str(\widehat{M}+\widehat{X})U_0\nn\\
&=&2(m+x_0)e^{i\varphi}-(m'+x_1)(e^{s_1}+e^{s_2}), \nn \\
\Str(\widehat{M}-\widehat{X})U^{-1}&=&\Str(\widehat{M}-\widehat{X})(U_0^{-1}+(U_0^{-1}G)^2U_0^{-1}+(U_0^{-1}G)^4U_0^{-1})\nn\\
&=&2(m-x_0)e^{-i\varphi}-(m'-x_1)(e^{-s_1}+e^{-s_2})+2[(m-x_0)e^{-2i\varphi-s_1}+(m-x_1)e^{-i\varphi-2s_1}]\alpha^*\alpha\nn\\
&&+2[(m-x_0)e^{-2i\varphi-s_2}+(m-x_1)e^{-i\varphi-2s_2}]\beta^*\beta\nn\\
&&+2[2(m-x_0)e^{-i\varphi}+(m'-x_1)(e^{-s_1}+e^{-s_2})]e^{-2i\varphi-s_1-s_2}\alpha^*\alpha\beta^*\beta,\nn\\
\Str U^{2}&=& \Str (U_0^2 +G^2)\\
&=&2e^{2i\varphi}-e^{2s_1}-e^{2s_1}+4(\alpha^*\alpha+\beta^*\beta),\nn\\
\Str U^{-2} &=&\Str U_0^{-2}+\Str (U_0^{-1}GU_0^{-1})^2+2\Str U_0^{-2}(U_0^{-1}G)^2+\Str [(U_0^{-1}G)^2U_0^{-1}]^2\nn\\
&&+2\Str (U_0^{-1}GU_0^{-1})^2(U_0^{-1}G)^2+2\Str U_0^{-2}(U_0^{-1}G)^4\nn\\
&=&2e^{-2i\varphi}-e^{-2s_1}-e^{-2s_1}\nn\\
&&+4(e^{-2i\varphi}+e^{-i\varphi-s_1}+e^{-2s_1})^2e^{-i\varphi-s_1}\alpha^*\alpha+4(e^{-2i\varphi}+e^{-i\varphi-s_2}+e^{-2s_2})e^{-i\varphi-s_2}\beta^*\beta\nn\\
&&+4[3+e^{-2i\varphi}+e^{-s_1-s_2}+2e^{-i\varphi-s_1}+2e^{-i\varphi-s_2}+e^{-2s_1}+e^{-2s_1}]e^{-2i\varphi-s_1-s_2}\alpha^*\alpha\beta^*\beta.\nn
\end{eqnarray}
We also need the $\Sdet^{(\nu+1)/2} U$  which can be expanded as
\begin{equation}
\Sdet^{(\nu+1)/2} U =e^{(\nu+1)(i\phi -(s_1+s_2)/2)}
\left [1 -(\nu+1) ( e^{-i\varphi-s_1}\alpha^*\alpha +  e^{-i\varphi-s_2}\beta^*\beta) + \nu(\nu+1)e^{-2i\varphi-s_1-s_2}\alpha^*\alpha \beta^* \beta    \right].
\end{equation}
Collecting everything we can integrate over the Grassmann variables which only selects the highest order polynomial in those. To keep the notation simple we define the following abbreviations
\be
t_{z1} &=& iL e^{-i\varphi-s_1}[(m-x_0)e^{-i\varphi} + (m'-x_1) e^{-s_1}], \nn \\
t_{z2} &=& iL e^{-i\varphi-s_2}[(m-x_0)e^{-i\varphi} + (m'-x_1) e^{-s_2}], \nn \\
t_{z12} &=& iL e^{-i\varphi-s_1-s_2}[2(m-x_0)e^{-i\varphi} + (m'-x_1) (e^{-s_1}+e^{-s_2})], \nn \\
d_1 &=& (\nu+1) e^{-i\varphi-s_1}, \nn \\
d_2 &=& (\nu+1) e^{-i\varphi-s_2}, \nn \\
d_{12} &=& \nu(\nu+1)  e^{-2i\varphi-s_1-s_2}, \nn \\
t_{a1}&=& 4a^2[1+e^{-i\varphi-s_1}(e^{-2i\varphi}+e^{-2s_1}+e^{-i\varphi-s_1})],\nn \\
t_{a2}&=& 4a^2[1+e^{-i\varphi-s_2}(e^{-2i\varphi}+e^{-2s_2}+e^{-i\varphi-s_2})],\nn \\
t_{a12} &=& 4a^2e^{-i2\varphi-s_1-s_2}(3e^{-2i\varphi}+e^{-s_1-s_2}+ e^{-2s_1}+e^{-2s_2}+
2e^{-i\varphi-s_1}+2e^{-i\varphi-s_2}).
\ee
The final result for the generating function is up to an overall constant
\be
Z&\propto& \int_{-\pi}^{\pi} \frac{d\varphi}{2\pi} \int_{-\infty}^{\infty}ds_1 \int_{-\infty}^{\infty} ds_2\left|\sinh\frac{s_1-s_2}{2}\right| e^{(\nu+2) i\varphi-(\nu-2)(s_1+s_2)/2} \exp\left[2Lm\sin\varphi+iLm'(\sinh s_1+\sinh s_2)\right]\nn\\
&&\times\exp\left[-2iLx_0\cos\varphi+iLx_1(\cosh s_1+\cosh s_2) +4a^2\cos2\varphi-2a^2(\cosh2s_1+\cosh2s_2)\right]\nn\\
&&\times[(d_1 - t_{z1} - t_{a1})(d_2  - t_{z2}- t_{a2})+d_{12}-d_1d_2 +t_{z12} +t_{a12} ].
\ee
This expression is explicitly shown in Eq.~\eqref{final} which immediately makes clear that the integral splits into a finite sum and each term factorizes into a compact one-fold integral and a coupled, non-compact two-fold integral.

\subsection{Properties of the Functions $S_{\mu,\alpha}$ and $\Phi_\mu$}
\label{app:sphi}

In this appendix we discuss properties of the integrals  $S_{\mu,\alpha}$ and $\Phi_\mu$
which were introduced in Eqs.~(\ref{non-comp-int}) and (\ref{comp-int}). Their properties
are in particular useful for the numerical evaluation of the integrals.

The integrals~\eqref{comp-int} and \eqref{non-comp-int} satisfy the following recurrence relations
\begin{eqnarray}
0&=&\mu \Phi_\mu+(x_0+m)\Phi_{\mu+1}+(x_0-m)\Phi_{\mu-1}-4a^2\Phi_{\mu+2}+4a^2\Phi_{\mu-2},\nn\\
0&=&\mu S_{\mu,\alpha}+(x_1+m')S_{\mu-1,\alpha+1}+(x_1-m')S_{\mu+1,\alpha+1}-4a^2(2S_{\mu-2,\alpha+2}-2S_{\mu+2,\alpha+2}-S_{\mu-2,\alpha}+S_{\mu+2,\alpha}).\label{rel}
\end{eqnarray}
These relations can be found by acting $\partial_\varphi$ and $(\partial_{s_1}+\partial_{s_2})$ onto the integrands which vanish under the integrals. They are useful to check the quality of the numerical evaluation of the integrals~\eqref{comp-int} and \eqref{non-comp-int}.

Two other useful relations of $S_{\mu,\alpha}$ concern the derivatives which are important to calculate the observables introduced in subsection~\ref{sec:observe}. The derivatives with respect to the quark mass $m'$ and the axial mass $x_1$ are given by
\begin{equation}\label{derivatives-S}
 \partial_{m'}S_{\mu,\alpha}=-S_{\mu+1,\alpha+1}-S_{\mu-1,\alpha+1},\qquad \partial_{x_1}S_{\mu,\alpha}=S_{\mu+1,\alpha+1}-S_{\mu-1,\alpha+1}.
\end{equation}
Moreover, there are also relations between a negative and positive index $\mu$,
\begin{equation}\label{rel-index}
\Phi_\mu(m,x_0,a)=(-1)^\mu \Phi_{-\mu}(-m,x_0,a)\quad {\rm and}\quad S_{\mu,\alpha}(m',x_1,a)=(-1)^\mu S_{-\mu,\alpha}(-m',x_1,a).
\end{equation}
These relations are helpful when computing these integrals.

\subsection{Real and Imaginary Part of $S_{\mu,\alpha}$}\label{app:Im}

We assume $\mu\geq0$ because of the relation~\eqref{rel-index}. This simplifies the computation of  the imaginary part of the integral~\eqref{non-comp-int}. This integral can 
 first of all be written in a more suitable version. 
For this purpose we introduce a Gaussian integral over an auxiliary variable $t$ to linearize the term $\cosh 2s=2\sinh^2 s+1$ to a $\sinh s$ term,
\begin{eqnarray}
S_{\mu,\alpha}(m',x_1,a)&=&\frac{(iL)^\mu}{\sqrt{8\pi a^2}}\int_{-\infty}^\infty ds\int_{1}^\infty dy\int_{-\infty}^\infty dt  e^{-\mu s} \frac{y^{\alpha+1}}{\sqrt{2y^2-1}}\nn\\
&&\times\exp\left[-\frac{y^2}{8a^2(2y^2-1)}(t-m')^2+2iLty\sinh s+2iLx_1y\cosh s-4a^2(2y^2-1)\right].\label{non-comp-int-a}
\end{eqnarray}
Hence the variable $t$ acts as an effective quark mass. The integral over $s$ is equal to a modified Bessel function of the second kind,
\begin{eqnarray}
S_{\mu,\alpha}(m',x_1,a)&=&\frac{1}{\sqrt{2\pi a^2}}\int_{1}^\infty dy\int_{-\infty}^\infty dt \frac{y^{\alpha+1}}{\sqrt{2y^2-1}}\frac{1}{(t-x_1)^\mu}\left(\sqrt{|t^2-x_1^2|}e^{i\phi}\right)^\mu\nn\\
&&\times\exp\left[-\frac{y^2}{8a^2(2y^2-1)}(t-m')^2-4a^2(2y^2-1)\right]K_{\mu}\left(2y\sqrt{|t^2-x_1^2|}e^{i\phi}\right)\label{non-comp-int-b}
\end{eqnarray}
with $\phi=-L\sign(x_1)\Theta(x_1^2-t^2)\pi/2$. The phase is important and reflects the transformation (only a complex shift in the variable $s$) to bring the integral into the form~\eqref{BesselK}.

The imaginary part of this integral consists of two contributions. The imaginary part has to be taken either of the term $1/(t-x_1)^\mu$ which results in the $(\mu-1)$st derivative of the Dirac delta function, $\delta^{(\mu-1)}(t-x_1)$, or of the term $\left(\sqrt{|t^2-x_1^2|}e^{i\phi}\right)^\mu K_{\mu}\left(2y\sqrt{|t^2-x_1^2|}e^{i\phi}\right)$ which can be dealt by  relation~\eqref{K-J-rel}. Thus, we find,
\begin{eqnarray}
\lim_{\varepsilon\to0}{\rm Im}\,S_{\mu,\alpha}(m',x_1+i\varepsilon,a)&=&(-1)^{\mu-1}\sqrt{\frac{\pi}{8 a^2}}\int_{1}^\infty dy\int_{-\infty}^\infty dt \frac{y^{\alpha+1}}{\sqrt{2y^2-1}}\exp\left[-\frac{y^2}{8a^2(2y^2-1)}(t-m')^2-4a^2(2y^2-1)\right]\nn\\
&&\hspace*{-3cm}\times\biggl[\frac{1}{y^\mu}\sum_{k=0}^{\mu-1}\frac{(\mu-k-1)!}{(\mu-1)!k!}[y^2(x_1^2-t^2)]^k\delta^{(\mu-1)}(t-x_1)-{\rm sign}\, x_1\,J_{\mu}\left(2y\sqrt{x_1^2-t^2}\right)\left(\frac{x_1+t}{x_1-t}\right)^{\mu/2}\Theta(x_1^2-t^2)\biggl].\nn\\ \label{non-comp-int-c}
\end{eqnarray}
The sum in the first term comes from the series representation of $K_\mu$, see Eq.~\eqref{BesselK-series}, since the  $(\mu-1)$st derivative of the remaining parts vanish at $t=x_1$.

The real part of the partition function $Z$ is needed to compute the chiral condensate $\Sigma(m)$. The real part of the expression~\eqref{non-comp-int-b} cannot be easily taken since the integrand has a pole of order $\mu$ at $t=0$. Therefore we integrate by parts and symmetrize the integrand with respect to $t\to-t$. Then the real part is given by
\begin{eqnarray}
{\rm Re}\,S_{\mu,\alpha}(m',x_1=0,a)&=&S_{\mu,\alpha}(m',x_1=0,a)\nn\\
&=&\frac{1}{\sqrt{2\pi a^2}(\mu-1)!}\int_{1}^\infty dy\int_{0}^\infty dt \frac{y^{\alpha+1}}{\sqrt{2y^2-1}}\exp\left[-4a^2(2y^2-1)\right]\label{non-comp-int-d}\\
&&\times\frac{1}{t}\partial_{t'}^{\mu-1}\left[|t'+t|^\mu\exp\left[-\frac{y^2}{8a^2(2y^2-1)}(t'+t-m')^2\right]K_{\mu}\left(2y|t'+t|\right)-\{t\to-t\}\right]_{t'=0}.\nn
\end{eqnarray}
Note that $\phi=0$ at $x_1=0$. Despite the modulus in $|t'+t|^\mu K_{\mu}\left(2y|t'+t|\right)$ the function is differentiable at $t'+t=0$ because the pole of the Bessel function is cancelled and the derivative of the logarithm yields a pole which cancels with the zero of the Bessel function $I_\mu$, cf. Eq.~\eqref{BesselK-series}.

\subsection{Explicit Expressions for Spectral Observables}
\label{app:explicit}

In this appendix we give explicit expressions for the spectral observables 
in terms of the functions  $S_{\mu,\alpha}$ and $\Phi_\mu$
which were introduced in Eqs. (\ref{non-comp-int}) and (\ref{comp-int}), and
we use the short-hand notations
\begin{equation}
\Phi_\mu=\Phi_\mu(m,\lambda,a),\ S_{\mu,\alpha}=S_{\mu,\alpha}(m,\lambda=0,a),\ {\rm and}\ \mathcal{S}_{\mu,\alpha}=\mathcal{S}_{\mu,\alpha}(m,\lambda,a).
\end{equation}

For the spectral density of the Hermitian Wilson Dirac operator we find
\begin{eqnarray} 
\rho_5(m,\lambda,a)&=&\left.\partial_{x_1} {\rm Im}\,Z_\nu(\widehat{M},\widehat{X},a)\right|_{\substack{\widehat{M}=m\eins_4\\ \widehat{X}=(\lambda+i\varepsilon)\eins_4\to\lambda\eins_4}}\nn\\
&=&16a^4\bigl[\Phi_{\nu-4}\bigl(\mathcal{S}_{\nu+1,1}-\mathcal{S}_{\nu-1,1}\bigl) +2\Phi_{\nu-3}\bigl(\mathcal{S}_{\nu+2,2}-\mathcal{S}_{\nu,2}\bigl)
+\Phi_{\nu-2}\bigl(4\mathcal{S}_{\nu+3,3}-4\mathcal{S}_{\nu+1,3}-\mathcal{S}_{\nu+3,1}+\mathcal{S}_{\nu+1,1})\nn\\
&&\hspace*{-1cm}
 +2\Phi_{\nu-1}\bigl(\mathcal{S}_{\nu+4,2}-\mathcal{S}_{\nu+2,2}+\mathcal{S}_{\nu,2}-\mathcal{S}_{\nu-2,2}\bigl)
 +\Phi_{\nu}\bigl(\mathcal{S}_{\nu+5,1}-\mathcal{S}_{\nu+3,1}+4\mathcal{S}_{\nu+1,3}-4\mathcal{S}_{\nu-1,3} -2\mathcal{S}_{\nu+1,1}+2\mathcal{S}_{\nu-1,1}\bigl)\nn\\
&&\hspace*{-1cm}
 +2\Phi_{\nu+1}\bigl(4\mathcal{S}_{\nu+2,4}-4\mathcal{S}_{\nu,4}-3\mathcal{S}_{\nu+2,2}+3\mathcal{S}_{\nu,2}\bigl)
 +\Phi_{\nu+2}\bigl(\mathcal{S}_{\nu-1,1}-\mathcal{S}_{\nu-3,1}\bigl)\bigl]\nn\\
&&+4a^2\bigl[\Phi_{\nu-2}\bigl((2\nu+1)\mathcal{S}_{\nu+1,1}-(2\nu-1)\mathcal{S}_{\nu-1,1}\bigl)
+2\Phi_{\nu-1}\bigl(\mathcal{S}_{\nu+2,0}+(\nu-1)\mathcal{S}_{\nu+2,2}-(\nu-1)\mathcal{S}_{\nu,2}\bigl)\nn\\
&&\hspace*{-1cm}
 +\Phi_{\nu}\bigl((2\nu+1)\mathcal{S}_{\nu+1,1}-(2\nu-1)\mathcal{S}_{\nu+3,1}+4\nu \mathcal{S}_{\nu+3,3}-4\nu \mathcal{S}_{\nu+1,3}\bigl)
 +2\Phi_{\nu+1}\bigl((\nu+3)\mathcal{S}_{\nu,2}-(\nu+1)\mathcal{S}_{\nu-2,2}-\mathcal{S}_{\nu,0}\bigl)\bigl]\nn\\
&&-8a^2(m-\lambda)\bigl[
 \Phi_{\nu-3}\bigl(\mathcal{S}_{\nu+1,1}-\mathcal{S}_{\nu-1,1}\bigl)+2\Phi_{\nu-2}\bigl(\mathcal{S}_{\nu+2,2} -\mathcal{S}_{\nu,2}\bigl)+2\Phi_{\nu-1}\bigl(\mathcal{S}_{\nu+3,3}-\mathcal{S}_{\nu+1,3}\bigl)\nn\\
&&\hspace*{-1cm}+\Phi_{\nu}\bigl(\mathcal{S}_{\nu,2}-\mathcal{S}_{\nu-2,2}+\mathcal{S}_{\nu+4,2}-\mathcal{S}_{\nu+2,2}\bigl)
 +\Phi_{\nu+1}\bigl(2\mathcal{S}_{\nu+1,3}-2\mathcal{S}_{\nu-1,3}-\mathcal{S}_{\nu+1,1}+\mathcal{S}_{\nu-1,1}\bigl)\bigl]\nn\\
&&+(m-\lambda)^2\bigl[\Phi_{\nu-2}\bigl(\mathcal{S}_{\nu+1,1}-\mathcal{S}_{\nu-1,1}\bigl)
 +2\Phi_{\nu-1}\bigl(\mathcal{S}_{\nu+2,2}-\mathcal{S}_{\nu,2}\bigl)
 +\Phi_{\nu}\bigl(\mathcal{S}_{\nu+3,1}-\mathcal{S}_{\nu+1,1}\bigl)\bigl]\nn\\
&&-2(m-\lambda)\bigl[\Phi_{\nu-1}\bigl((\nu+1) \mathcal{S}_{\nu+1,1}-\nu \mathcal{S}_{\nu-1,1}\bigl)
 +\Phi_{\nu}\bigl(\mathcal{S}_{\nu+2,0}+\nu \mathcal{S}_{\nu+2,2}-\nu \mathcal{S}_{\nu,2}\bigl)\bigl]\nn\\
&&+\nu\Phi_{\nu}\bigl((\nu+3)\mathcal{S}_{\nu+1,1}-(\nu+1)\mathcal{S}_{\nu-1,1}\bigl).\label{rho5-result}
\end{eqnarray}
 The analytical result for the chiral condensate for the quenched theory is given by
\begin{eqnarray}
\Sigma(m,a)&=&\left.\partial_{m'} {\rm Re}\,Z_\nu(\widehat{M},\widehat{X},a)\right|_{\substack{\widehat{M}=m\eins_4\\ \widehat{X}=i\varepsilon\eins_4\to0}}\nn\\
&=&16a^4\bigl[\Phi_{\nu-4}\bigl(S_{\nu+1,1}+S_{\nu-1,1}\bigl)
 +2\Phi_{\nu-3}\bigl(S_{\nu+2,2}+S_{\nu,2}\bigl)
 +\Phi_{\nu-2}\bigl(4S_{\nu+3,3}+4S_{\nu+1,3}-S_{\nu+3,1}-S_{\nu+1,1})\nn\\
&&\hspace*{-1cm}+2\Phi_{\nu-1}(S_{\nu+4,2}+S_{\nu+2,2}+S_{\nu,2}+S_{\nu-2,2})
 +\Phi_{\nu}(S_{\nu+5,1}+S_{\nu+3,1}+4S_{\nu+1,3}+4S_{\nu-1,3}-2S_{\nu+1,1}-2S_{\nu-1,1})\nn\\
&&\hspace*{-1cm}+2\Phi_{\nu+1}(4S_{\nu+2,4}+4S_{\nu,4}-3S_{\nu+2,2}-3S_{\nu,2})
 +\Phi_{\nu+2}(S_{\nu-1,1}+S_{\nu-3,1})\bigl]\nn\\
&&+4a^2\bigl[\Phi_{\nu-2}\bigl((2\nu+1)S_{\nu+1,1}+(2\nu-1)S_{\nu-1,1}\bigl)+2\Phi_{\nu-1}(S_{\nu+2,0}+(\nu-1)S_{\nu+2,2}+(\nu-1)S_{\nu,2})\nn\\
&&\hspace*{-1cm}+\Phi_{\nu}\bigl(4\nu S_{\nu+3,3} +4\nu S_{\nu+1,3}-(2\nu-1)S_{\nu+3,1}-(2\nu+1)S_{\nu+1,1}\bigl)
 +2\Phi_{\nu+1}\bigl((\nu+3)S_{\nu,2}+(\nu+1)S_{\nu-2,2}-S_{\nu,0}\bigl)\bigl]\nn\\
&&-8a^2m\bigl[\Phi_{\nu-3}\bigl(S_{\nu+1,1}+S_{\nu-1,1}\bigl)
 +2\Phi_{\nu-2}\bigl(S_{\nu+2,2}+S_{\nu,2}\bigl)
 +2\Phi_{\nu-1}\bigl(S_{\nu+3,3}+S_{\nu+1,3}\bigl)\nn\\
 &&\hspace*{-1cm}+\Phi_{\nu}\bigl(S_{\nu,2}+S_{\nu-2,2}+S_{\nu+4,2}+S_{\nu+2,2}\bigl)+\Phi_{\nu+1}\bigl(2S_{\nu+1,3}+2S_{\nu-1,3}-S_{\nu+1,1}-S_{\nu-1,1}\bigl)\bigl]\nn\\
&&+m^2\bigl[\Phi_{\nu-2}\bigl(S_{\nu+1,1}+S_{\nu-1,1}\bigl)+2\Phi_{\nu-1}\bigl(S_{\nu+2,2}+S_{\nu,2}\bigl) +\Phi_{\nu}\bigl(S_{\nu+3,1}+S_{\nu+1,1}\bigl)\bigl]\nn\\
&&-2m\bigl[\Phi_{\nu-1}\bigl((\nu+1) S_{\nu+1,1}+\nu S_{\nu-1,1}\bigl)+\Phi_{\nu}\bigl(S_{\nu+2,0}+\nu S_{\nu+2,2}+\nu S_{\nu,2}\bigl)\bigl]\nn\\
&&+\nu\Phi_{\nu}((\nu+3)S_{\nu+1,1}+(\nu+1)S_{\nu-1,1})\label{sigma-result}.
\end{eqnarray}
The distribution of chirality over the real eigenvalues can be 
written  as 
\begin{eqnarray}
\rho_\chi(m,a)&=&-\left.\frac{1}{\pi}\partial_{m} {\rm Im}\,Z_\nu(\widehat{M},\widehat{X},a)\right|_{\substack{\widehat{M}=m\eins_4\\ \widehat{X}=(\lambda+i\varepsilon)\eins_4\to\lambda\eins_4}}\nn\\
&=&16a^4\bigl[\Phi_{\nu-4}\bigl(\mathcal{S}_{\nu+1,1}+\mathcal{S}_{\nu-1,1}\bigl)
 +2\Phi_{\nu-3}\bigl(\mathcal{S}_{\nu+2,2}+\mathcal{S}_{\nu,2}\bigl)
 +\Phi_{\nu-2}\bigl(4\mathcal{S}_{\nu+3,3}+4\mathcal{S}_{\nu+1,3}-\mathcal{S}_{\nu+3,1}-\mathcal{S}_{\nu+1,1})\nn\\
&&\hspace*{-1cm}+2\Phi_{\nu-1}(\mathcal{S}_{\nu+4,2}+\mathcal{S}_{\nu+2,2}+\mathcal{S}_{\nu,2}+\mathcal{S}_{\nu-2,2})
 +\Phi_{\nu}(\mathcal{S}_{\nu+5,1}+\mathcal{S}_{\nu+3,1}+4\mathcal{S}_{\nu+1,3}+4\mathcal{S}_{\nu-1,3}-2\mathcal{S}_{\nu+1,1} -2\mathcal{S}_{\nu-1,1})\nn\\
&&\hspace*{-1cm}+2\Phi_{\nu+1}(4\mathcal{S}_{\nu+2,4}+4\mathcal{S}_{\nu,4}-3\mathcal{S}_{\nu+2,2}-3\mathcal{S}_{\nu,2})
 +\Phi_{\nu+2}(\mathcal{S}_{\nu-1,1}+\mathcal{S}_{\nu-3,1})\bigl]\nn\\
&&+4a^2\bigl[\Phi_{\nu-2}\bigl((2\nu+1)\mathcal{S}_{\nu+1,1}+(2\nu-1)\mathcal{S}_{\nu-1,1}\bigl)+2\Phi_{\nu-1}(\mathcal{S}_{\nu+2,0}+(\nu-1)\mathcal{S}_{\nu+2,2}+(\nu-1)\mathcal{S}_{\nu,2})\nn\\
&&\hspace*{-1cm}+\Phi_{\nu}\bigl(4\nu \mathcal{S}_{\nu+3,3} +4\nu \mathcal{S}_{\nu+1,3}-(2\nu-1)\mathcal{S}_{\nu+3,1}-(2\nu+1)\mathcal{S}_{\nu+1,1}\bigl)
 +2\Phi_{\nu+1}\bigl((\nu+3)\mathcal{S}_{\nu,2}+(\nu+1)\mathcal{S}_{\nu-2,2}-\mathcal{S}_{\nu,0}\bigl)\bigl]\nn\\
&&-8a^2m\bigl[\Phi_{\nu-3}\bigl(\mathcal{S}_{\nu+1,1}+\mathcal{S}_{\nu-1,1}\bigl)
 +2\Phi_{\nu-2}\bigl(\mathcal{S}_{\nu+2,2}+\mathcal{S}_{\nu,2}\bigl)
 +2\Phi_{\nu-1}\bigl(\mathcal{S}_{\nu+3,3}+\mathcal{S}_{\nu+1,3}\bigl)\nn\\
 &&\hspace*{-1cm}+\Phi_{\nu}\bigl(\mathcal{S}_{\nu,2}+\mathcal{S}_{\nu-2,2}+\mathcal{S}_{\nu+4,2} +\mathcal{S}_{\nu+2,2}\bigl)+\Phi_{\nu+1}\bigl(2\mathcal{S}_{\nu+1,3}+2\mathcal{S}_{\nu-1,3} -\mathcal{S}_{\nu+1,1}-\mathcal{S}_{\nu-1,1}\bigl)\bigl]\nn\\
&&+m^2\bigl[\Phi_{\nu-2}\bigl(\mathcal{S}_{\nu+1,1}+\mathcal{S}_{\nu-1,1}\bigl) +2\Phi_{\nu-1}\bigl(\mathcal{S}_{\nu+2,2}+\mathcal{S}_{\nu,2}\bigl) +\Phi_{\nu}\bigl(\mathcal{S}_{\nu+3,1}+\mathcal{S}_{\nu+1,1}\bigl)\bigl]\nn\\
&&-2m\bigl[\Phi_{\nu-1}\bigl((\nu+1) \mathcal{S}_{\nu+1,1}+\nu \mathcal{S}_{\nu-1,1}\bigl)+\Phi_{\nu}\bigl(\mathcal{S}_{\nu+2,0}+\nu \mathcal{S}_{\nu+2,2}+\nu \mathcal{S}_{\nu,2}\bigl)\bigl]\nn\\
&&+\nu\Phi_{\nu}((\nu+3)\mathcal{S}_{\nu+1,1}+(\nu+1)\mathcal{S}_{\nu-1,1}).\label{rhochi-result}
\end{eqnarray}
Though all three expression look quite complicated they are a finite sum of two kinds of integrals, only. This simplifies the numerical evaluation a lot.

\section{The Continuum Limit $a\to0$}\label{app:cont-limit}

To be self-consistent we briefly review the exact continuum limit (see subsection~\ref{app:exact-limit}) and the Gaussian orthogonal random matrix ensemble of finite matrix  size (see subsection~\ref{app:GOE}) which should describe the spectral broadening of the former zero modes into the real axis quite well at small lattice spacing $|a|\ll1$. In particular we wish to show how to extract the known results for the spectral observables from our calculations.

\subsection{Exact limit $a\to0$}\label{app:exact-limit}

In the continuum limit $a\to0$ the partition function takes the form
\be 
Z_\nu(\widehat{M},\widehat{X},a=0)&=&(m-x_0)^2\Phi_{\nu-2}S_{\nu,0}+2(m-x_0)(m'-x_1)\Phi_{\nu-1}S_{\nu+1,1}-2\nu(m-x_0)\Phi_{\nu-1}S_{\nu,0}\nn\\
&&+(m'-x_1)^2\Phi_{\nu}S_{\nu+2,0}-2\nu(m'-x_1)\Phi_{\nu}S_{\nu+1,1}+(\nu+1)\nu\Phi_{\nu}S_{\nu,0}.\label{step0}
\ee
with
\begin{eqnarray}
\Phi_\mu(m,x_0,a)&=&\left(\frac{m-x_0}{m+x_0}\right)^{\mu/2}I_{\mu}\left(2\sqrt{m^2-x_0^2}\right)\label{comp-int-a0}
\end{eqnarray}
and
\begin{equation}
S_{\mu,\alpha}(m',x_1,a)=2\frac{1}{(m'-x_1)^\mu}\int_1^\infty dy y^\alpha \left(\sqrt{|m'^2-x_1^2|}e^{i\phi}\right)^{\mu}K_{\mu}\left(2y\sqrt{|m'^2-x_1^2|}e^{i\phi}\right)\label{non-comp-int-a0}
\end{equation}
with $\phi=-L\sign(x_1)\Theta(x_1^2-m'^2)\pi/2$. We have to be careful about the phase $e^{i\phi}$ since it may change the sign of the result while it is unimportant for the phase of $\sqrt{m^2-x_0^2}$ in the compact, analytic integral. The reason is the cut along the negative real line of $K_\mu$. The index $\alpha$ only takes the values $\alpha=0,1$ which simplifies the calculation a lot.

In the first step we simplify the partition function by employing the relations
\begin{eqnarray}
 \mu\Phi_\mu+(x_0+m)\Phi_{\mu+1}+(x_0-m)\Phi_{\mu-1}&=&0,\nn\\
 z K_{\mu+1}(z)+(z\partial_z-\mu)K_{\mu}(z)&=&0,\nn\\
K_{\mu+1}(z)+K_{\mu-1}(z)+2\partial_z K_{\mu}(z)&=&0 .
\end{eqnarray}
The second and third relation is needed for expressing $S_{\nu+1,1}$ and $S_{\nu+2,0}$ in terms of $S_{\nu,0}$ and $K_\nu$ respectively,
\begin{eqnarray}
(m'-x_1)S_{\nu+1,1}(m',x_1,a=0)&=&\frac{\nu+1}{2}S_{\nu,0}(m',x_1,a=0)+\frac{\left(\sqrt{|m'^2-x_1^2|}e^{i\phi}\right)^\nu}{(m'-x_1)^\nu}K_{\nu}\left(2\sqrt{|m'^2-x_1^2|}e^{i\phi}\right),\nn\\
(m'-x_1)^2S_{\nu+2,0}(m',x_1,a=0)&=&-(m'^2-x_1^2)S_{\nu,0}(m',x_1,a=0)\nn\\
&&+2\frac{\left(\sqrt{|m'^2-x_1^2|}e^{i\phi}\right)^{\nu+1}}{(m'-x_1)^\nu}K_{\nu+1}\left(2\sqrt{|m'^2-x_1^2|}e^{i\phi}\right)
\end{eqnarray}
resulting from integration by parts. Then the quenched partition function reads
\be 
Z_\nu(\widehat{M},\widehat{X},a=0)&=&2\frac{\left(\sqrt{|m'^2-x_1^2|}e^{i\phi}\right)^\nu (m-x_0)^{\nu/2}}{(m'-x_1)^\nu(m+x_0)^{\nu/2}}\left[[(m^2-x_0^2)-(m'^2-x_1^2)]I_{\nu}\left(2\sqrt{m^2-x_0^2}\right)\right.\nn\\
&&\times\int_1^\infty dy K_{\nu}\left(2y\sqrt{|m'^2-x_1^2|}e^{i\phi}\right)+\sqrt{m^2-x_0^2}I_{\nu+1}\left(2\sqrt{m^2-x_0^2}\right)K_{\nu}\left(2\sqrt{|m'^2-x_1^2|}e^{i\phi}\right)\nn\\
&&\left.+\sqrt{|m'^2-x_1^2|}e^{i\phi}I_{\nu}\left(2\sqrt{m^2-x_0^2}\right)K_{\nu+1}\left(2\sqrt{|m'^2-x_1^2|}e^{i\phi}\right)\right]. \label{step1}
\ee
The last two terms also appear in QCD with three colors. In this way one can easily check the normalization $Z_\nu(m\eins_{4},x\eins_{4},a=0)=1$.

In the second step we take the derivative with respect to $m'$ or $x_1$ and set $m=m'$ and $x_0=x_1=x$, yielding
\be 
\left.\partial_{m'}Z_\nu(\widehat{M},\widehat{X},a=0)\right|_{\substack{\widehat{M}=m\eins_4\\\widehat{X}=x\eins_4}}&=&-\frac{\nu x}{m^2-x^2}-4mI_{\nu}\left(2\sqrt{|m^2-x^2|}e^{i\phi}\right)\int_1^\infty dy K_{\nu}\left(2y\sqrt{|m^2-x^2|}e^{i\phi}\right)\nn\\
&&\hspace*{-5cm}+\frac{2m}{\sqrt{|m^2-x^2|}e^{i\phi}}\partial_y\left[2\sqrt{|m^2-x^2|}e^{i\phi}I_{\nu+1}\left(2\sqrt{|m^2-x^2|}e^{i\phi}\right)K_{\nu}\left(y\right)+y I_{\nu}\left(2\sqrt{|m^2-x^2|}e^{i\phi}\right)K_{\nu+1}\left(y\right)\right]_{y=2\sqrt{|m^2-x^2|}e^{i\phi}}.\nn\\
&=&\frac{\nu }{x-m}-4mI_{\nu}\left(2\sqrt{|m^2-x^2|}e^{i\phi}\right)\int_1^\infty dy K_{\nu}\left(2y\sqrt{|m^2-x^2|}e^{i\phi}\right)\nn\\
&&\hspace*{-5cm}-4m\left[I_{\nu+1}\left(2\sqrt{|m^2-x^2|}e^{i\phi}\right)K_{\nu-1}\left(2\sqrt{|m^2-x^2|}e^{i\phi}\right)+ I_{\nu}\left(2\sqrt{|m^2-x^2|}e^{i\phi}\right)K_{\nu}\left(2\sqrt{|m^2-x^2|}e^{i\phi}\right)\right] \label{step2-a}
\ee
and
\be 
\left.\partial_{x_1}Z_\nu(\widehat{M},\widehat{X},a=0)\right|_{\substack{\widehat{M}=m\eins_4\\\widehat{X}=x\eins_4}}&=&\frac{\nu m}{m^2-x^2}+4xI_{\nu}\left(2\sqrt{|m^2-x^2|}e^{i\phi}\right)\int_1^\infty dy K_{\nu}\left(2y\sqrt{|m^2-x^2|}e^{i\phi}\right)\nn\\
&&\hspace*{-5cm}-\frac{2x}{\sqrt{|m^2-x^2|}e^{i\phi}}\partial_y\left[2\sqrt{|m^2-x^2|}e^{i\phi}I_{\nu+1}\left(2\sqrt{|m^2-x^2|}e^{i\phi}\right)K_{\nu}\left(y\right)+y I_{\nu}\left(2\sqrt{|m^2-x^2|}e^{i\phi}\right)K_{\nu+1}\left(y\right)\right]_{y=2\sqrt{|m^2-x^2|}e^{i\phi}}\nn\\
&=&\frac{\nu }{m-x}+4xI_{\nu}\left(2\sqrt{|m^2-x^2|}e^{i\phi}\right)\int_1^\infty dy K_{\nu}\left(2y\sqrt{|m^2-x^2|}e^{i\phi}\right)\nn\\
&&\hspace*{-5cm}+4x\left[I_{\nu+1}\left(2\sqrt{|m^2-x^2|}e^{i\phi}\right)K_{\nu-1}\left(2\sqrt{|m^2-x^2|}e^{i\phi}\right)+ I_{\nu}\left(2\sqrt{|m^2-x^2|}e^{i\phi}\right)K_{\nu}\left(2\sqrt{|m^2-x^2|}e^{i\phi}\right)\right].\label{step2b}
\ee
Up to an overall sign, Equation~\eqref{step2-a} is equal to 
the chiral condensate $\Sigma$ when setting $x=0$ (implying $\phi=0$), i.e. the chiral condensate without zero modes is
\begin{equation}\label{sigma-chiGOE}
\Sigma_{\rm chGOE}^{(\nu)}(m)=4m\left[I_{\nu+1}\left(2|m|\right)K_{\nu-1}\left(2|m|\right)+ I_{\nu}\left(2|m|\right)K_{\nu}\left(2|m|\right)+I_{\nu}\left(2|m|\right)\int_1^\infty dy K_{\nu}\left(2y|m|\right)\right].
\end{equation} 
This result is in agreement with the expression in \cite{damgaard-chgoe}, but
has the advantage that there is no need to distinguish even and odd $\nu$.
For odd $\nu$ the Bessel function $K_\nu$ can be written as a total derivative
allowing us to evaluate the integral exactly,
\be
K_{2k+1}(x) = -2 \frac d{dx}[K_{2k}(x) - K_{2k-2}(x) + \cdots +(-1)^{k}K_2(x) 
+(-1)^{k+1} \frac 12 K_0(x)].   
\ee
For $\nu = 2k+1$,  this results in the condensate
\be
\Sigma(m) &=&  -\frac \nu m +4I_\nu(2m)[K_{2k}(2m) - K_{2k-2}(2m) 
+ \cdots +(-1)^{k}K_2(2m)] +(-1)^{k+1}I_\nu(2m) K_0(2m)\nn \\ &&
-4mI_{\nu+1}(2m)K_{\nu-1}(2m)- 4mI_{\nu+1}(2m)K_{\nu-1}(2m),
\ee
which up to a rescaling $m \to 2m$ and $\Sigma \to \Sigma/2$ agrees with 
the result in \cite{damgaard-chgoe}. For even $\nu$ we can also use the 
recursion relation for Bessel functions but we are left with an integral over
$K_0$. 
\be
K_{2k}(x) = 2(-1)^{\nu/2} \frac d{dx}\left [ \sum_{k=0}^{\nu/2-1} (-1)^kK_{2k+1}(x)
\right ] +(-1)^{\nu/2}K_0(x)  . 
\ee
This results in
\be
\Sigma(m) &=&  -\frac \nu m -4mI_{\nu+1}(2m)K_{\nu-1}(2m)- 4mI_{\nu+1}(2m)K_{\nu-1}(2m)\nn\\
&& +4(-1)^{\nu/2}I_\nu(2m)\sum_{k=0}^{\nu/2-1} (-1)^kK_{2k+1}(x)
-2m (-1)^{\nu/2}I_\nu(2m)  \int_1^\infty dy  K_0(2my).
\ee
The integral over $K_0$ can be expressed into modified Struve functions, and
it can be numerically shown that it agrees with Eq.~(8) of \cite{damgaard-chgoe}.

Let us underline that the valence quark mass dependence
of the chiral condensate for ${\rm chGSE}$ is also derived in \cite{damgaard-chgoe}. The authors again obtain  separate
expressions for even and odd $\nu$ which can be simplified to
$-2x K_{N_{\rm f}+2\nu}(2x) \int_0^1 dy I_{N_{\rm f}+2\nu}(2xy)$ up to rescaling.

From the result~\eqref{step2-a} we can immediately deduce 
the well-known result for the distribution of the chiralities over the real eigenvalues which is a Dirac delta function in the continuum limit,
\begin{eqnarray}\label{rho-chi-a0}
\rho_{\chi}(m,a=0)=\frac{1}{\pi}{\rm Im}\left.\partial_{m'}Z_\nu(\widehat{M},\widehat{X},a=0)\right|_{\substack{\widehat{M}=m\eins_4\\\widehat{X}=-i\varepsilon\eins_4\to0}}=\nu\delta(m).
\end{eqnarray}
Only the first term of Eq.~\eqref{step2-a} contributes to this results  since the remaining parts are real for $m>x=0$.

The spectral density of the Hermitian Dirac operator $D_5$ can be obtained by setting $x=\lambda+\imath \varepsilon$ and taking the imaginary part of Eq.~\eqref{step2b} in the limit $\varepsilon\to0$.
We have to distinguish two cases,  $|m|> |x| $  or $|m|< |x|$. In the first case all terms are real and vanish while in the latter case we have to apply the relations~\eqref{index-rel}, \eqref{K-J-rel}, and \eqref{calc-K-J-b} yielding,
\begin{eqnarray}\label{rho-5-a0}
\rho_{5}(\lambda,m,a=0)&=&\frac{1}{\pi}{\rm Im}\left.\partial_{x_1}Z_\nu(\widehat{M},\widehat{X},a=0)\right|_{\substack{\widehat{M}=m\eins_4\\\widehat{X}=(\lambda+i\varepsilon)\eins_4\to\lambda\eins_4}}\nn\\
&=&\nu \delta(m-\lambda)+\frac{|\lambda|}{\sqrt{\lambda^2-m^2}} J_\nu(2\sqrt{\lambda^2-m^2})\left[1-\int_0^{2\sqrt{\lambda^2-m^2}}dyJ_\nu(y)\right]\Theta(|\lambda|-|m|)\nn\\
&&+2|\lambda|\left[ J_\nu^2(2\sqrt{\lambda^2-m^2})- J_{\nu+1}(2\sqrt{\lambda^2-m^2}) J_{\nu-1}(2\sqrt{\lambda^2-m^2})\right]\Theta(|\lambda|-|m|),
\end{eqnarray}
cf. \cite{GOE-level}. Note that Eqs.~\eqref{K-J-rel} and \eqref{calc-K-J-b} are multiplied with $n=-{\rm sign}(\lambda)$ because of the phase $\phi=-\sign(x_1)\Theta(x_1^2-m'^2)\pi/2$ for $L=1$.

\subsection{Spectral Observables of the Gaussian Orthogonal Ensemble}\label{app:GOE}

We choose the abbreviation $\widehat{Y}=(\widehat{M}-\widehat{X})/(4a)=\diag(y_0,y_0,y_1,y_1)$. The partition function~\eqref{part-GOE-SUSY} is equal to a partition function of $\nu\times\nu$ real symmetric random matrices distributed by a Gaussian. Let $\Sym(\nu)$ be the set of these real symmetric matrices. 
In terms of these matrices, the partition function reads,
\begin{eqnarray}
Z_{\GOE}^{(\nu)}\left(\widehat{Y}\right)&=&\frac{1}{(2\pi)^{\nu(\nu-1)/4}\pi^{\nu/2}}\int_{\Sym(\nu)} dH \exp[-\tr H^2]\frac{\det(H-y_0\eins_{\nu})}{\det(H-y_1\eins_{\nu})}.\label{app-GOE-a}
\end{eqnarray}
For the cases $\nu=0,1$ they take particular simple forms,
\begin{eqnarray}
Z_{\GOE}^{(0)}\left(\widehat{Y}\right)&=&1\label{app-GOE-nu0}
\end{eqnarray}
and
\begin{eqnarray}
Z_{\GOE}^{(1)}\left(\widehat{Y}\right)&=&\frac{1}{\sqrt{\pi}}\int_{-\infty}^\infty dE \frac{E-y_0}{E-y_1}e^{-E^2}=1+\sqrt{\pi}(y_0-y_1)\frac{{\rm erf}(iy_1)}{i}e^{-y_1^2}-iL\sqrt{\pi}(y_1-y_0)e^{-y_1^2}.\label{app-GOE-nu1}
\end{eqnarray}
In the latter equation we assumed that the imaginary part of $x_1={\rm Re}\, x_1+iL\varepsilon$ ($L=\pm1$) is infinitesimal small, $\varepsilon\to0$. The error function ${\rm erf}(z)=2\int_0^z dz' e^{-z'^2}/\sqrt{\pi}$ is real despite the imaginary unit.

For $\nu>1$ the result directly following from the representation~\eqref{app-GOE-a} are not that trivial as the ones in Eqs.~\eqref{app-GOE-nu0} and \eqref{app-GOE-nu1}. 
Therefore we start for those cases from the supersymmeric representation~\eqref{part-GOE-SUSY} which, in the parametrization~\eqref{para-sigma}, reads as
\begin{eqnarray}
Z_{\GOE}^{(\nu)}\left(\widehat{Y}\right)&=&\frac{1}{\pi}\int_{-\infty}^\infty du\int_{-\infty}^\infty dv_1\int_{-\infty}^\infty dv_2|v_1-v_2|e^{-(2u^2+v_1^2+v_2^2)}\frac{(y_0-iu)^\nu}{(y_1-v_1)^{\nu/2}(y_1-v_2)^{\nu/2}}\nn\\
&&\times\left[1-\frac{\nu}{4}\frac{2y_1-v_1-v_2}{(y_0-iu)(y_1-v_1)(y_1-v_2)}+\frac{\nu(\nu-1)}{16}\frac{1}{(y_0-iu)^2(y_1-v_1)(y_1-v_2)}\right].\label{app-GOE-b}
\end{eqnarray}
We have already integrated over the orthogonal matrix $\widetilde{O}$ and the four Grassmann variables $\eta,\eta^*,\chi,\chi^*$. Moreover, we rescaled the remaining variables $(u,v_1,v_2)\to 4 a(u,v_1,v_2)$. The normalization constant is fixed by $y_0=y_1=i\varepsilon\to\infty$.

The integral over $u$ is equal to Hermite polynomials,
\begin{equation}\label{Hermite}
\int_{-\infty}^\infty du e^{-2u^2} (y_0-iu)^\mu=\sqrt{\frac{\pi}{2}}2^{-\mu}H_{\mu}\left(2y_0\right),
\end{equation}
where we have chosen the monic normalization $H_{\mu}\left(z\right)=z^\mu+\ldots$ Then the partition function is equal to
\begin{eqnarray}
Z_{\GOE}^{(\nu)}\left(\widehat{Y}\right)&=&H_{\nu}\left(2y_0\right)\mathcal{I}_{\nu,0}(y_1)+ H_{\nu-1}\left(2y_0\right)\mathcal{I}_{\nu,1}(y_1)+\nu(\nu-1)H_{\nu-2}\left(2y_0\right)\mathcal{I}_{\nu+2,0}(y_1)\label{app-GOE-c}
\end{eqnarray}
with
\begin{eqnarray}\label{double-int}
 \mathcal{I}_{\mu,\alpha}(y_1)&=&\frac{2^{-\mu}}{\sqrt{2\pi}}\partial_{y_1}^\alpha\int_{-\infty}^\infty dv_1\int_{-\infty}^\infty dv_2|v_1-v_2|e^{-(v_1^2+v_2^2)}\frac{1}{(y_1-v_1)^{\mu/2}(y_1-v_2)^{\mu/2}}.
\end{eqnarray}
The latter twofold integral exits due to the non-vanishing imaginary part 
of $y_1=m-{\rm Re}\, x_1-iL\varepsilon$. This integral can be brought 
into a form which is easily integrable, and the spectral observables 
we are interested in can be readily extracted. 
To find such a representation it is suitable to understand 
the integral~\eqref{double-int} in terms of  
an integral over a $2\times2$ real symmetric matrix $H$, 
\begin{eqnarray}\label{double-int-b}
 \mathcal{I}_{\mu,\alpha}(y_1)&\propto&\partial_{y_1}^\alpha\int_{\Sym(2)} \frac{dH e^{-\tr H^2}}{{\det}^{\mu/2}(y_1\eins_2-H)},
\end{eqnarray}
where $v_1$ and $v_2$ are the eigenvalues of $H$. The determinant 
in Eq.~\eqref{double-int-b}
can   be written as an integral over  $2\times2$ positive definite real symmetric matrix $H'$
\begin{eqnarray}\label{double-int-c}
\frac {1}{ {\det}^{\mu/2}(y_1\eins_2-H)}&\propto&e^{iL\pi\mu/2}\int_{\Sym_+(2)} dH' \exp[-2iL\tr H'(y_1\eins_2-H)]{\det}^{(\mu-3)/2} H'.
\end{eqnarray}
The index $\mu$ has to be larger than $1$ and thus $\nu\geq2$ which is okay since we know explicit results for $\nu=0,1$.
The phase in front of this integral is important as well. The integral over $H$ yields a Gaussian integral in $H'$. After diagonalizing $H'=O'\diag(E_1,E_2)O'^T$ with $O'\in\Ort(2)$ and $E_1,E_2>0$ we find
\begin{equation}\label{double-int-d}
 \mathcal{I}_{\mu,\alpha}(y_1)=\frac{2^{\mu+\alpha-3}e^{iL\pi(\mu-\alpha)/2}}{(\mu-2)!}\int_0^\infty dE_1\int_0^\infty dE_2|E_1-E_2| (E_1E_2)^{(\mu-3)/2}(E_1+E_2)^\alpha\exp[-(E_1^2+E_2^2)-2iLy_1(E_1+E_2)].
\end{equation}
Already at this point we can set the imaginary part of $y_1$ exactly to zero. 
We order the eigenvalues $E_1>E_2$ and change to polar coordinates $E_1=r\cos\varphi$ and $E_2=r\sin\varphi$ with $r>0$ and $0\leq\varphi\leq\pi/4$,
\begin{eqnarray}
 \mathcal{I}_{\mu,\alpha}(y_1)&=&\frac{2^{(\mu+3\alpha)/2}e^{iL\pi(\mu-\alpha)/2}}{(\mu-2)!}\int_0^\infty dR\int_0^{\pi/4} d\varphi R^{\mu+\alpha-1}\sin^{(\mu-3)/2}(2\varphi) \cos\left(\varphi+\frac{\pi}{4}\right) \cos^\alpha\left(\varphi-\frac{\pi}{4}\right)\nn\\
 &&\times\exp\left[-R^2-\sqrt{8}iLy_1R\cos\left(\varphi-\frac{\pi}{4}\right)\right]\nn\\
 &=&\frac{2^{(\mu+3\alpha-2)/2}}{(\mu-2)!}\int_0^{\pi/4} d\varphi\sin^{(\mu-3)/2}(2\varphi) \cos\left(\varphi+\frac{\pi}{4}\right)\cos^\alpha\left(\varphi-\frac{\pi}{4}\right)\nn\\
 &&\times\left[\Gamma\left(\frac{\mu+\alpha}{2}\right)e^{iL\pi(\mu-\alpha)/2}M\left(\frac{\mu+\alpha}{2},\frac{1}{2};-2y_1^2\cos^2\left(\varphi-\frac{\pi}{4}\right)\right)\right.\nn\\
 &&\left.+\sqrt{8}\Gamma\left(\frac{\mu+\alpha+1}{2}\right)e^{iL\pi(\mu-\alpha-1)/2}y_1\cos\left(\varphi-\frac{\pi}{4}\right) M\left(\frac{\mu+\alpha+1}{2},\frac{3}{2};-2y_1^2\cos^2\left(\varphi-\frac{\pi}{4}\right)\right)\right].\label{double-int-e}
\end{eqnarray}
We have employed Kummer's confluent hypergeometric function~\cite{Abramowitz}
\begin{equation}\label{Kummer}
 M\left(a,b;z\right)=\sum_{j=0}^\infty\frac{\Gamma[a+j]\Gamma[b]}{\Gamma[a]\Gamma[b+j]}\frac{z^j}{j!}.
\end{equation}
We underline that only one of the two terms in Eq.~\eqref{double-int-e} is real and the other one imaginary depending on the parity of $\mu$. In random matrix theory this subtle difference between even and odd matrix dimension for real matrices is well known \cite{mehta}.

Let us define the abbreviations
\begin{equation}\label{I-int-1}
\mathcal{I}_{\mu,\alpha}^{(1)}(y_1)=\int_0^{\pi/4} d\varphi\sin^{(\mu-3)/2}(2\varphi) \cos\left(\varphi+\frac{\pi}{4}\right)\cos^\alpha\left(\varphi-\frac{\pi}{4}\right)M\left(\frac{\mu+\alpha}{2},\frac{1}{2};-2y_1^2\cos^2\left(\varphi-\frac{\pi}{4}\right)\right),
\end{equation}
and
\begin{equation}\label{I-int-2}
\mathcal{I}_{\mu,\alpha}^{(2)}(y_1)=y_1\int_0^{\pi/4} d\varphi\sin^{(\mu-3)/2}(2\varphi) \cos\left(\varphi+\frac{\pi}{4}\right)\cos^{\alpha+1}\left(\varphi-\frac{\pi}{4}\right)M\left(\frac{\mu+\alpha+1}{2},\frac{3}{2};-2y_1^2\cos^2\left(\varphi-\frac{\pi}{4}\right)\right).
\end{equation}
Then the real part of the first derivative in $y_1$ and setting $y_1\to m/4a$ yields the contribution of the chiral condensate which results from this finite dimensional Gaussian orthogonal ensemble, i.e.
\begin{eqnarray}
 &&\Sigma_{\GOE}^{(\nu>1)}\left(\frac{m}{4a}\right)=-\left.\partial_{m'}{\rm Re}\,Z_{\GOE}^{(\nu)}\left(\widehat{Y}\right)\right|_{\widehat{Y}=(m/4a)\eins_4}\label{Sigma-GOE}\\
 &=&\frac{2^{\nu/2}\Gamma[(\nu+2)/2]}{(\nu-2)! a}\cos\left(\frac{\pi\nu}{2}\right)\left[H_{\nu}\left(\frac{m}{2a}\right)\mathcal{I}_{\nu,1}^{(2)}\left(\frac{m}{4a}\right)+H_{\nu-1}\left(\frac{m}{2a}\right)\mathcal{I}_{\nu,2}^{(1)}\left(\frac{m}{4a}\right)-(\nu+2)H_{\nu-2}\left(\frac{m}{2a}\right)\mathcal{I}_{\nu+2,1}^{(2)}\left(\frac{m}{4a}\right)\right]\nn\\
 &&-\frac{2^{(\nu-1)/2}\Gamma[(\nu+3)/2]}{(\nu-2)! a}\sin\left(\frac{\pi\nu}{2}\right)\left[\frac{1}{\nu+1}H_{\nu}\left(\frac{m}{2a}\right)\mathcal{I}_{\nu,1}^{(1)}\left(\frac{m}{4a}\right)-4H_{\nu-1}\left(\frac{m}{2a}\right)\mathcal{I}_{\nu,2}^{(2)}\left(\frac{m}{4a}\right)-H_{\nu-2}\left(\frac{m}{2a}\right)\mathcal{I}_{\nu+2,1}^{(1)}\left(\frac{m}{4a}\right)\right].\nn 
\end{eqnarray}
Note that only one of the two terms contribute because either $\cos\left(\pi\nu/2\right)$ or $\sin\left(\pi\nu/2\right)$ does not vanish depending on the parity of $\nu$. The ``chiral condensate" for the case $\nu=0$ vanishes,
\begin{eqnarray}
 \Sigma_{\GOE}^{(\nu=0)}\left(\frac{m}{4a}\right)=0,\label{Sigma-GOE-nu0}
\end{eqnarray}
while the one for $\nu=1$ can be read off from Eq.~\eqref{app-GOE-nu1},
\begin{eqnarray}
 \Sigma_{\GOE}^{(\nu=1)}\left(\frac{m}{4a}\right)=\sqrt{\frac{\pi}{16a^2}}\frac{{\rm erf}(im/4a)}{i}\exp\left(-\frac{m^2}{16a^2}\right).\label{Sigma-GOE-nu1}
\end{eqnarray}
Other representations of the partition function and the chiral condensate in terms of Hermite polynomials and
its Cauchy transform exist in the literature \cite{mehta}.
 However we chose the representation Eq.~\eqref{Sigma-GOE} which can be easily 
evaluated  numerically and the cases even and odd $\nu$ can be discussed on the same footing.

The level density is obtained from the imaginary part of the first derivative in $y_1$,
\begin{eqnarray}
 &&\rho_{\GOE}^{(\nu>1)}\left(\frac{m}{4a}\right)=-\frac{1}{\pi}\left.\partial_{m'}{\rm Im}\,Z_{\GOE}^{(\nu)}\left(\widehat{Y}\right)\right|_{\widehat{Y}=(m/4a)\eins_4}\label{level-GOE}\\
 &=&\frac{2^{\nu/2}\Gamma[(\nu+2)/2]}{(\nu-2)! a}\sin\left(\frac{\pi\nu}{2}\right)\left[H_{\nu}\left(\frac{m}{2a}\right)\mathcal{I}_{\nu,1}^{(2)}\left(\frac{m}{4a}\right)+H_{\nu-1}\left(\frac{m}{2a}\right)\mathcal{I}_{\nu,2}^{(1)}\left(\frac{m}{4a}\right)-(\nu+2)H_{\nu-2}\left(\frac{m}{2a}\right)\mathcal{I}_{\nu+2,1}^{(2)}\left(\frac{m}{4a}\right)\right]\nn\\
 &&+\frac{2^{(\nu-1)/2}\Gamma[(\nu+3)/2]}{(\nu-2)! a}\cos\left(\frac{\pi\nu}{2}\right)\left[\frac{1}{\nu+1}H_{\nu}\left(\frac{m}{2a}\right)\mathcal{I}_{\nu,1}^{(1)}\left(\frac{m}{4a}\right)-4H_{\nu-1}\left(\frac{m}{2a}\right)\mathcal{I}_{\nu,2}^{(2)}\left(\frac{m}{4a}\right)-H_{\nu-2}\left(\frac{m}{2a}\right)\mathcal{I}_{\nu+2,1}^{(1)}\left(\frac{m}{4a}\right)\right].\nn
\end{eqnarray}
The exceptional cases for $\nu=0,1$ are
\begin{eqnarray}
 \rho_{\GOE}^{(\nu=0)}\left(\frac{m}{4a}\right)=0,\label{rho-GOE-nu0}
\end{eqnarray}
and
\begin{eqnarray}
 \rho_{\GOE}^{(\nu=1)}\left(\frac{m}{4a}\right)=\frac{1}{\sqrt{16\pi a^2}}\exp\left(-\frac{m^2}{16a^2}\right).\label{rho-GOE-nu1}
\end{eqnarray}
These expressions are a good approximation for the level density of the real eigenvalues of the non-Hermitian Wilson Dirac operator at very small lattice spacing $|a|\ll1$ which are the former zero modes of the continuum QCD Dirac operator.

\section{Computation of the Saddle Point Solutions~\eqref{saddle-lambda0-ch} and \eqref{saddle-sol-ch}}\label{app:saddlepointsol}

We have to solve the rational equation
\begin{equation}\label{saddlepoint-b}
p(V)=\frac{m}{2} (V-V^{-1}) +\frac{\lambda}{2}(V+V^{-1}) -2a^2 (V^2 - V^{-2})=0.
\end{equation}
with real coefficients. In particular each eigenvalue $z$ of $V$ satisfies exactly the same equation. Because the rational function $p(z)$ has real coefficients it has either no, two or four real zero points satisfying Eq.~\eqref{saddlepoint-b}. All other zero points are complex conjugate pairs. Since $p(z\to0)\to2a^2z^{-2}\to+\infty$ while $p(z\to\pm\infty)\to-2a^2z^{2}\to-\infty$ we have at least two real zero points. The question is: What happens with the other two zero points? We are in particular interested the complex solutions since they are important for the level density $\rho_5$ when we have the general situation with $\lambda\neq0$.

Splitting the eigenvalue $z$ of $V$ into a radial and angular part, i.e. $z=re^{i\varphi}$ with $r\geq0$ and $\varphi\in]-\pi,\pi[$, we obtain two equations,
\begin{eqnarray}
\left(\frac{m+\lambda}{2}r\cos\varphi-\frac{m-\lambda}{2}r^{-1}\cos\varphi -2a^2 (r^2 - r^{-2})(2\cos^2\varphi-1)\right)&=&0,\nn\\ \left(\frac{m+\lambda}{2}r+\frac{m-\lambda}{2}r^{-1} -4a^2 (r^2 + r^{-2})\cos\varphi\right)\sin\varphi&=&0,\label{saddlepoint-c}
\end{eqnarray}
corresponding to the real and imaginary part of the saddle point equation, respectively. The second equality is either solved by $\sin\varphi=0$ implying a real solution or the cosine can be expressed in terms of the radius $r>0$.

Let us assume $\sin\varphi\neq0$. Then the equation solved by the radius is
\begin{eqnarray}
\sinh^3 2\vartheta+\left(1-\frac{m^2-\lambda^2}{(8a^2)^2}\right)\sinh 2\vartheta+\frac{2 m\lambda}{(8a^2)^2}&=&0\label{saddlepoint-d}
\end{eqnarray}
with $r=e^\vartheta$. This equation has always one real solution which is
\begin{equation}\label{saddle-sol}
\sinh 2\vartheta=-\left(\frac{m\lambda}{(8a^2)^2}+\sqrt{\frac{m^2\lambda^2}{(8a^2)^4}-\frac{1}{27}\left(\frac{m^2-\lambda^2}{(8a^2)^2}-1\right)^3}\right)^{1/3}-\left(\frac{m\lambda}{(8a^2)^2}-\sqrt{\frac{m^2\lambda^2}{(8a^2)^4}-\frac{1}{27}\left(\frac{m^2-\lambda^2}{(8a^2)^2}-1\right)^3}\right)^{1/3},
\end{equation}
where we have to take the following root $x^{1/3}=\sign (x) |x|^{1/3}$ if $x$ is real. There are two additional real solutions in the region with
\begin{equation}\label{saddle-sol-reg}
\frac{m^2\lambda^2}{(8a^2)^4}<\frac{1}{27}\left(\frac{m^2-\lambda^2}{(8a^2)^2}-1\right)^3 \ \Leftrightarrow\ \frac{\lambda^2}{8a^2}<\left[\left(\frac{m}{8a^2}\right)^{2/3}-1\right]^3
\end{equation}
which implies $\lambda^2<m^2-(8a^2)^2$ and can only appear if the quark mass satisfies $|m|>8a^2$. However in this region the cosine of the angle $\varphi$ has to be larger than $1$ implying that it has to be complex which should not be the case. This can be seen in the following short computation.

The other two solutions are given by 
\begin{eqnarray}
\sinh 2\vartheta_n&=&-\sign(m\lambda)\left[e^{i 2n\pi/3}\left(\frac{|m\lambda|}{(8a^2)^2}+i\,\sign(m\lambda)\sqrt{\frac{1}{27}\left(\frac{m^2-\lambda^2}{(8a^2)^2}-1\right)^3-\frac{m^2\lambda^2}{(8a^2)^4}}\right)^{1/3}\right.\nn\\
&&\left.+e^{-i 2n\pi/3}\left(\frac{|m\lambda|}{(8a^2)^2}-i\,\sign(m\lambda)\sqrt{\frac{1}{27}\left(\frac{m^2-\lambda^2}{(8a^2)^2}-1\right)^3-\frac{m^2\lambda^2}{(8a^2)^4}}\right)^{1/3}\right]\label{saddle-calc-a}
\end{eqnarray}
with $n=1,2$. Hence each solution is bounded as
\begin{eqnarray}\label{saddle-calc-b}
\frac{1}{\sqrt{3}}\sqrt{\frac{m^2-\lambda^2}{(8a^2)^2}-1}\leq \sign(m\lambda)\sinh 2\vartheta_1\leq \sqrt{\frac{m^2-\lambda^2}{(8a^2)^2}-1} &\Rightarrow&\frac{1}{\sqrt{3}}\sqrt{\frac{m^2-\lambda^2}{(8a^2)^2}+2}\leq \cosh 2\vartheta_1\leq\sqrt{\frac{m^2-\lambda^2}{(8a^2)^2}},\nn\\
0\leq \sign(m\lambda)\sinh 2\vartheta_2\leq \frac{1}{\sqrt{3}}\sqrt{\frac{m^2-\lambda^2}{(8a^2)^2}-1} &\Rightarrow&1\leq \cosh 2\vartheta_0\leq\frac{1}{\sqrt{3}}\sqrt{\frac{m^2-\lambda^2}{(8a^2)^2}+2}.
\end{eqnarray} 
We combine the two equalities in Eq.~\eqref{saddlepoint-c} such that we have for the angle
\begin{eqnarray}
|\cos2\varphi|=\left|\frac{(m+\lambda)^2r^2-(m-\lambda)^2r^{-2}}{32a^4(r^4-r^{-4})}\right|=\frac{m^2+\lambda^2}{(8a^2)^2\cosh 2\vartheta_n}+\frac{2|m\lambda|}{(8a^2)^2|\sinh 2\vartheta_n|}.\label{saddle-calc-c}
\end{eqnarray}
This allows us to make the following estimations
\begin{eqnarray}
n=1:& &|\cos2\varphi|\geq\frac{m^2+\lambda^2}{8a^2\sqrt{m^2-\lambda^2}}+\frac{2|m\lambda|}{8a^2\sqrt{m^2-\lambda^2-(8a^2)^2}}\geq\frac{(|m|+|\lambda|)^2}{8a^2\sqrt{m^2-\lambda^2}},\nn\\
n=2:& &|\cos2\varphi|\geq\frac{\sqrt{3}(m^2+\lambda^2)}{8a^2\sqrt{m^2-\lambda^2+2(8a^2)^2}}+\frac{\sqrt{12}|m\lambda|}{8a^2\sqrt{m^2-\lambda^2-(8a^2)^2}}\geq\frac{\sqrt{3}(|m|+|\lambda|)^2}{8a^2\sqrt{m^2-\lambda^2+2(8a^2)^2}},
\label{saddle-calc-d}
\end{eqnarray}
In both cases the right hand side is a function which is monotonously increasing in  $|\lambda|$ implying that $|\lambda|=0$ is the minimum. The remaining function in $|m|$ is also monotonously increasing such that we have a minimum at the values at $|m|=8a^2$ which results in
\begin{eqnarray}
|\cos2\varphi|&\geq&1\label{saddle-calc-e}
\end{eqnarray}
meaning that those solutions for $\sinh2\vartheta$ are forbidden.

Hence we have only a complex solution if
\begin{equation}\label{region-sol}
\frac{|\lambda|}{8a^2}>\left[\left(\frac{ |m|}{8a^2}\right)^{2/3}-1\right]^{3/2}
\end{equation}
which is
\begin{equation}\label{sol-a}
U_0=i L\,  e^{\vartheta+L i \varphi}\eins_4\ {\rm with}\ \varphi=\arccos\left[\frac{m\cosh\vartheta+\lambda\sinh\vartheta}{8a^2\cosh2\vartheta}\right]
\end{equation}
and $\vartheta$ is given by Eq.~\eqref{saddle-sol}.
The sign $L$ is fixed because of the non-compact double integral over $s_1$ and $s_2$. The infinitesimal imaginary shift $i L\varepsilon$ in $\lambda$ and the singularity at $e^{s_1},e^{s_2}=0$ prevents to shift the contour to the solution $z=e^{\vartheta-L i \varphi}$ in the thermodynamic limit. 

The region~\eqref{region-sol} implies a spectral gap of $D_5$ in the interval
\begin{equation}
\lambda\in\left[-\left( |m|^{2/3}-(8a^2)^{2/3}\right)^{3/2},\left( |m|^{2/3}-(8a^2)^{2/3}\right)^{3/2}\right],
\end{equation}
 cf. Fig.~\ref{fig:thermo}.a). It is closed if $|m|\leq 8a^2$ such that we enter a new phase which is the Aoki phase for lattice QCD with two colors.
 
 The situation of the spectrum looks slightly different at $\lambda=0$. In this case the saddle point equation takes the simple form $p(z)=2a^2(z-z^{-1})(z+z^{-1}-m/(4a^2))=0$ where we can readily read off the solutions. Plugging these solutions into the exponent~\eqref{partition-thermo} we have to choose the maximum which is
\begin{equation}\label{saddle-lambda0}
U_0=\left\{\begin{array}{cl} \displaystyle iL\frac{m}{8a^2}+\sqrt{1-\left(\frac{m}{8a^2}\right)^2}, & |m|<8a^2, \\ iL\,\sign\, m\eins_4, & |m|>8a^2. \end{array}\right.
\end{equation}
Again we would obtain always two solutions but only one is accessible in the thermodynamic limit due to the infinitesimal increment $i L\varepsilon$.

\section{Density of $\rho_5(m-1,\lambda=0, a\to 0)$}
\label{app:a=0}

In this appendix we consider the continuum limit of $\rho_5(m=0,\lambda= 0, a)$
and show that the convergence to the continuum limit is non-uniform.
The level density reduces in this situation to
\begin{eqnarray}
\rho_5(m=0,\lambda=0,a)&=&16a^4\bigl[2\Phi_{4}\mathcal{S}_{1,1} -2\Phi_{3}\bigl(\mathcal{S}_{2,2}-\mathcal{S}_{0,2}\bigl)
+4\Phi_{2}\bigl(\mathcal{S}_{3,3}-\mathcal{S}_{1,3})\nn\\
&&+2\Phi_{1}\bigl(4\mathcal{S}_{2,4}-4\mathcal{S}_{0,4}-\mathcal{S}_{2,2}+2\mathcal{S}_{0,2}-\mathcal{S}_{4,2}\bigl)
 +\Phi_{0}\bigl(\mathcal{S}_{5,1}- \mathcal{S}_{3,1}+8\mathcal{S}_{1,3} -4\mathcal{S}_{1,1}\bigl)\bigl]\nn\\
&&+4a^2\bigl[2\Phi_{1}\bigl(2\mathcal{S}_{0,2}-\mathcal{S}_{0,0}-\mathcal{S}_{2,0}\bigl)
+\Phi_{0}\bigl(\mathcal{S}_{1,1}+\mathcal{S}_{3,1}\bigl)\bigl].\label{rho5-result-a}
\end{eqnarray}
Thereby we used the relations~\eqref{rel-index} between positive and negative indices. Considering Eq.~\eqref{comp-int}, the compact integrals $\Phi_\mu$ vanish for $\mu$ an odd integer if the quark mass and the axial mass are zero. Hence we have
\begin{equation}
\rho_5(m=0,\lambda=0,a)=16a^4\bigl[2\Phi_{4}\mathcal{S}_{1,1}
+4\Phi_{2}\bigl(\mathcal{S}_{3,3}-\mathcal{S}_{1,3})
 +\Phi_{0}\bigl(\mathcal{S}_{5,1}-\mathcal{S}_{3,1}+8\mathcal{S}_{1,3} -4\mathcal{S}_{1,1}\bigl)\bigl]+4a^2\Phi_{0}\bigl(\mathcal{S}_{1,1}+\mathcal{S}_{3,1}\bigl).\label{rho5-result-b}
\end{equation}
Thus we have only non-compact integrals $\mathcal{S}_{\mu,\alpha}$ with the index $\alpha=1,3$ and $\mu$ odd and positive. The integral over $y$, see Eq.~\eqref{non-comp-int}, can be integrated for both indices $\alpha=1,3$,
\begin{equation}\label{non-comp-1-a}
\mathcal{S}_{\mu,1}=\frac{(-1)^{(\mu-1)/2}}{2\pi (8a^2)}\int_{-\infty}^\infty ds \frac{e^{-\mu s}}{\cosh 2s}\exp[-4 a^2 \cosh 2s],
\end{equation}
and
\begin{equation}\label{non-comp-3-a}
\mathcal{S}_{\mu,3}-\mathcal{S}_{\mu,1}=\frac{(-1)^{(\mu-1)/2}}{2\pi (8a^2)^2}\int_{-\infty}^\infty ds \frac{e^{-\mu s}}{\cosh^2 2s}\exp[-4 a^2 \cosh 2s].
\end{equation}
We can omit the exponential in the limit $a\to0$ if $\mu$ is small enough, i.e.
\begin{equation}\label{non-comp-1-b}
\mathcal{S}_{\mu,1}\overset{|a|\ll1}{\approx}\frac{(-1)^{(\mu-1)/2}}{2\pi (8a^2)}\int_{-\infty}^\infty ds \frac{e^{-\mu s}}{\cosh 2s}=\frac{(-1)^{(\mu-1)/2}}{32a^2\cos(\mu\pi/4)},\ {\rm for}\ |\mu|<2,
\end{equation}
and
\begin{equation}\label{non-comp-3-b}
\mathcal{S}_{\mu,3}-\mathcal{S}_{\mu,1}\overset{|a|\ll1}{\approx}\frac{(-1)^{(\mu-1)/2}}{2\pi (8a^2)^2}\int_{-\infty}^\infty ds \frac{e^{-\mu s}}{\cosh^2 2s}=\frac{(-1)^{(\mu-1)/2}\mu}{8 (8a^2)^2\sin(\mu\pi/4)},\ {\rm for}\ |\mu|<4.
\end{equation}
For larger $\mu$ the exponential becomes crucial since it guarantees the integrability. Before performing the other integrals let us see what remains to be calculated,
\begin{equation}
\rho_5(m=0,\lambda=0,a)\overset{|a|\ll1}{\approx}16a^4\left(\mathcal{S}_{5,1}-\mathcal{S}_{3,1}+\frac{1}{ 2^{11/2}a^4}\right)+4a^2\left(\frac{1}{2^{9/2}a^2}+\mathcal{S}_{3,1}\right).\label{rho5-result-c}
\end{equation}
Here we used the fact that $\Phi_\mu\propto |a|^\mu$ and $\Phi_0\approx1$ for $|a|\ll1$. Thus we have still to calculate the asymptotics of $\mathcal{S}_{3,1}$ and $\mathcal{S}_{5,1}$ which can be done by splitting the integrals,
\begin{eqnarray}
\mathcal{S}_{3,1}&=&-\frac{1}{2\pi (8a^2)}\int_{-\infty}^\infty ds \frac{e^{-3 s}}{\cosh 2s}\exp[-4 a^2 \cosh 2s]\nn\\
&=&\frac{1}{2\pi (8a^2)}\int_{-\infty}^\infty ds \frac{e^{ s}}{\cosh 2s}\exp[-4 a^2 \cosh 2s]-\frac{1}{\pi (8a^2)}\int_{-\infty}^\infty ds e^{-s}\exp[-4 a^2 \cosh 2s]\nn\\
&\overset{|a|\ll1}{\approx}&\frac{1}{2^{9/2}a^2}-\frac{K_{1/2}(4a^2)}{\pi (8a^2)}\nn\\
&\overset{|a|\ll1}{\approx}&\frac{1}{2^{9/2}a^2}-\frac{1}{2^{9/2}\sqrt{\pi} a^3}+\mathcal{O}(a^{-1})\label{non-comp-1-3}
\end{eqnarray}
and
\begin{eqnarray}
\mathcal{S}_{5,1}&=&\frac{1}{2\pi (8a^2)}\int_{-\infty}^\infty ds \frac{e^{-5 s}}{\cosh 2s}\exp[-4 a^2 \cosh 2s]\nn\\
&=&-\frac{1}{2\pi (8a^2)}\int_{-\infty}^\infty ds \frac{e^{-s}}{\cosh 2s}\exp[-4 a^2 \cosh 2s]+\frac{1}{\pi (8a^2)}\int_{-\infty}^\infty ds e^{-3s}\exp[-4 a^2 \cosh 2s]\nn\\
&\overset{|a|\ll1}{\approx}&-\frac{1}{2^{9/2}a^2}+\frac{K_{3/2}(4a^2)}{\pi (8a^2)}\nn\\
&\overset{|a|\ll1}{\approx}&\frac{1}{2^{9/2}a^2}+\frac{1}{2^{9/2}\sqrt{\pi} a^3}+\frac{1}{2^{13/2}\sqrt{\pi} a^5}+\mathcal{O}(a^{-1})\label{non-comp-1-5}
\end{eqnarray}
 which yields the limit~\eqref{rho5-result-d}.

\end{document}